\documentclass[natbib]{emulateapj}
\submitted{MNRAS, accepted}
\usepackage{graphicx}
\usepackage{rotating}
\usepackage{amsmath}
\usepackage[usenames]{color}
\usepackage[colorlinks=true, pdfstartview=FitV, linkcolor=OrangeRed, citecolor=blue, urlcolor= red]{hyperref}

\shortauthors{Oh et al.} \shorttitle{2D Bayesian automated tilted-ring fitting
of disk galaxies in large H{\sc i} galaxy surveys: {\sc 2dbat}}

\begin{document}

\newcommand{\C}{\ensuremath{\mathfrak{C}}}
\newcommand{\VNFW}{\ensuremath{V_{200}}} \newcommand{\Vh}{\ensuremath{V_h}}
\newcommand{\kms}{\ensuremath{\mathrm{km}\,\mathrm{s}^{-1}}}
\newcommand{\kmskpc}{\ensuremath{\mathrm{km}\,\mathrm{s}^{-1}}\,\mathrm{kpc}^{-1}}
\newcommand{\cm}{\ensuremath{\mathrm{cm}^{-2}}}
\newcommand{\kmsnospace}{\ensuremath{\mathrm{km}\,\mathrm{s}^{-1}}}
\newcommand{\barkms}{\ensuremath{\mathrm{km}\,\mathrm{s}^{-1}\,\mathrm{kpc}^{-1}}}
\newcommand{\kmsMpc}{\ensuremath{\mathrm{km}\,\mathrm{s}^{-1}\,\mathrm{Mpc}^{-1}}}
\newcommand{\etal}{et al.} \newcommand{\LCDM}{$\Lambda$CDM}
\newcommand{\MLmax}{\ensuremath{\Upsilon_{max}}}
\newcommand{\Lsun}{\ensuremath{\rm{{L}_{\odot}}}}
\newcommand{\Msun}{\ensuremath{\rm{{M}_{\odot}}}}
\newcommand{\mass}{\ensuremath{\rm{{\cal M}}}}
\newcommand{\magsq}{\ensuremath{\mathrm{mag}\,\mathrm{arcsec}^{-2}}}
\newcommand{\Lsundens}{\ensuremath{\rm L_{\odot}\,\mathrm{pc}^{-2}}}
\newcommand{\surfdens}{\ensuremath{\rm M_{\odot}\,\mathrm{pc}^{-2}}}
\newcommand{\cubedens}{\ensuremath{M_{\odot}\,\mathrm{pc}^{-3}}}

%\long\def\Ignore#1{\relax} \long\def\Comment#1{{\footnotesize #1}}

%-----------------------------------------------------% % Title and Affiliation

\title{2D Bayesian automated tilted-ring fitting of disk galaxies in large H{\sc
i} galaxy surveys: {\sc 2dbat}}

\author{Se-Heon Oh\altaffilmark{1,2,3}, Lister Staveley-Smith\altaffilmark{2,3}, Kristine Spekkens\altaffilmark{4}, Peter Kamphuis\altaffilmark{5} \&
B\"arbel S. Koribalski\altaffilmark{6}} \email{seheonoh@kasi.re.kr}

\altaffiltext{1}{Korea Astronomy and Space Science Institute, Daejeon 305-348, Korea}
\altaffiltext{2}{International Centre for Radio Astronomy Research (ICRAR),
University of Western Australia, 35 Stirling Highway, Perth, WA 6009, Australia}
\altaffiltext{3}{ARC Centre for All-sky Astrophysics (CAASTRO)}
\altaffiltext{4}{Department of Physics, Royal Military College of Canada, P.O. Box 17000, Station Forces, Kingston, Ontario, K7K 7B4, Canada}
\altaffiltext{5}{National Centre for Radio Astrophysics, TIFR, Ganeshkhind, Pune 411007, India}
\altaffiltext{6}{CSIRO Astronomy \& Space Science, Australia Telescope National Facility, PO Box 76, Epping, NSW 1710, Australia}

%%-----------------------------------------------------%
\begin{abstract}
We present a novel algorithm based on a Bayesian method for 2D
tilted-ring analysis of disk galaxy velocity fields.
Compared to the conventional algorithms based on a chi-squared minimisation
procedure, this new Bayesian-based algorithm suffers less from local minima of
the model parameters even with highly multi-modal posterior
distributions. Moreover, the Bayesian analysis, implemented via Markov Chain Monte Carlo
(MCMC) sampling, only requires broad ranges of posterior distributions of the
parameters, which makes the fitting procedure fully automated. This feature will be essential when performing
kinematic analysis on the large number of resolved galaxies expected to be detected in neutral hydrogen (H{\sc i})
surveys with the Square Kilometre Array (SKA) and its pathfinders. 
The so-called `2D Bayesian Automated Tilted-ring fitter' ({\sc 2dbat}) implements Bayesian fits of 2D
tilted-ring models in order to derive rotation curves of galaxies. We explore
{\sc 2dbat} performance on (a) artificial H{\sc i} data cubes built based on
representative rotation curves of intermediate-mass and massive spiral galaxies, and
(b) Australia Telescope Compact Array (ATCA) H{\sc i} data from the Local Volume H{\sc i}
Survey (LVHIS). We find that {\sc 2dbat} works best for well-resolved galaxies with
intermediate inclinations ($20^{\circ} < i < 70^{\circ}$), complementing three-dimensional
techniques better suited to modelling inclined galaxies.
\end{abstract} 

\section{Introduction} \label{intro}

Observational studies of mass distributions in disk galaxies provide a critical
clue to understanding their formation and evolution \citep{2001ARA&A..39..137S}.
This can be achieved by observing the motions of kinematic tracers in the
galaxies, such as stars, gas (H{\sc i}, H$\alpha$, CO etc.),
planetary nebulae (PNe), open and globular clusters etc., which are normally
gravitationally bound to their host systems
\citep{1980ApJ...238..471R,1987A&A...176...34H,1993ApJ...414..454C,2001ARA&A..39..137S,2008PhDT........32H}.
Compared to the other kinematic tracers, neutral hydrogen (H{\sc
i}) is in general more uniformly distributed in the disks of galaxies. It also has
a larger extent, typically several times the Holmberg radius {\rm
$R_{26.5}$} \citep{1994A&AS..107..129B} even at an H{\sc i} column density above $\geq10^{19}\,{\rm atoms}\,{\rm cm}^{-2}$ in irregulars and spirals \citep{1981AJ.....86.1791B,1981AJ.....86.1825B,1981A&A...102..134H}.
For this reason, H{\sc i} has been
	widely used as a tracer in kinematic studies of both resolved and unresolved
	galaxies, including analyses of the rotation curves of disk galaxies (e.g., \citealt{1978PhDT.......195B}), their angular momentum distributions, and the Tully-Fisher relation \citep{1977A&A....54..661T}.

The usefulness of H{\sc i} as a kinematic tracer in galaxy dynamics will be further enhanced by the upcoming Square Kilometre Array (SKA) pathfinders, such as the Australian SKA Pathfinder (ASKAP; \citealt{2008ExA....22..151J}),
APERTIF on the Westerbork Synthesis Radio Telescope (WSRT), and the South African
MeerKAT telescope. These will open a new golden age for H{\sc i}-related
science, providing unprecedented flow of high quality data in tandem with observations at other wavelengths.
For example, the `Widefield ASKAP L-band Legacy All -sky Blind surveY'
(WALLABY; \citealt{2012PASA...29..359K}) is a top-ranked ASKAP all-sky H{\sc
i} 21cm spectral line survey expected to detect up to 500,000 galaxies ($z<0.26$),
all spectrally resolved, including $\sim$5,000 well-resolved galaxies ($>5$ beams across the
major axis) out to 200 Mpc (\citealt{2012MNRAS.426.3385D}; see also \citealt{2015MNRAS.452.2680S} and \citealt{2016MNRAS.460.2143W}).
This affords the possibility of deriving
kinematic parameters and rotation curves for statistically meaningful samples
out to this distance. 
One of the main science goals of WALLABY and other widefield H{\sc i} surveys is
to provide kinematic parameters for large numbers of resolved galaxies for the first time.
Such data will provide
stringent observational constraints on the evolution of mass in and
around galaxies and the link between the effects of environment, mass distribution and
other fundamental galaxy properties like halo mass and angular momentum.

Several standard methods for deriving galaxy kinematics can be classified by the
dimension of data analyzed: (1) 1D spectroscopy (e.g.,
\citealt{2001MNRAS.323..285B,2001AJ....122.2381M} etc.); (2) 2D velocity fields
(e.g., \citealt{1974ApJ...193..309R}; \citealt{2006MNRAS.366..787K};
\citealt{Spekkens_2007}; \citealt{2010MNRAS.404.1733S}); (3) 3D spectral line data (e.g.,
\citealt{2007A&A...468..731J}; \citealt{2015MNRAS.451.3021D};
\citealt{2015AJ....150...92B}; \citealt{2017MNRAS.464...21P}). 
The 1D approach, despite being observationally straightforward, is most affected by
observational systematics which can lead to large systematic uncertainties in the derived kinematics.

Some observational systematics, including beam smearing and projection effects,
can be reduced in 3D approaches which use the full available information without any
compression. There are several publicly available codes for 3D fitting of
kinematic models to a data cube, such as TiRiFiC \citep{2007A&A...468..731J}, $^{\rm
3D}${\sc barolo} \citep{2015MNRAS.451.3021D}, GalPak$^{\rm 3D}$ \citep{2015AJ....150...92B},
and GALACTUS \citep{2017MNRAS.464...21P}. Of particular usefulness is their potential ability to
model kinematic asymmetries, peculiarities or extraplanar gas disk and warps even in
nearly edge-on or face-on galaxies (e.g.,
\citealt{2011A&A...526A.118H,2011ApJ...740...35Z,2013MNRAS.434.2069K}).

However, the higher degree of flexibility in 3D kinematic models and larger
number of free parameters to be fitted requires a significant amount of
processing time, and often makes the fitting procedure too sensitive to
inhomogeneous distributions of gaseous or stellar components in galaxies
\citep{2007A&A...468..731J}. In this respect, 2D methods where a 3D galaxy kinematic
model is projected onto the plane of an infinitely thin disk have an advantage
over the 3D approaches in terms of their relatively simple parameterization. For
example, for well-resolved galaxies with intermediate inclinations (e.g.,
$30^{\circ} < i < 70^{\circ}$), 2D methods are found to provide reliable fits
comparable to those from a 3D analysis and with lower computational expense
(\citealt{2015MNRAS.452.3139K}). 
In practice, 2D methods have been adopted to derive kinematic properties of galaxies from many H{\sc i} and optical studies, such
as WHISP\footnote{The Westerbork H{\sc i} Survey of Irregular and Spiral Galaxies} \citep{2002ASPC..276...84V}, FIGGS\footnote{Faint Irregular Galaxies GMRT Survey} \citep{2008MNRAS.386.1667B}, THINGS\footnote{The H{\sc i} Nearby Galaxy Survey} \citep{Walter_2008}, LITTLE
THINGS\footnote{Local Irregulars That Trace Luminosity Extremes, The H{\sc i}
Nearby Galaxy Survey} \citep{Hunter_2012}, LVHIS\footnote{The Local Volume H{\sc
i} Survey; \url{http://www.atnf.csiro.au/research/LVHIS}} (\citealt{Koribalski_2010}; Koribalski et al. submitted), VLA-ANGST\footnote{Very Large Array - ACS Nearby Galaxy Survey Treasury} \citep{2012AJ....144..123O}, and SAMI\footnote{The Sydney-AAO Multi-Object Integral-Field Spectrograph} \citep{2012MNRAS.421..872C}.
 
Owing to its efficient and reliable performance, the 2D method
will be also used as a standard tool for the
kinematic analysis of the resolved galaxies in WALLABY. However, existing 2D implementations
require time-intensive supervision on a galaxy-by-galaxy
basis, which is no longer feasible for a large number of
galaxies. To improve on this, we have been developing an automated pipeline that
applies either a 2D or 3D tilted-ring model, depending on the galaxy geometry (e.g.,
inclination) and data quality (e.g., S/N, angular resolution etc.). We refer the reader
to Fig. 1 in \cite{2015MNRAS.452.3139K} for a flow-chart of the WALLABY kinematic pipeline.
However, fitting algorithms based on a $\chi^{2}$ minimisation often
suffer in being unable to efficiently find the global minima of models with a large number of free
parameters. This results in lower accuracy, poorer error estimation, and creates difficulties in
automation.

In an effort to develop an automated pipeline for
deriving the kinematics of resolved galaxies in future SKA pathfinder
galaxy surveys, this paper describes a newly developed algorithm for 2D
tilted-ring analysis based on a Bayesian Markov Chain Monte Carlo (MCMC)
technique, which we call the 2D Bayesian Automated Tilted-ring fitter ({\sc
2dbat}\footnote{{\sc 2dbat} is downloadable from
\url{https://github.com/seheonoh/2dbat}}).
This better allows us to quantify the kinematic geometry of galaxy disks,
and derive high-quality rotation curves that can be used for mass modeling of
baryons and dark matter halos.
It is anticipated that {\sc 2dbat} and the `Fully Automated TiRiFiC' ({\sc fat})
algorithm described by \cite{2015MNRAS.452.3139K} will form the backbone of the
WALLABY kinematic analysis pipeline.

The structure of this paper is as follows. The conventional way of performing a
2D tilted-ring analysis and its limitations are discussed in
Section~\ref{The_previous_approach}. A new 2D tilted-ring fitting algorithm
based on a Bayesian MCMC method is described in
Section~\ref{2D_TR_Bayesian_MCMC}, followed by a description of the software
which implements the algorithm in Section~\ref{The_software}. A performance
test of the software using both artificial and sample galaxies from Australia
Telescope Compact Array (ATCA) observations are discussed in
Section~\ref{performance_test}. Lastly, the main results of this paper and
conclusions are summarized in Section~\ref{concolusions}.

\section{The standard approach}\label{The_previous_approach}

\subsection{2D tilted-ring models} \label{2D_tilted_ring_model} Since its first
introduction by \cite{1974ApJ...193..309R} aimed at describing the systematic
distribution of H{\sc i} in disk galaxies, 2D tilted-ring analysis has been
widely used as a standard tool for deriving galaxy rotation curves
and investigating large and small scale kinematic structures and the properties of
gaseous components in and around galaxies
\citep[etc.]{1978PhDT.......195B,deBlok_1997,deBlok_2008,Oh_2011a}.

This approach models a galaxy's disk with a set of concentric ellipses, each with its own
kinematic centre ($x_{\rm C}$, $y_{\rm C}$), systemic velocity ($v_{\rm SYS}$),
position angle ($\phi$),
inclination ($i$), radial expansion velocity ($v_{\rm EXP}$) and rotation velocity
($v_{\rm ROT}$). The line-of-sight (LOS) velocity $v_{\rm LOS}({x,y})$ of the disk at a sky position of
($x, y$) is given by \citep{1974ApJ...193..309R,Begeman_1989}:
\begin{align}\label{eq:1}
v_{\rm LOS}(x, y) &= v_{\rm SYS} \nonumber \\
                &+ \sin i \{v_{\rm ROT}(r) \cos\theta + v_{\rm EXP}(r) \sin\theta\}
\end{align}
\noindent where $v_{\rm ROT}$ is the rotational velocity, $v_{\rm EXP}$ is the
expansion velocity and $v_{\rm SYS}$ is the systemic velocity of
the disk. The sky position ($x$, $y$) in a rectangular coordinate system can
be converted to ($r$, $\theta$) in a polar coordinate system using following
relations \citep{Begeman_1989}, \begin{equation}\label{eq:2}
	\cos\,\theta = \frac{-(x-x_{C})\,{\sin}\,\phi
+ (y-y_{C})\,{\cos}\,\phi}{r}, \end{equation}
\begin{equation}\label{eq:3} {\sin}\,\theta
	= \frac{-(x-x_{C})\,{\cos}\,\phi - (y-y_{C})\,
{\sin}\,\phi}{r\,{\cos}\,i}, \end{equation} \noindent
where $r$ is the radial distance from the centre ($x_{\rm C}$, $y_{\rm C}$) and $\theta$ is
the azimuthal angle measured counter-clockwise from the major axis in the plane
of the disk. As per standard convention, $\phi$ is the angle measured counter-clockwise
from the north to the semi-major axis of the receding half of the disk.
By fitting the 2D tilted-ring model to a velocity field extracted from
spectral line observations, the parameters are derived for each ring which are
then used to construct a model velocity field of the disk.

\subsection{Fitting procedures and limitations}
\label{Fitting_procedures_and_limitations} One of the publicly available
software implementations of the 2D tilted-ring approach is {\sc rotcur} in GIPSY which
has been widely used for deriving rotation curves of resolved galaxies,
particularly from H{\sc i} observations \citep{Begeman_1989}.  Based on
a least-squares fitting algorithm, {\sc rotcur} finds the best fit of a 2D
tilted-ring model to a given velocity field by minimising the smallest velocity
residuals. 

In general {\sc rotcur} is implemented in a heavily supervised manner on a galaxy-by-galaxy basis.
The user must guide the fit through the available parameter space,
typically adopting an approach like the one as described in \cite{Oh_2009_phd}:
(1) Estimation of initial values of
tilted-ring parameters ($x_{\rm C}$, $y_{\rm C}$, $v_{\rm SYS}$, $\phi$, $i$,
$v_{\rm EXP}$, $v_{\rm ROT}$): for the
geometrical parameters ($x_{\rm C}$, $y_{\rm C}$, $\phi$, $i$), ellipse fits can be performed
on either the velocity field itself or other moment maps (e.g., moment 0 and 2)
as well as ancillary optical or infrared images. Initial
estimates of the kinematic parameters ($v_{\rm SYS}$, $v_{\rm EXP}$, $v_{\rm
ROT}$) can be approximated after inspection of the velocity field.  (2) Determination of the kinematic centre
and systemic velocity ($x_{\rm C}$, $y_{\rm C}$, $v_{\rm SYS}$): in principle, all the ring parameters are
allowed to vary radially in the tilted-ring analysis. However, in practice,
constant representative kinematic centre position and systemic velocity are
often adopted. The radially averaged values of the
parameters can be derived from an initial fitting of the model to the velocity
field made with all ring parameters free.  (3) Derivation of the kinematic
position angle and inclination ($\phi$, $i$): a fit can be made after fixing the
derived ($x_{\rm C}$, $y_{\rm C}$, $v_{\rm SYS}$) except for the other ring parameters.  Unlike the
kinematic centre and systemic velocity ($x_{\rm C}$, $y_{\rm C}$, $v_{\rm
SYS}$), $\phi$ and $i$ often
vary with galaxy radius due to various dynamical structures present in galaxies,
such as bars and warps which could be modeled by radial variations of $\phi$ and
$i$, respectively (e.g., \citealt{2000ApJ...533..850S}). Assuming that any radial
variation in $\phi$ and $i$, if it exists, is more or less continuous, not showing
abrupt jumps or drops over the radius of a galaxy, we should be able to fit a simple analytic
function \-- for example a low-order polynomial can be used to model the initial tilted-ring
fit results. The parameters of these polynomials are iterated
consecutively until the mean differences between the successive models or single values of the ring
parameters are less than the limits provided.  (4) Derivation of the final rotation
curves: in the last step, after fixing all the ring parameters except for
$v_{\rm ROT}$
with the derived single values of ($x_{\rm C}$, $y_{\rm C}$, $v_{\rm SYS}$) and
models ($\phi$, $i$), we
perform the fitting and derive the final rotation curves.

The 2D tilted-ring analysis based on a least-squares fitting algorithm is often
sensitive to initial estimates of the ring parameters, and gets trapped in local
minima. In addition, models for $\phi$ and $i$ are usually derived manually,
and are dependent on subjective model choices.  Consequently, this requires the user to
monitor the fit quality, making it difficult to fully automate the
fitting procedure. It is therefore not desirable to use 2D
tilted-ring fitting algorithm in such a conventional way for the
kinematic analysis of a large number of galaxies.

\section{Automated 2D tilted-ring fitting of disk galaxies in a Bayesian
framework}\label{2D_TR_Bayesian_MCMC}

\subsection{A new algorithm} \label{A_new_algorithm}

In an effort towards the automated kinematic analysis of detections from large
H{\sc i} galaxy surveys, we present a novel
algorithm which enables us to perform robust 2D tilted-ring
analysis in a fully automated manner. In this Section, we describe our new
approach based on a Bayesian MCMC technique.

As given in Eqs.~\ref{eq:1}, ~\ref{eq:2} and ~\ref{eq:3}, $\phi(r)$ and
$i(r)$ are needed when deriving the LOS model velocity
at a projected sky position ($x, y$). Specifically, Eqs.~\ref{eq:2} and ~\ref{eq:3} imply that:
\begin{align}\label{eq:4}
r &= \Biggl[\biggl\{-(x-x_{C}) \, \sin\phi + (y-y_{C}) \, \cos\phi\biggr\}^2 \nonumber \\ 
 &+ \biggl\{\frac{(x-x_{C}) \, \cos\phi\ + (y-y_{C}) \, \sin\phi}{\cos i} \biggr\}^2\Biggr]^{\!1/2}.
\end{align}
If $\phi$ and $i$ are independent of $r$, the latter can be directly derived from
Eq.~\ref{eq:4}. However, if not, adequate
functional forms that provide a sufficient approximation to the radial
variations of $\phi$ and $i$ should be assumed. As discussed earlier, kinematic
$\phi$ and $i$ can vary with galaxy radius due to dynamical structures in
galaxies including lopsidedness, warps, bars, spiral arms, and non-circular motions.
The combined effect of such structures tends to result in random
variations of $\phi$ and $i$ which are not necessarily described by any specific
functional form. To remove any unphysical discontinuities of $\phi$ and $i$ and
regularise their radial variations, we use the {\it basis spline}
\citep{1978pgts.book.....D}, also called the `B-spline'. This is a piecewise radial polynomial function of degree $n$ 
where the order $n$ is less than the number of
rings in the tilted-ring model. The radial extent of the galaxy is broken up into some
number of intervals where each interval has two endpoints, called `breakpoints'.
For continuity and smoothness, these breakpoints are converted to
`knots' which constitute a knot vector 
\begin{equation} \label{eq:5} \displaystyle t = \{t_{0}, t_{1}, ...,
    t_{n+k-1}\},
\end{equation} \noindent where $n$ is the number of basis splines of order $k$.
The $n$ B-splines are defined by

\begin{eqnarray} \label{eq:6}
    \displaystyle B_{m,1}(r) &=& \left\{\begin{array}{rl} 1 & \qquad t_{m} \le r
< t_{m+1} \\ 0 & \qquad \rm{otherwise}
\end{array} \right.
\end{eqnarray}

\begin{eqnarray} \label{eq:7}
B_{m,k}(r) &=& \frac{r-t_{m}}{t_{m+k-1} - t_{m}}B_{m,k-1}(r) \nonumber \\
					&+& \frac{t_{m+k} - r}{t_{m+k} - t_{m+1}} B_{m+1,k-1}(r)
\end{eqnarray} \noindent where $m=0, 1, ... n-1$. Constant, linear, quadratic,
and cubic B-splines are given by $k=1, 2, 3$ and $4$, respectively.  The models
of $\phi$ and $i$ used in the new algorithm are given by expanding the B-spline functions
as follows, \begin{equation} \label{eq:8} \phi(r) = \sum_{l=1}^{U}
c^{\phi}_{u}B^{\phi}_{l, k}(r),  \end{equation} 
\begin{equation} \label{eq:9}
	i(r) = \sum_{m=1}^{V} c^{i}_{v}B^{i}_{m, k}(r), \end{equation} \noindent where $U$ and $V$
are the numbers of B-splines, and $c^{\phi}_{m}$ and $c^{i}_{m}$
are the coefficients of the B-splines for $\phi$ and $i$, respectively.
Similarly, the expansion velocity, $v_{\rm EXP}$, can also be modeled by the expansion
of $W$ B-spline functions,
\begin{equation} \label{eq:10}
v_{\rm EXP}(r) = \sum_{n=1}^{W} c^{v_{\rm EXP}}_{w}B^{v_{\rm EXP}}_{n,k}(r).
\end{equation}

If the models of kinematic $\phi$ and $i$ given in Eq.~\ref{eq:8} and \ref{eq:9}
are inserted into Eq.~\ref{eq:4}, the deprojected galaxy radius at a sky position
of ($x, y$) is given as follows,
\begin{equation} \label{eq:11}
r = r(x, y, x_{\rm C}, y_{\rm C}, c^{\phi}, c^{i}).
\end{equation}
This is a non-linear equation which can be solved numerically
given the parameters using a Newton-Rapson, bisection, false position or Brent
method \citep{numerical_recipe}.

Next, for the purpose of ensuring continuity of $\phi(r)$ and $r$,
we need to assume a model rotation velocity $v^{\rm MODEL}_{\rm ROT}(r)$ at the derived
galaxy radius $r$ to construct a 2D model velocity field $v^{\rm MODEL}$($x,
y$). For this, we use the rotation velocity of the Einasto halo model
\citep{1965TrAlm...5...87E,1968PTarO..36..414E,2010MNRAS.402...21N}. This empirical model has been
widely adopted for taking the density profiles of halos not only in \LCDM\
simulations but also in observations (e.g., \citealt{2004MNRAS.349.1039N,2011AJ....142..109C}). Compared to
both the pseudo-isothermal (e.g., \citealt{1991MNRAS.249..523B}) and Navarro, Frenk \& White (NFW; \citealt{NFW_1996}) halo models, which have two free parameters and which are usually used for a disk-halo
decomposition of disk galaxies (\citealt{1985ApJ...299...59C,1991MNRAS.249..523B,1994AJ....107..543M,deBlok_1997,deBlok_2008} etc.), it often provides better descriptions of the density profiles
by having a third parameter, the so-called {\it Einasto} index $n$
which quantifies the degree of curvature of the profile \citep{2004MNRAS.349.1039N,2005MNRAS.358.1325C,2005MNRAS.362...95M}. In addition, it also has been used to describe a wide range of rotation curve shapes of galaxies from bulge-less dwarfs to bulge-dominated disk galaxies \citep{2010MNRAS.406.2493G,2011AJ....142..109C}.

The Einasto mass profile is given as,
\begin{equation} \label{eq:12}
	M_{\rm E}(r) = 4\pi n r^{3}_{-2} \rho_{-2} e^{2n} (2n)^{-3n} \gamma
	(3n, \frac{r}{r_{-2}}),
\end{equation}
\noindent where $r$ is the galaxy radius, and $\rho_{-2}$ is the density at the radius $r_{-2}$ where the
logarithmic density slope is $-2$. $\gamma$ is the lower incomplete gamma function
given by,
\begin{equation} \label{eq:13}
	\gamma (3n, x) = \int_{0}^{x} dt\,e^{-t} t^{3n-1}.
\end{equation}
Assuming spherical symmetry of the model, the Einasto halo rotation curve can be
computed by
\begin{align}
	v_{\rm E}(r) &= \sqrt{\frac{GM_{\rm E}(r)}{r}} \nonumber \\
					  &= \sqrt{4\pi G n \frac{r^{3}_{-2}}{r} \rho_{-2} e^{2n}
(2n)^{-3n} \gamma (3n, \frac{r}{r_{-2}})},
	\label{eq:14}
\end{align}
\noindent where $G$ the gravitational constant.

Lastly, the model LOS velocity $v^{\rm MODEL}$($x, y$) at a projected sky
position of ($x$, $y$) is given by inserting the model $\phi$, $i$, $v_{\rm EXP}$ and
$v_{\rm E}$ together into Eq.~\ref{eq:1} as follows,
\begin{equation}
	\label{eq:15} 
	\displaystyle v^{\rm MODEL} = v^{\rm MODEL}(x_{\rm C}, y_{\rm
	C}, {v_{\rm SYS}}, c^{\phi},
c^{i}, c^{v_{\rm EXP}}, n, r_{-2}, \rho_{-2}). \end{equation}

This 2D model velocity field defined with given tilted-ring parameters
is fitted to the observed velocity field of a galaxy.
Unlike the conventional 2D tilted-ring fit, which is done `ring-by-ring',
this new method fits all the available pixels of a given velocity at the
same time. From this, the tilted-ring parameters given in Eq.~\ref{eq:15} are derived. 
In the last step, after fixing all of the ring parameters derived except for
$v_{\rm ROT}$ and $v_{\rm EXP}$ (usually set to zero), the 2D tilted-ring model in Eq.~\ref{eq:1}
is fitted again `ring-by-ring' to each ellipse defined with the derived ring
parameters, and the final $v_{\rm ROT}$ and $v_{\rm EXP}$ (if fitted) are derived.
Therefore, the Einasto model only provides a mechanism for finding a smooth form for the variation
of inclination and position angle. The final rotation curve is not be an Einasto profile.

\subsection{Bayesian model fitting} \label{Bayesian_model_fitting}

We use a Bayesian MCMC technique to efficiently sample the high-dimensional
parameter space of the proposed 2D tilted-ring model given in Eq.~\ref{eq:15},
and fit it to all the available data points of a given velocity field at the
same time. Consider a case where model parameters $\Theta$ are estimated
by applying a statistical model that is described by a probability density
function $p(y|\Theta)$ to the observed data, $y: \{y_{1}, y_{2}, ...\}$.  According
to probability theory, Bayesian parameter estimation deals with the model
parameters $\Theta$ as random variables whose distributions are defined with
information available about the data. By using such information, the so-called
priors of the model parameters, uncertainties in the model are taken into
consideration \citep{Sivia_2006}. In a Bayesian framework using MCMC techniques,
the final model parameters are expressed as probability distributions. 

In general Bayesian parameter estimation consists of three main parts: (1) the
probability distribution of model parameters which is referred to as the `prior
distribution', $p(\Theta|g)$. The prior distribution represents the observer's
beliefs about the model parameters; (2) the statistical function, the so-called
`likelihood function', $p(y|\Theta, g)$ which is the probability of the data
given the model parameters; (3) the posterior distribution of the model
parameters given the data ($y$) and the model to fit ($g$) which is the product
of the prior distribution and the likelihood function:
\begin{equation} \label{eq:16} p(\Theta|y, g) = \frac{p(y|\Theta, g) \times
p(\Theta|g)}{p(y|g)}, \end{equation} \noindent where $p(y|g)$ is a normalization
factor called the evidence. 

In order to make a Bayesian fit of the proposed 2D titled-ring model given in
Eq.~\ref{eq:15} to a velocity field, we use a log-likelihood function for a
Student-t distribution:
\begin{align}\label{eq:17}
{\rm log}\,L &= \sum_{t=1}^{N}w_{t}\,{\rm log}\biggl[\frac{\Gamma(\frac{\nu + 1}{2})}{\sqrt{\pi(\nu-2)}\Gamma(\frac{\nu}{2})}\biggr] \\ \nonumber
&- \frac{1}{2}\sum_{t=1}^{N}w_{t}\,{\rm log}\sigma_{t}^{2}  \\ \nonumber
&- \frac{\nu+1}{2}\sum_{t=1}^{N}w_{t}\,{\rm log}\biggl[1+\frac{\epsilon_{t}^{2}}{\sigma_{t}^{2}(\nu-2)}\biggr]
\end{align}

\noindent where $\epsilon_{t} = v_{t}^{\rm LOS} - v_{t}^{\rm MODEL}$, $N$ is the number of total data points to fit,
$\nu$ ($>$ 2) is the number of degrees of freedom, and $\Gamma$ is the gamma function. The value of $\sigma_{t}$, which is a
free parameter, sets the overall scaling of the distribution.
The Student-t distribution can have a wider wing and lower peak than the normal distribution.
It approaches the normal distribution as $\nu$ increases. 
To make the fit of our 2D tilted-ring model as insensitive to any outliers as possible we use a small value $\nu=3$
in this work.

The weight $w_{t}$ is mainly for compensating for the smaller contribution of the pixels near the kinematic centre than the outer region in the 2D analysis. It also includes the effect of the LOS velocity error, $v^{\rm LOS-error}$.
In addition, the pixels in a ring are weighted by $\rm |cos(\theta)|^{\rm q}$ to give more weight around the major axis in the fit where $\rm q=0, 1$ or $2$. We adopt:
\begin{equation} \label{eq:18}
w_{t} = \frac{l_{\rm outermost}}{l_{t}} \times \frac{|\cos(\theta_{t})|^{\rm q}}{v^{\rm LOS-error}_{t}},
\end{equation}
\noindent where $l_{\rm outermost}$ and $l_{t}$ are the perimeters of the outermost ellipse and the one where the pixel ($t$) lies which are defined by the derived ring parameters ($x_{\rm C}$, $y_{\rm C}$, $\phi$ and $i$).

In Eq.~\ref{eq:16}, the evidence $p(y|g)$ can be calculated using the law of
total probability given by \begin{equation} \label{eq:19} p(y|g)
= \sum_{i} p(y|\Theta, g_{i})\,p(\Theta|g_{i}). \end{equation} In a Bayesian
analysis, calculating the evidence is the most time consuming step, and MCMC
techniques are often used to sample the model parameters from the posterior
distribution. This allows us to estimate the evidence efficiently. 
There are several existing MCMC sampling algorithms, such as Gibbs
sampler \citep{Geman:1984:SRG:2286442.2286617,Casella/George:1992},
Metropolis-Hastings \citep{metropolis53,Hastings:1970:MCS}, and nested sampling
\citep{Skilling04,sivia2006data}. Conventional samplers, such as Metropolis-Hastings
and Gibbs sampling often reach convergence to
stationary solutions very slowly if the posterior distribution is highly
multi-modal.  However, nested sampling has been found to be robust and
efficient in parameter estimation and model selection even with highly
multi-modal posteriors \citep{2008MNRAS.384..449F,2009MNRAS.398.1601F}.
Moreover, it has been found to be efficient in calculating the evidence,
allowing posterior inference as a by-product \citep{Skilling04}. This enables us
to perform Bayesian parameter estimation and model selection simultaneously.

We use the {\sc multinest} library which implements the nested
sampling algorithm \citep{2008MNRAS.384..449F,2009MNRAS.398.1601F}.
It has been successfully applied as a robust Bayesian inference tool for several
problems in particle physics and astrophysics, such as particle physics
phenomenology (e.g., \citealt{2010PhRvD..81i5012A}), gravitational wave
astronomy (e.g., \citealt{2009CQGra..26u5003F}), exoplanet detection (e.g.,
\citealt{2011MNRAS.415.3462F}), and absorption line detection \citep{2012PASA...29..221A}.
We adopt {\sc multinest} as the Bayesian inference engine for {\sc 2dbat}.

\section{The software} \label{The_software}

{\sc 2dbat} performs the Bayesian fitting of the 2D tilted-ring model in
Eq.~\ref{eq:15} to velocity fields of galaxies via MCMC. 
We use a version where importance nested sampling (NIS) is supported (see
\citealt{2013arXiv1306.2144F} for the complete description of the algorithm).
One of the most important advantages of {\sc 2dbat} is that only
broadly defined ranges of the parameters are required for the priors,
which makes the fitting procedure fully automated. {\sc 2dbat} is written
in ANSI C, including additional libraries like {\sc multinest} \citep{2008MNRAS.384..449F,2009MNRAS.398.1601F}, CFITSIO \citep{1999ASPC..172..487P}, GNU Scientific Library (GSL)
and some routines from Numerical Recipes \citep{numerical_recipe}.
In the following Sections, we describe the main
layout of {\sc 2dbat} and its supplementary features for improving
the fit quality.

% Fig.1. Sampling of velocity field
\begin{figure*} \epsscale{1.0}
	\includegraphics[angle=0,width=1.0\textwidth,bb=50 270 520 720,clip=]{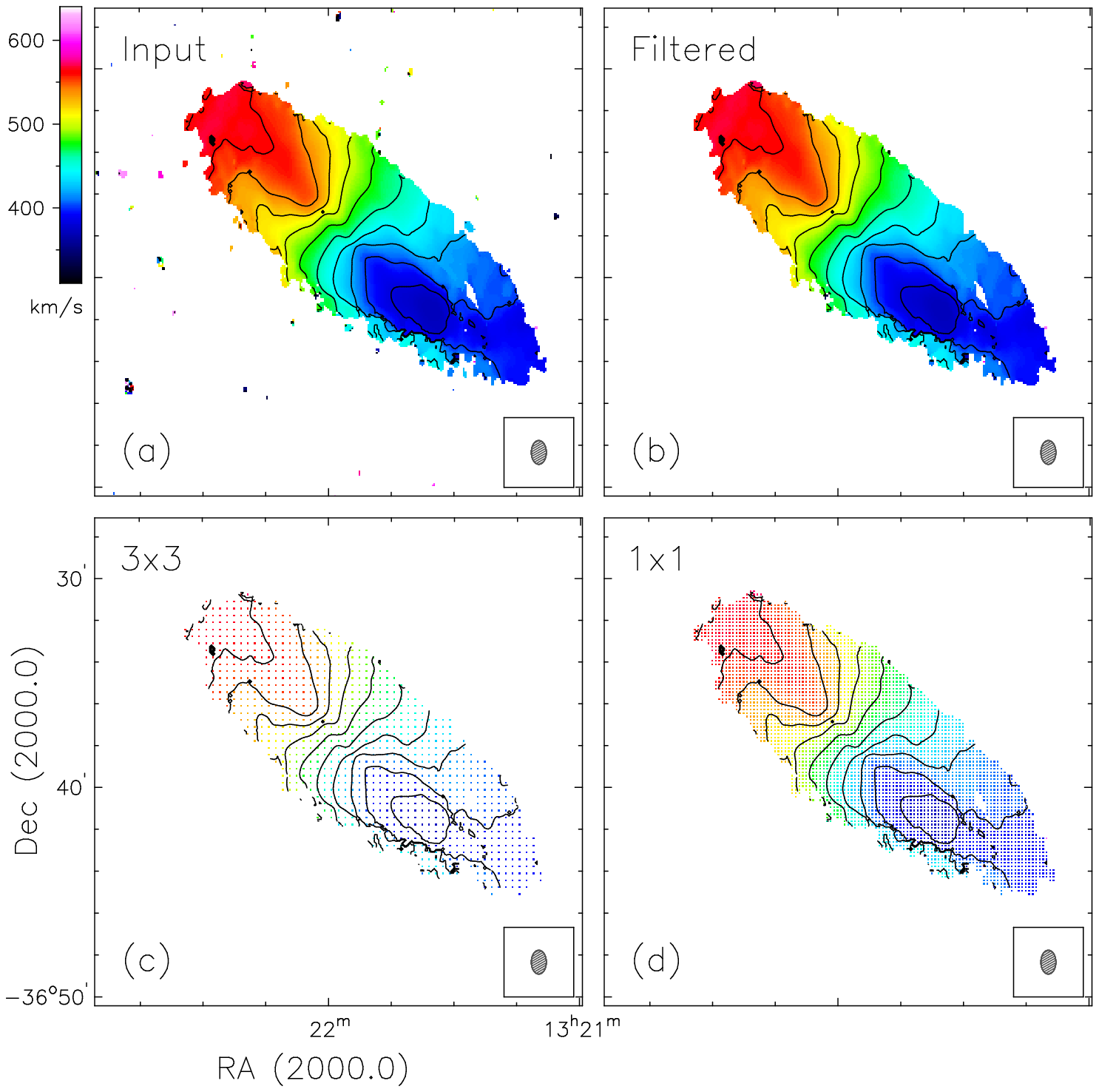} \caption{An example of the pixel sampling mode in {\sc 2dbat} (LVHIS galaxy NGC 5102): (a) non-filtered and non-sampled ATCA H{\sc i} Hermite $h_3$ velocity field. The contours are spaced by 20 \kms, and the synthesized beam is $68.23\,\arcsec\times43.00\,\arcsec$; (b) the largest connected velocity field passed through the connected-component labelling (CCL) algorithm described in Section~\ref{CCL}; (c) a sampled velocity field with a grid spacing of $3\times3$ pixels; (d) a sampled velocity field with a grid spacing of $1\times1$ pixels. The beam size is indicated by the ellipse in the bottom-right corner of each panel.
\label{Fig01}} \end{figure*}

\subsection{Main layout} \label{Layout} Through the following three main
steps, {\sc 2dbat} automatically extracts the ring parameters for the 2D
tilted-ring model in Eq.~\ref{eq:15} given a degree of regularisation.

\subsubsection{Mask outlying pixels in the input 2D maps}\label{CCL}
Outlying pixels that are sporadically distributed in the velocity field and thus have
very low likelihood affect the Bayesian fitting in a way that increases the
multimodality of the posterior distributions of the parameters
\citep{dawi:1973}. Despite their insignificant contribution to the global
kinematics of a galaxy, these outliers often result in larger uncertainties and
longer execution time in the fitting process. It is therefore desirable to
remove such outliers to minimise their impact on the execution time and the fit
quality. To this end, we use the connected-component labelling (CCL) algorithm
which finds the largest connected area in a 2D image by masking isolated pixels
 \citep{Cormen:2009:IAT:1614191}. Using a two-pass procedure, the
largest connected region is extracted: (a) in the first pass, scanning from
left-to-right and top-to-bottom of the velocity field, successive integers in
increasing order starting from number one are temporarily assigned to pixels
depending on their connectivity of neighbour pixels. (b) in the second pass,
the temporary labels are replaced by the smallest label of its equivalence
class, and the connected area with the smallest label is found. We refer to
\cite{Cormen:2009:IAT:1614191} for a full description of the algorithm. An
example of the extracted largest connected area of the ATCA Hermite $h_3$
velocity field of the LVHIS galaxy NGC 5102 (HIPASS J1321-36) (Koribalski et al. submitted)
via the CCL algorithm is shown in the panel (b) of Fig.~\ref{Fig01}.

% Fig.2. Flow chart of ROTCUR procedure
\begin{figure*} \epsscale{1.0}
\includegraphics[angle=0,width=1.0\textwidth,bb=-5 -10 640
400,clip=]{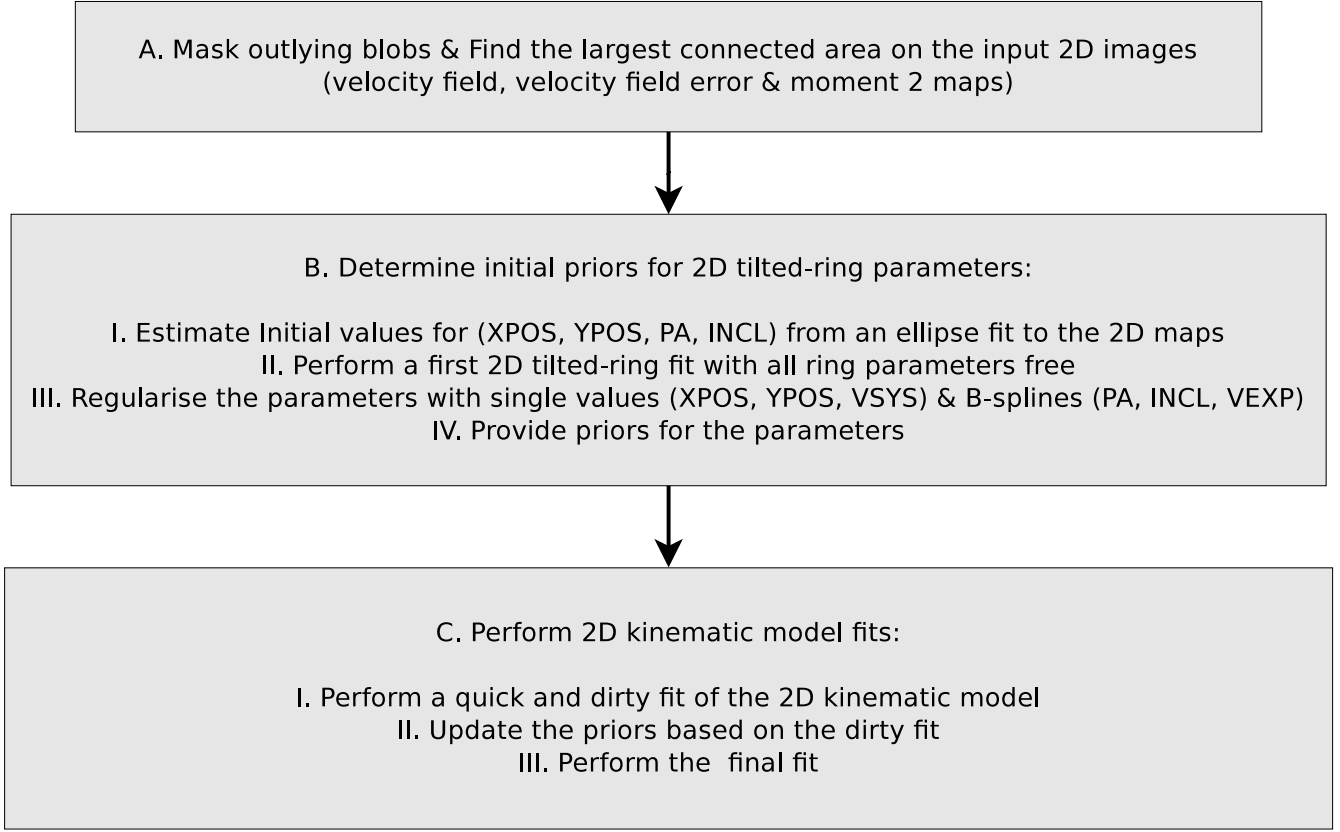} \caption{The main layout of {\sc 2dbat}. See Section~\ref{Layout} for more details.
\label{Fig02}}
\end{figure*}

\subsubsection{Optimal range of priors for the ring parameters}
{\sc multinest} requires several input parameters, such as the number of live points $N$, the sampling
efficiency $f$, the tolerance level $\epsilon$, and the prior distributions,
which all influence the robustness and efficiency of the 2D tilted-ring fitting.
The fit quality and the execution time are particularly dependent on the assumed
prior distributions of the ring parameters. The prior ranges should cover all
the possible values for each of the parameters while being narrowed down in an
optimal way. Bayesian fit with optimal prior distributions does not only reduce
the execution time but it also avoids any overshooting, leading to more robust
results.

We fit an ellipse to the input velocity field to derive a {\it rough} parameterisation
of the galaxy's disk on which the estimation of initial values for the
geometrical parameters, such as centre position ($x^{\rm ellipse}_{\rm C}$,
$y^{\rm ellipse}_{\rm C}$), position angle ($\phi^{\rm ellipse}$), inclination
($i^{\rm ellipse}$) and the length of semi-major axis ($L$$_{\rm smx}$) is
based. The ellipse fit can also be made to the other moment maps (i.e.,
moment 0 or 2) if needed. We then perform a first tilted-ring analysis with all ring
parameters allowed to vary freely by fitting a model LOS velocity given in Eq.~\ref{eq:1} to the
successive ellipses defined with the ellipse fit above, sub-divided into rings of a 
minimum width of one beam. The fit results are used for the
regularisation of the ring parameters on which their uniform prior distributions
are based. We regularise the ring parameters as a function of galaxy radius by
performing error-weighted averaging of ($x_{\rm C}$, $y_{\rm C}$, $v_{\rm SYS}$) and fitting of the
B-spline functions ($0^{\rm th}$, $1^{\rm st}$, $2^{\rm nd}$ or $3^{\rm rd}$
order) given in Eqs.~\ref{eq:8}, \ref{eq:9} and \ref{eq:10} for $\phi$, $i$ and
$v_{\rm EXP}$, respectively. The regularisation is done by only using the rings
satisfying all the following criteria:
\begin{equation}\label{eq:20}
\overline{\delta^{\rm parameter}_{m}}
	- 5\sigma^{\rm parameter}_{m} < \delta^{\rm parameter}_{m} < \overline{\delta^{\rm parameter}_{m}}
+ 5\sigma^{\rm parameter}_{m}, \end{equation}
\noindent where $\delta^{\rm parameter}_{m}$ are the statistical uncertainties of the
fitted parameters (i.e., $x_{\rm C}$, $y_{\rm C}$, $v_{\rm SYS}$, $\phi$, $i$, and $v_{\rm EXP}$) in each ring ($m=1,...,n$ where
$n$ is the number of the rings) derived from the first tilted-ring analysis
above, and $\overline{\delta^{\rm parameter}}$ and $\sigma^{\rm parameter}$ are
their mean and standard deviation values. We note that this outlier removal process does not
much affect the final Bayesian fit results while just providing initial estimates of
the priors.

\subsubsection{Performing the Bayesian fit of the tilted-ring model}
First, we carry out a dirty but quick Bayesian fit of 2D tilted-ring model with
a smaller number of live points (e.g., $N=50$) and less-conservative tolerance (e.g., 0.3) and
sampling efficiency (0.8) assuming the initial, conservative uniform priors of the ring
parameters given in Table~\ref{table:1}. These initial prior distributions are further
tuned in accordance with the results of the dirty fit.
We then apply the full Bayesian model fit with the given parameter setup
and the degree of the regularisation. Lastly, we derive the final rotation
velocity $v_{\rm ROT}$ by fitting the LOS model velocities to the receding, approaching
and both sides of the tilted-rings defined with the parameters from the full Bayesian
fitting. 

{\sc 2dbat} results include (1) an ascii text file containing the
rotation curves and the fitted ring parameters, (2) standard
posterior sample files by {\sc multinest}, (3) model velocity fields constructed
using the best fits of the 2D tilted-ring analysis, (4) residual maps between the
input and model velocity fields and (5) a weighted 2D error map of the
velocity field which is described in the following Section. A schematic
flowchart describing the main layout of {\sc 2dbat} is shown in Fig.~\ref{Fig02}.

%\begin{sidewaystable*}[!htbp]
\begin{table}
\caption{Initial uniform prior distributions of parameters}
\label{table:1}
\centering
\begin{tabular}{@{} lcc @{}}
\hline
\hline
Parameter    & Min &  Max \\
(1) & (2) & (3) \\
\hline
$x_{\rm C}$ & $x^{\rm TR}_{\rm C} - 0.5L_{\rm smx}$ & $x^{\rm TR}_{\rm C} + 0.5L_{\rm smx}$ \\
$y_{\rm C}$ & $y^{\rm TR}_{\rm C} - 0.5L_{\rm smx}$ & $y^{\rm TR}_{\rm C} + 0.5L_{\rm smx}$ \\ 
$v_{\rm SYS}$ & $v^{\rm TR}_{\rm SYS} - \sigma^{v_{\rm LOS}}$ & $v^{\rm TR}_{\rm SYS} + \sigma^{v_{\rm LOS}}$ \\
$c^{\phi}_{u} \{u=1,...U\}$ & $c^{\phi\_{\rm TR}}_{{\it u}} - 5\sigma^{c^{\phi\_{\rm TR}}}_{{\it u}}$ & $c^{\phi\_{\rm TR}}_{{\it u}} + 5\sigma^{c^{\phi\_{\rm TR}}}_{{\it u}}$ \\
$c^{i}_{v} \{v=1,...V\}$ & $c^{i\_{\rm TR}}_{{\it v}} - 5\sigma^{c^{i\_{\rm TR}}}_{{\it v}}$  & $c^{i\_{\rm TR}}_{{\it v}} + 5\sigma^{c^{i\_{\rm TR}}}_{{\it v}}$ \\
$c^{v_{\rm EXP}}_{w} \{w=1,...W\}$ & $c^{v_{\rm EXP\_TR}}_{{\it w}} - 5\sigma^{c^{v_{\rm EXP\_TR}}}_{{\it w}}$ & $c^{v_{\rm EXP\_TR}}_{{\it w}} + 5\sigma^{c^{v_{\rm EXP\_TR}}}_{{\it w}}$ \\
$n$ & 0 & 3$n^{\rm TR}$ \\
$r_{-2}$ & 0 & 3$r^{\rm TR}_{-2}$ \\
$\rho_{-2}$ & 0 & 3$\rho^{\rm TR}_{-2}$ \\
\hline
\end{tabular}
\begin{minipage}{85mm}
   \scriptsize{
{\bf (1):} Parameters of 2D tilted-ring model given in Eq.~\ref{eq:15}; 
{\bf (2)(3):} Default boundaries of the uniform priors of the parameters: Super-script `TR' indicates the values derived from the intial tilted-ring fit; $\rm L_{\rm smx}$ (the length of the semi-major axis); $\sigma^{v_{\rm LOS}}$ (1$\sigma$ of the $v_{\rm LOS}$ distribution); $n^{\rm TR}$, $r^{\rm TR}_{-2}$, and $\rho^{\rm TR}_{-2}$ are derived by fitting the Einasto rotation velocity given in Eq.~\ref{eq:14} to the initial rotation velocity $v_{\rm ROT}^{\rm TR}$. See Section~\ref{Layout} for more details.}
\end{minipage}
\end{table}

\subsection{Error estimation} \label{error_estimation} We adopt the standard
deviations of the posterior distributions of the parameters as their errors
except for $v_{\rm ROT}$. As discussed above, in the last step of the algorithm, only the
final rotation velocity $v_{\rm ROT}$ is fitted to the tilted-rings defined with the
other ring parameters derived from the last Bayesian fit.
Therefore, the uncertainties of the other ring parameters are not fully incorporated
in the standard deviation of $v_{\rm ROT}$ derived in the final fit. As discussed in
\cite{deBlok_2008} (see also \citealt{Swaters_1999}), such formal standard
deviations of $v_{\rm ROT}$ ($\rm \sigma^{{\it v}_{\rm ROT}}_{\it m}$ where $m$ is a ring
number from $1$ to $n$) do not represent the true physical uncertainties, and are usually
much smaller than the dispersions of LOS velocities along the rings. Following \cite{Swaters_1999}
and \cite{deBlok_2008}, {\sc 2dbat} also provides three types of uncertainties for $v_{\rm ROT}$:
(1) the error in $v_{\rm E}$, $\sigma_{v_{\rm E}}$, (2) $\sigma_{\rm asym}$ (a pseudo-$1\sigma$
uncertainty due to asymmetries as one fourth of the difference between the approaching and receding
side velocities; see \citealt{deBlok_2008}), and
(3) $\sigma_{\rm LOS}$ (the average velocity dispersion along the rings). $\sigma_{v_{\rm E}}$ is
associated with the errors of the Einasto halo profile, $n$, $r_{-2}$ and $\rho_{-2}$, which is given by

\begin{align}\label{eq:26}
\sigma_{v_{\rm E}} &= \Biggl[{(\sigma_{n}} \frac{\partial v_{\rm
E}}{\partial n})^{2} + (\sigma_{r_{\rm -2}} \frac{\partial v_{\rm
E}}{\partial r_{\rm -2}})^{2} + (\sigma_{\rho_{\rm -2}} \frac{\partial v_{\rm
E}}{\partial \rho_{\rm -2}})^{2} \\ \nonumber
&+ 2 \frac{\partial v_{\rm E}}{\partial n} \frac{\partial v_{\rm E}}{\partial r_{-2}}\sigma_{n r_{-2}}
+ 2 \frac{\partial v_{\rm E}}{\partial r_{-2}} \frac{\partial v_{\rm E}}{\partial \rho_{-2}}\sigma_{r_{-2}\rho_{-2}} \\ \nonumber
&+ 2 \frac{\partial v_{\rm E}}{\partial n} \frac{\partial v_{\rm E}}{\partial \rho_{-2}}\sigma_{n \rho_{-2}}\Biggr]^{\!1/2}
\end{align}

\noindent where $\sigma_{n}$, $\sigma_{r_{-2}}$, and $\sigma_{\rho_{-2}}$ are the standard deviations of
the fitted $n$, $r_{-2}$ and $\rho_{-2}$ of the Einasto halo rotation velocity from the Bayesian analysis.
$\sigma_{n r_{-2}}$, $\sigma_{r_{-2}\rho_{-2}}$, and $\sigma_{n \rho_{-2}}$ are the covariances between the parameters.
The full derivation of the error propagation for the Einasto halo model is given in the Appendix. 
From this, one can define the uncertainties in the rotation curves by adding either two of the above three uncertainties or even all of them in quadrature as a 'very' conservative error budget.

% Fig.3. an example of 2dbat result for a model galaxy
\begin{figure*} \epsscale{1.0}
\includegraphics[angle=0,width=1.0\textwidth,bb=105 172 520 604,clip=]{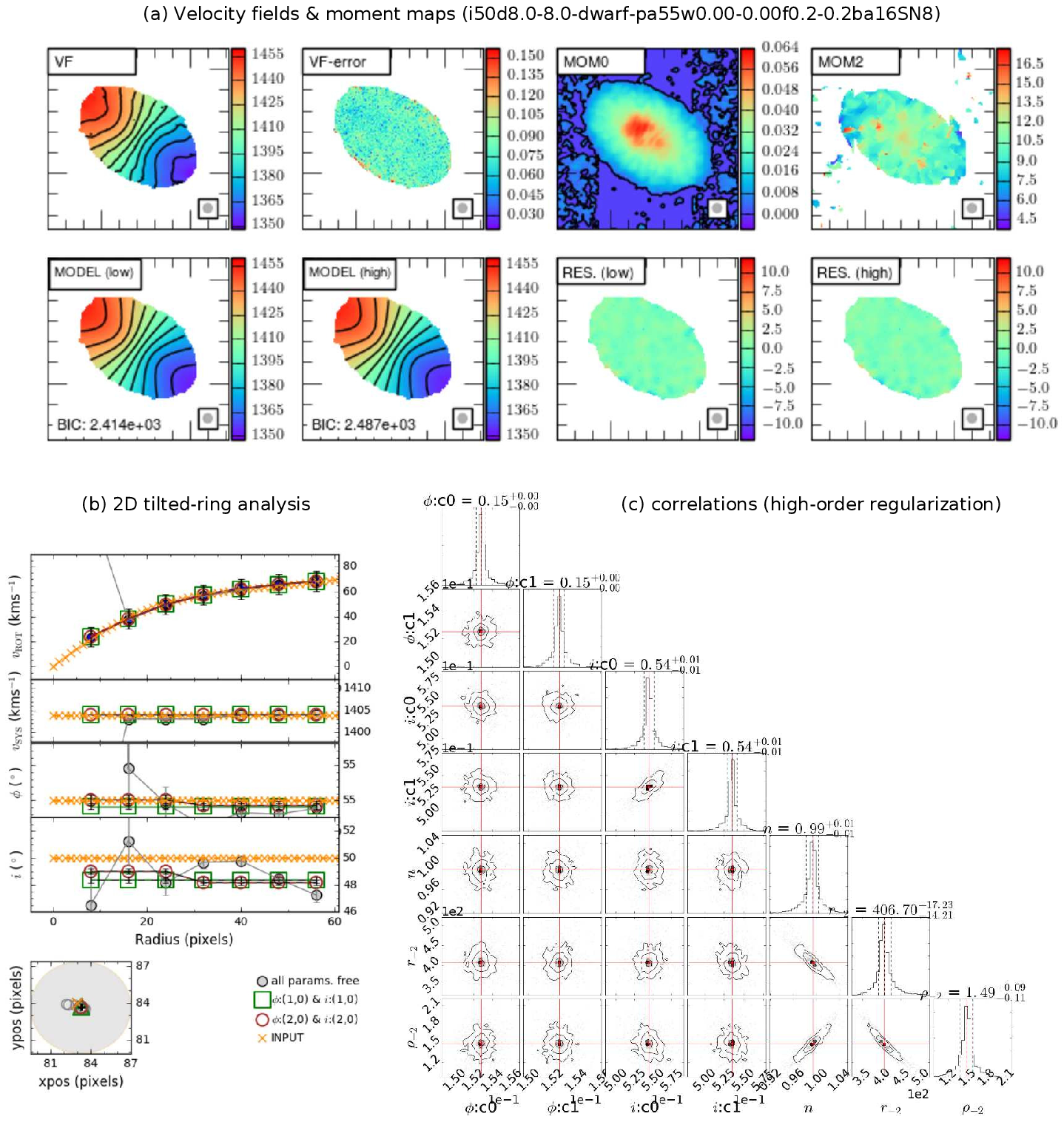}
\caption{{\sc 2dbat} analysis for an artificial {\it dwarf} galaxy:
{\bf (a)} hermite $h_3$ velocity field, error, moment maps, model and residual
velocity fields. Contours are spaced by 10 \kms\ in the velocity fields,
4 \kms\ in both the moment 2 and residual velocity fields, and 0.1 $\rm mJy\,beam^{-1}$ in the moment 0. The beam size is indicated by the ellipse
in the bottom-right corner of each panel.
The BIC values derived from the Bayesian fits for the model velocity fields are denoted in the panels of model velocity fields, respectively;
{\bf (b)} 2D tilted-ring analysis: the rotation curves derived using the Hermite
$h_3$ velocity field in the two regularisation modes (green
open squares: constant, brown open circles: higher-order B-splines, 
grey filled circles: fit results with all the ring parameters free,
orange cross mark: the input ring parameters used for contructing the model data cube). 
See Section~\ref{model_fit_results} for more details;
{\bf (c)} correlations of the ring parameters: the black
contours and histograms show the posterior constraints from the Bayesian
analysis. The best fits are indicated by red lines. 
\label{Fig03}}
\end{figure*}

%%%%%%%%%%%%%%%%%%%%%%%%%%%%%%%%%%%%%%%%%%%%%%%%%%%%%%%%%%%%%%%%%%%%
%%%%%%%%%%%%%%%%%%%%%%%%%%%%%%%%%%%%%%%%%%%%%%%%%%%%%%%%%%%%%%%%%%%%
%%%%%%%%%%%%%%%%%%%%%%%%%%%%%%%%%%%%%%%%%%%%%%%%%%%%%%%%%%%%%%%%%%%%
%%%%%%%%%%%%%%%%%%%%%%%%%%%%%%%%%%%%%%%%%%%%%%%%%%%%%%%%%%%%%%%%%%%%
%\section{{\sc 2dbat} fit results of artificial galaxies}
%: Fig.4. /mnt/g0/research/KASI/projects/wallaby/peter_cubes/Test_Galaxies/i50d8.0-8.0-NGC891s-pa55w0.00-0.00f0.2-0.2ba4SN8
\begin{figure*} \epsscale{1.0}
\includegraphics[angle=0,width=1.0\textwidth,bb=105 172 520 604, clip=]{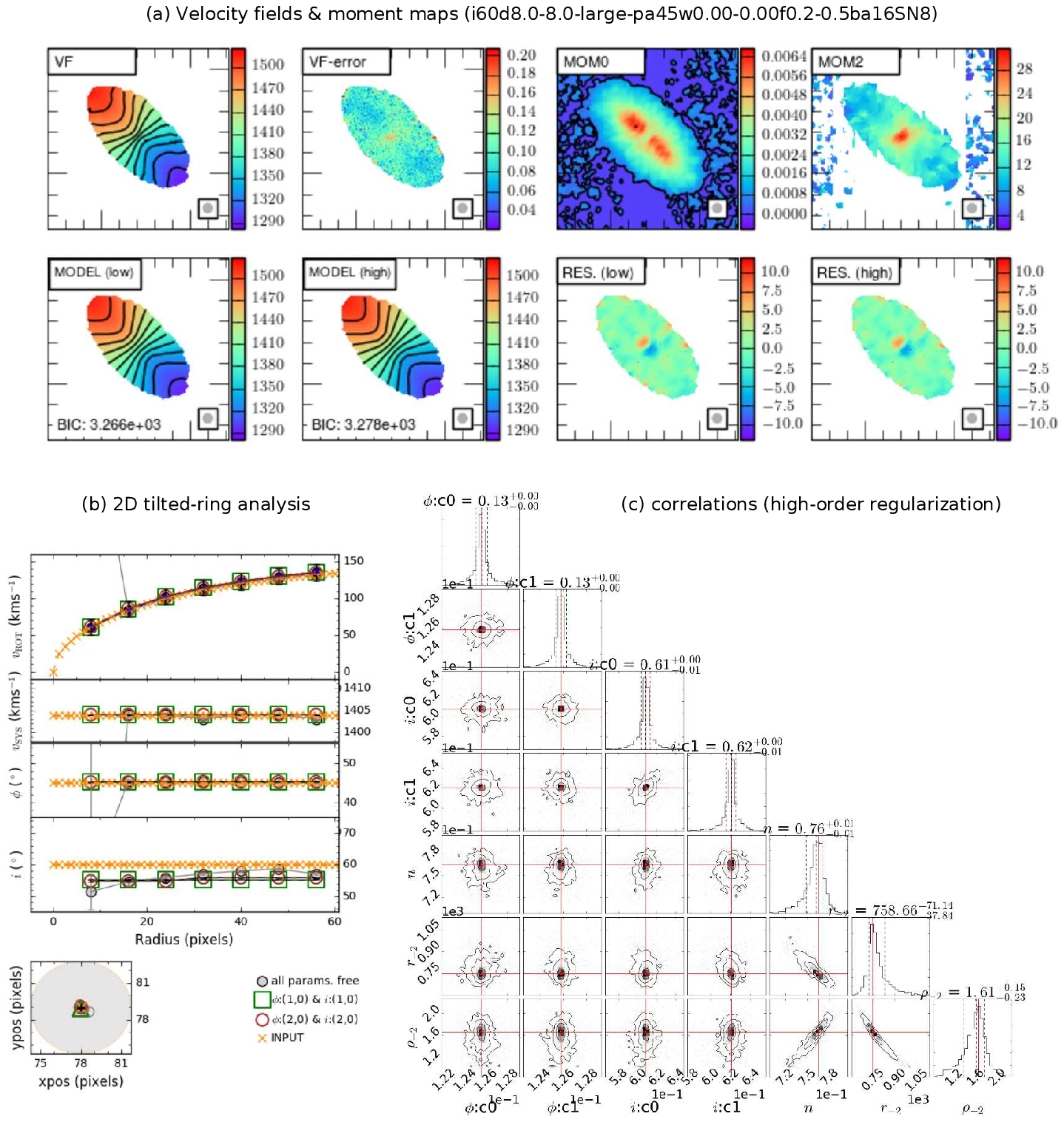}
\caption{{\sc 2dbat} analysis for an artificial {\it intermediate-mass} galaxy:
{\bf (a)} hermite $h_3$ velocity field, error, moment maps, model and residual
velocity fields. Contours are spaced by 20 \kms\ on the velocity fields,
4 \kms\ on both the moment 2 and residual velocity fields, and 0.1 $\rm mJy\,beam^{-1}$ on the moment 0. The beam size is indicated by the ellipse
in the bottom-right corner of each panel.
The BIC values derived from the Bayesian fits for the model velocity fields are denoted in the panels of model velocity fields, respectively;
{\bf (b)} 2D tilted-ring analysis\-- the rotation curves derived using the Hermite
$h_3$ velocity field in the two regularisation modes (green
open squares: constant, brown open circles: higher-order B-splines, 
grey filled circles: fit results with all the ring parameters free,
orange cross mark: the input ring parameters used for contructing the model data cube). 
See Section~\ref{model_fit_results} for more details;
{\bf (c)} correlations of the ring parameters\-- the black
contours and histograms show the posterior constraints from the Bayesian
analysis. The best fits are indicated by red lines. 
\label{Fig04}}
\end{figure*}

%: Fig.5. /mnt/g0/research/KASI/projects/wallaby/peter_cubes/Test_Galaxies/i50d8.0-8.0-NGC891s-pa55w0.00-0.00f0.2-0.2ba8SN8
\begin{figure*} \epsscale{1.0}
\includegraphics[angle=0,width=1.0\textwidth,bb=105 172 520 604, clip=]{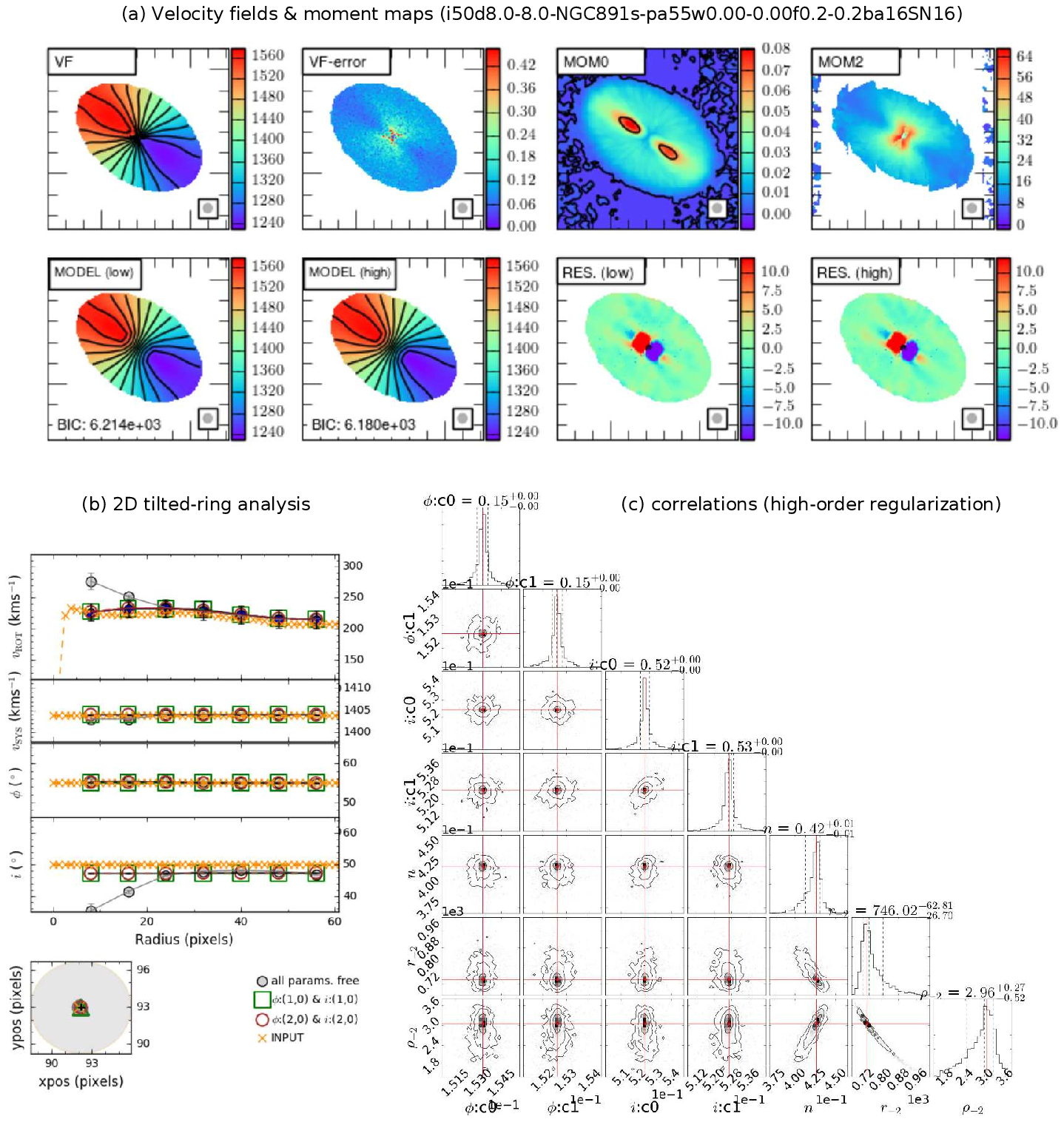}
\caption{{\sc 2dbat} analysis for an artificial {\it massive} galaxy:
{\bf (a)} hermite $h_3$ velocity field, error, moment maps, model and residual
velocity fields. Contours are spaced by 30 \kms\ in the velocity fields,
20 \kms\ in both the moment 2 and residual velocity fields, and 0.1 $\rm mJy\,beam^{-1}$ in the moment 0. The beam size is indicated by the ellipse
in the bottom-right corner of each panel.
The BIC values derived from the Bayesian fits for the model velocity fields are denoted in the panels of model velocity fields, respectively;
{\bf (b)} 2D tilted-ring analysis: the rotation curves derived using the Hermite
$h_3$ velocity field in the two regularisation modes (green
open squares: constant, brown open circles: higher-order B-splines, 
grey filled circles: fit results with all the ring parameters free,
orange cross mark: the input ring parameters used for contructing the model data cube). 
See Section~\ref{model_fit_results} for more details;
{\bf (c)} correlations of the ring parameters: the black
contours and histograms show the posterior constraints from the Bayesian
analysis. The best fits are indicated by red lines. 
\label{Fig05}}
\end{figure*}

\subsection{Improving the processing time} \label{processing_time} As presented
in Fig.~\ref{Fig01}, {\sc 2dbat} provides a pixel sampling mode in which the
velocity field is sampled with a grid spacing in units of pixels supplied by the
user. Although some spatial information is lost (see the panels (c) and (d) in Fig.~\ref{Fig01}), this option is useful for
reducing the processing time which increases significantly with the number of
pixels to be fitted in the Bayesian analysis. As will be shown in Fig.~\ref{lvhis1}, 
for well-resolved galaxies, the fit results
derived with sampling options where the grid spacing is comparable to or less
than the size of one beam are in general agreement with that derived with the
full resolution while improving the execution time significantly.

In addition {\sc 2dbat} has been developed to fully support the built-in
Message-Passing Interface (MPI) routines in {\sc multinest} by which the
Bayesian analysis can be parallelized. This enables us to improve the processing
time significantly on either a multi-core single or cluster system.

\section{Performance test and discussion} \label{performance_test}

In this Section, we test the performance of {\sc 2dbat} using
real data from LVHIS \citep{Koribalski_2010} as well as artificial
galaxies resembling rotation curves of intermediate-mass and massive spiral
galaxies which were also used to test {\sc fat} in \cite{2015MNRAS.452.3139K}.

\subsection{Artificial galaxies} \label{model_galaxies} We first apply {\sc 2dbat} to
the 52 artificial galaxies in \cite{2015MNRAS.452.3139K} to assess its performance
by comparing the recovered ring parameters with those used for constructing the
model galaxies.  \cite{2015MNRAS.452.3139K} built the data cubes of the model
galaxies using two representative rotation curves of intermediate-mass and
massive spiral galaxies as well as a solid body-like rotation curve of dwarf
galaxies.  They mimic the surface brightness of the galaxies by
locally perturbing an exponential profile with a scale length of 10 kpc which is
decreased by a factor of 20 depending on the size of the galaxies. The data
cubes for the three base model galaxies are constructed by (1) distributing the
flux based on the surface brightness profiles over the velocity ranges spaced by
a channel resolution of 4 \kms, (2) adding white noise, and (3) smoothing them
with a Gaussian beam with FWHM of 30\arcsec. The beam size is comparable to that
of the core of ASKAP at 21cm. In addition, warps are included in the cubes by radially varying the
angular momentum vector of the initial disk. Lastly, by varying the size,
inclination, position angle, velocity dispersion, angular momentum vector,
scale height, S/N, and rotation curve of the three base model galaxies, they ended
up with 52 model data cubes. 

We extract the velocity fields from the artificial data cubes to which 2D kinematic tilted-ring models
are fitted using {\sc 2dbat}. For this, we fit a third-order
Gauss-Hermite polynomial to individual velocity profiles of the data cubes. As
discussed in \cite{Oh_2011a}, this allows us to derive more reliable central
velocities of the profiles even with significant asymmetries compared to other
types of velocity fields, such as moment 1, single Gaussian, or peak velocity
fields. As examples, we present the extracted Hermite $h_3$ velocity fields of artificial dwarf, intermediate-mass,
and massive galaxies together with their moment maps (moment 0 and 2) in
panel (a) of Figs.~\ref{Fig03} to \ref{Fig05}. We note that their zeroth moment maps are not used
for the {\sc 2dbat} analysis but only for showing the integrated intensity of H{\sc i} in the galaxies.

\subsubsection{Fit results} \label{model_fit_results}

We run {\sc 2dbat} on the extracted Hermite $h_3$ velocity fields of the 52
artificial galaxies to derive their 2D tilted-ring parameters given the degree
of regularisation in a fully automated manner.
For $\phi$ and $i$ (or $v_{\rm EXP}$ which is set to zero in this work) whose radial changes 
can be regularised by B-splines in the 2D galaxy kinematics model,
we use two different regularisation modes (constant or high order) which can be
specified by the number of knots and spline order ($n$, $k$) as described in Section~\ref{A_new_algorithm}. For constant $\phi$ and $i$, we set $\phi^{\rm SPLINE}$($n=1$, $k=0$)
and $i^{\rm SPLINE}$($n=1$, $k=0$), respectively. Meanwhile, for the high orders of $\phi$ and $i$, we
set $\phi^{\rm SPLINE}$($n=2$, $k=0$) and $i^{\rm SPLINE}$($n=2$, $k=0$), depending on the
complexity of their radial variations.
The fitting setup of the {\sc 2dbat} runs adopted for this test is given
in Table~\ref{table:2}. For each artificial galaxy, we apply {\sc 2dbat} using the two
regularisation modes, resulting in 104 rotation curves in total of the 52 model
galaxies. Instead of showing all the fit results of the artificial galaxies, we present
those of three representative dwarf, intermediate-mass and massive galaxies
in Figs~\ref{Fig03} to \ref{Fig05}.

\begin{table}
\caption{Parameter setup for the performance test of {\sc 2dbat}}
\label{table:2}
\centering
\begin{tabular}{@{} lc @{}}
\hline
\hline
Parameter    & Variation \\
(1) & (2) \\
\hline
\multicolumn{2}{c}{{\sc sampling}} \\
\hline
Ring width & 1 beam \\
({\sc \rm RA}$_{\rm grid}$, {\sc RA}$_{\rm grid}$) & (0.3 beam, 0.3 beam) \\
({\sc \rm Dec}$_{\rm grid}$, {\sc Dec}$_{\rm grid}$) & (0.3 beam, 0.3 beam) \\
({\sc \rm RA}$^{\rm D}_{\rm grid}$, {\sc RA}$^{\rm D}_{\rm grid}$) & (0.3 beam, 0.3 beam) \\
({\sc \rm Dec}$^{\rm D}_{\rm grid}$, {\sc Dec}$^{\rm D}_{\rm grid}$) & (0.3 beam, 0.3 beam) \\
\\
\multicolumn{2}{c}{{\sc regularisation}} \\
\hline
$\phi$ (knots, spline order $k$) & (1, $k=0, 1, 2$ and $3$) \\
$i$ (knots, spline order $k$) & (1, $k=0$ and $1$) \\
$v_{\rm EXP}$ (knots, spline order $k$) & fixed to zero \\
\\
\multicolumn{2}{c}{{\sc weight}} \\
\hline
$|\cos\theta|^{t}$ & $t=1$ \\
Free angle around the minor axis & $10^{\circ}$ \\
%$v^{\rm LOS}_{\rm w-error}$ & 1 channel (4 \kms) or free \\
\\
\multicolumn{2}{c}{{\sc multinest parameters}} \\
\hline
{\it N} & 200 (50 for dirty fits) \\
{\it efr} & 0.8 \\
{\it tol} & 0.1 (0.3 for dirty fits) \\
maximum iteration & $\infty$ \\
\hline
\end{tabular}
\begin{minipage}{75mm}
   \scriptsize{
{\bf (1):} Parameters of {\sc 2dbat} and {\sc multinest}. The radial velocities within the free angle are discarded. Super-script 'D' indicates the parameters for the intial 'dirty' Bayesian fit. See Section~\ref{Layout} for more details.}
\end{minipage}
\end{table}

For each galaxy, we show (1) moment maps + velocity fields, (2) 2D tilted-ring
analysis, and (3) correlations between the ring parameters derived. As an
example, Fig.~\ref{Fig03} shows {\sc 2dbat} fit results for a well-sampled ($\sim$$7$
beams across the semi-major axis) intermediate mass galaxy with
$\phi=55^{\circ}$ and $i=50^{\circ}$. In the top panel (a), its moment
maps and input Hermite $h_3$ velocity field are presented together with the
residual maps and the model velocity fields which are derived using the fit
results with the two regularisation setup (i.e., constant or high-order $\phi$ and
$i$). The corresponding Bayesian Information Criteria (BIC) statistics values from the fits in the two regularisation modes
are also denoted on the model velocity fields, respectively.

In all of the panels for 2D tilted-ring analysis, the ring parameters and rotation curves derived using {\sc 2dbat}
are plotted as open squares, circles, and grey dots connected by
solid lines. The grey dots indicate the fit results made with all ring
parameters free. These unsupervised fits allow us to check how the radial
scatter of the individual ring parameters behaves in general.
We also overplot the {\it input} ring parameters and rotation curves as indicated by thick dashed lines in
the figures, respectively. The correlation panel (c) shows the marginalized posterior
distributions of the ring parameters derived in the high-order
regularisation setup of $\phi$ (cubic) and $i$ (linear) as adopted in the test.

\begin{figure*} \epsscale{1.0}
\includegraphics[angle=0,width=1.0\textwidth,bb=30 10 470 430,
clip=]{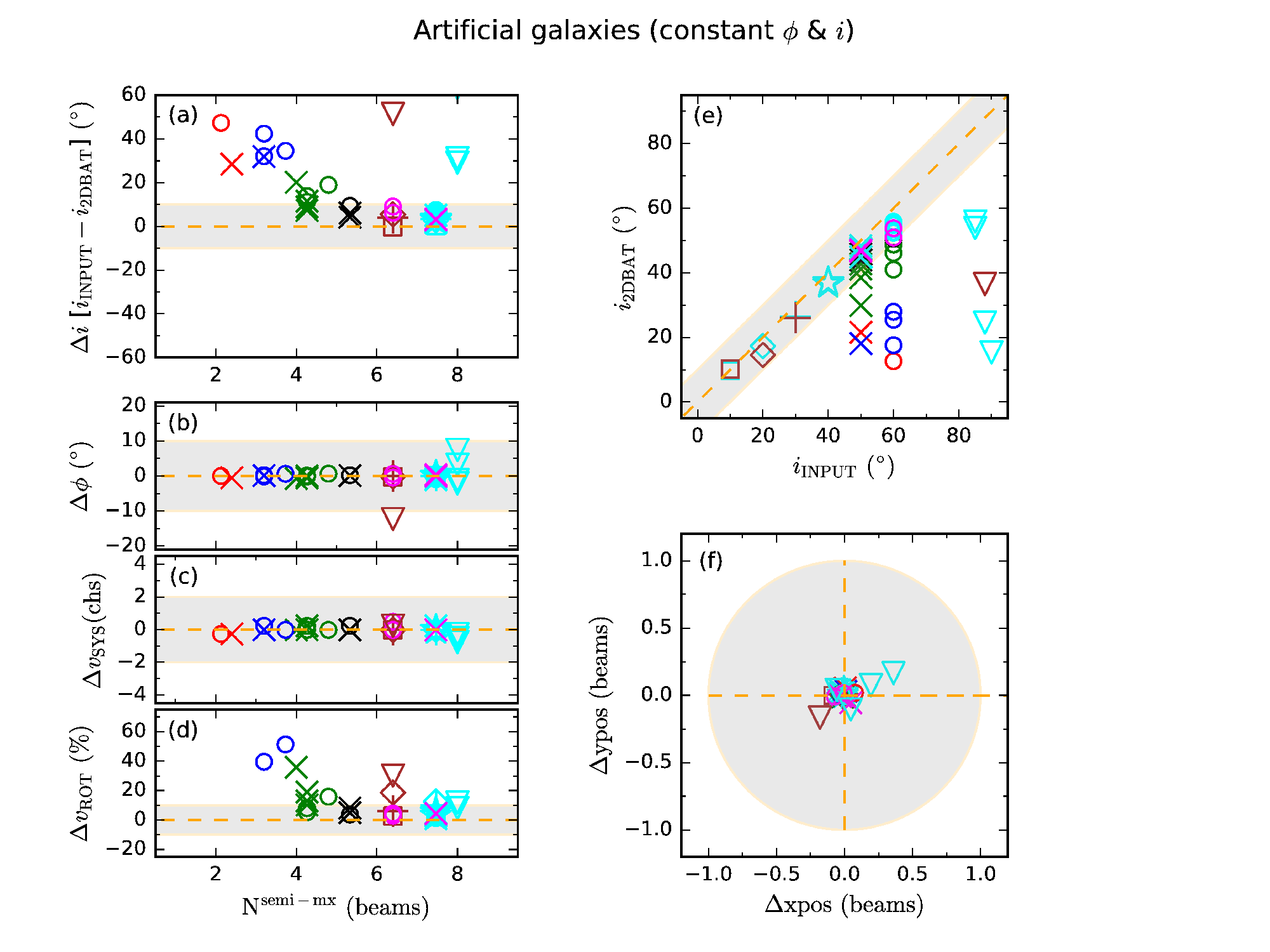} \caption{Comparison of the fit results between
    {\sc 2dbat} and the input ones (constant $\phi$ and {\it i}): (a) inclination $i$ offset against the number of
    resolved elements $\rm N^{\rm semi-mx}$ across the semi-major axis. Different symbols denote the
    range of the derived inclinations ($20^{\circ}\--70^{\circ}$) in steps of
    $10^{\circ}$. The grey shaded region indicates $\pm10^{\circ}$ offset;
    (b) position angle offset, $\Delta \phi$ ($\phi_{\rm INPUT}-\phi_{\rm
    2DBAT}$) in degrees against $\rm N^{\rm semi-mx}$; (c) Systemic velocity
    offset $\Delta v_{\rm SYS}$ ($v_{\rm SYS}^{\rm INPUT}-v_{\rm SYS}^{\rm
    2DBAT}$) in units of the channel resolution against $\rm N^{\rm semi-mx}$;
    (d) rotation velocity offset $\Delta v_{\rm ROT}$ against $\rm N^{\rm
    semi-mx}$. $\Delta v_{\rm ROT}$ is calculated as
    $(v_{\rm ROT}^{\rm INPUT}-v_{\rm ROT}^{\rm 2DBAT})/v_{\rm ROT}^{\rm Max}$ where $v_{\rm
    ROT}^{\rm Max}$ is the input maximum rotation velocities;
    (e) one-to-one comparison of the inclinations between the input values and those derived
    using {\sc 2dbat}. The grey shaded region indicates
    $\pm10^{\circ}$ offset with respect to the line of equality (dashed line);
    (f) centre position offset in beam size. The grey shaded region
    indicates one beam size centred on the input kinematic centre positions.
\label{Fig06}}
\end{figure*}

\begin{figure*} \epsscale{1.0}
\includegraphics[angle=0,width=1.0\textwidth,bb=30 10 470 430,
clip=]{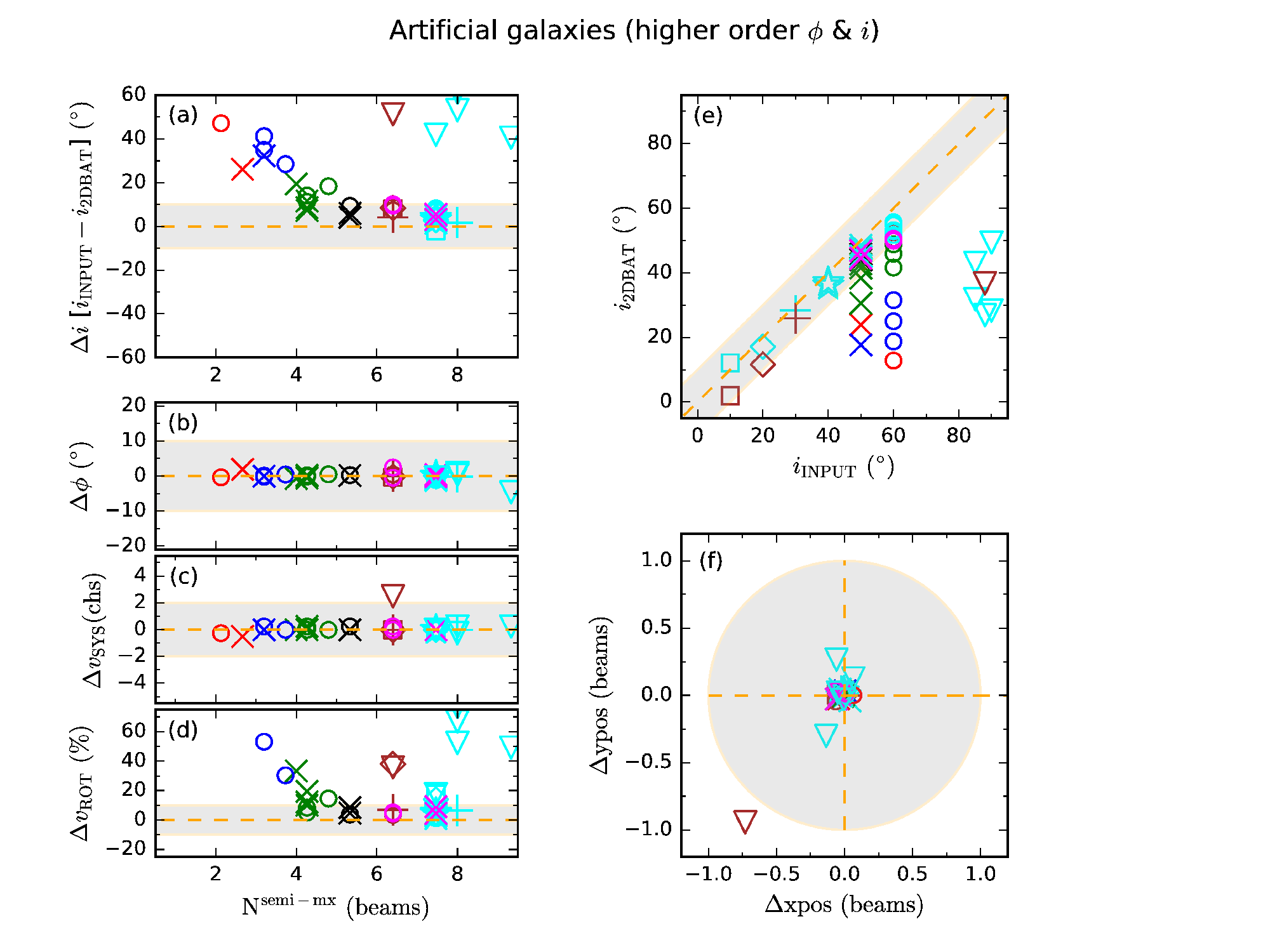} \caption{Comparison of the fit results between
    {\sc 2dbat} and the input ones (higher order $\phi$ and {\it i}): (a) inclination $i$ offset against the number of
    resolved elements $\rm N^{\rm semi-mx}$ across the semi-major axis. Different symbols denote the
    range of the derived inclinations ($20^{\circ}\--70^{\circ}$) in steps of
    $10^{\circ}$. The grey shaded region indicates $\pm10^{\circ}$ offset;
    (b) position angle offset, $\Delta \phi$ ($\phi_{\rm INPUT}-\phi_{\rm
    2DBAT}$) in degrees against $\rm N^{\rm semi-mx}$; (c) Systemic velocity
    offset $\Delta v_{\rm SYS}$ ($v_{\rm SYS}^{\rm INPUT}-v_{\rm SYS}^{\rm
    2DBAT}$) in units of the channel resolution against $\rm N^{\rm semi-mx}$;
    (d) rotation velocity offset $\Delta v_{\rm ROT}$ against $\rm N^{\rm
    semi-mx}$. $\Delta v_{\rm ROT}$ is calculated as
    $(v_{\rm ROT}^{\rm INPUT}-v_{\rm ROT}^{\rm 2DBAT})/v_{\rm ROT}^{\rm Max}$ where $v_{\rm
    ROT}^{\rm Max}$ is the input maximum rotation velocities;
    (e) one-to-one comparison of the inclinations between the input values and those derived
    using {\sc 2dbat}. The grey shaded region indicates
    $\pm10^{\circ}$ offset with respect to the line of equality (dashed line);
    (f) centre position offset in beam size. The grey shaded region
    indicates one beam size centred on the input kinematic centre positions.
\label{Fig07}}
\end{figure*}

In the following sections, we compare (1) the derived ring parameters, (2) rotation curves, and
(3) model velocity fields produced using the best fit results with those that were used to
build the model galaxies. From this, we examine how well {\sc 2dbat} is able to recover the input ring
parameters of the artificial galaxies in the 2D tilted-ring parameter space. 
Figs.~\ref{Fig06} and \ref{Fig07} show the comparison between the model's input and
{\sc 2dbat}'s output parameters derived
assuming constant and high-order regularisations of $\phi$ and $i$ of all the artificial galaxies,
respectively. As shown in the figures, the fit results derived in the two different regularisation
modes are robust and largely consistent with each other within the scatter.

% Fig.6. 2DBAT MODEL results : constant pa + incl 

$\bullet$ $\Delta i:$ The inclination difference, $\Delta i$ ($i_{\rm INPUT}
- i_{\rm 2DBAT}$) of the model galaxies is shown in the panels (a) against the
number of resolved elements across the semi-major axis, $\rm N^{\rm semi-mx}$, together with a direct
one-to-one comparison between them as given in the panels (e).
The input inclination values are shown by different symbols
in steps of $10^{\circ}$, from $10^{\circ}$ to $90^{\circ}$ in the panels (e) of
Figs.~\ref{Fig06} and \ref{Fig07}. $\Delta i$ values are grouped into bins from 2
to 8 beams as represented with different colours in the same panels.
On the whole, the derived inclinations by {\sc 2dbat} are in good agreement with the input ones
within $\sim$$10^{\circ}$ for the galaxies with $10 < i < 70^{\circ}$ with more than four resolution
elements across the semi-major axis.

Interestingly, {\sc 2dbat} provides reliable estimates of inclinations even for
face-on-like galaxies with inclinations of $10$ and $20^{\circ}$ as long as their velocity fields are
well-sampled (i.e., $\rm N^{\rm semi-mx} > 5$), regular and symmetric in shape.
However, the fit results show a trend of increasing $\Delta i$ towards
decreasing $\rm N^{\rm semi-mx}$ as colour-coded in the panels (a).
The inclination offset is most prominent at the smallest $\rm N^{\rm semi-mx}
< 3$. The majority of the galaxies with $\rm N^{\rm semi-mx} < 5$ show large
inclination offsets ($> 10^{\circ}$) although they have intermediate
inclinations with $50$ or $60$ degrees.
This can be mainly attributed to beam smearing. 
As discussed earlier, it usually results in more parallel iso-velocity contours across
the velocity field which are modelled using a lower inclination in
the 2D fit. This is the case of the galaxies with $\rm N^{\rm semi-mx} < 5$ in
panel (a) which seem to be significantly affected by the beam smearing. A similar trend of increasing
$\Delta i$ with decreasing $\rm N^{\rm semi-mx}$ is also found in \cite{2015MNRAS.452.3139K} where the same artificial galaxies are used for the tilted-ring
analysis in a three-dimensional way although it is only significant at $\rm
N^{\rm semi-mx} < 3$. Meanwhile, regardless of $\rm N^{\rm semi-mx}$, for
edge-on-like galaxies with $i > 80^{\circ}$, the offsets are very large ($>
40^{\circ}$), due to less data points to be fitted in
given rings and the projection effect of line-of-sight velocities in the
galaxies. This shows that either the lower sampling of velocity fields or the
higher degeneracy between $v_{\rm ROT}$ and $i$, or both are mainly responsible
for such large inclination offsets in face-on, and edge-on galaxies or even in galaxies
with intermediate inclinations but with poor sampling. Together with beam smearing,
this is a fundamental limitation of 2D tilted-ring analysis.

% Fig.7. 2DBAT MODEL results : varying pa + incl

\begin{table*}
\caption{Observational properties of LVHIS sample galaxies}
\label{table:3}
\scriptsize
   \centering
   \begin{tabular}{@{} llccrrrrrr @{}}
   \hline
\hline
      HIPASS ID    & NED ID &  $\alpha$ (J2000) & $\delta$ (J2000) & $v_{\rm SYS}$ & $\phi$ & $i$ & D & $\Theta_{\rm MX}$ & Fig\# \\
           &  & (hh:mm:ss) & (dd:mm:ss) & (\kms) &  ($^\circ$) & ($^\circ$) & (Mpc) & \\
        (1) & (2) & (3) & (4) & (5) & (6) & (7) & (8) & (9) & (10) \\
  \hline
HIPASS J1441-62&           &14 41 37 &-62 44 38 & 672$\pm$8& 	 & &  & 1.75 & \ref{lvhis1}\\
HIPASS J1305-40&CEN 06           &13 05 02 &-40 06 30 & 617$\pm$4&	  &60&6.1 & 2.38 & \ref{lvhis2}\\
HIPASS J0320-52&NGC 1311         &03 20 05 &-52 11 34 & 568$\pm$5& 40 &79$\pm$1&5.3 & 3.69 & \ref{lvhis3}\\
HIPASS J1337-39&           &13 37 30 &-39 52 56 & 492$\pm$4& 	 &36&4.8 & 4.29 & \ref{lvhis4}\\
HIPASS J1219-79&IC 3104          &12 19 04 &-79 42 55 & 429$\pm$4&	 45 &59$\pm$6&2.6 & 4.48 & \ref{lvhis5}\\
HIPASS J1047-38&ESO 318-G013     &10 47 39 &-38 51 45 & 711$\pm$7& 75 & $\ge 88$&  & 4.65 & \ref{lvhis6}\\
HIPASS J1428-46&UKS 1424-460     &14 28 06 &-46 18 32 & 390$\pm$2& 	&73&3.4 & 4.88 & \ref{lvhis7}\\
HIPASS J1620-60&ESO 137-G018     &16 20 56 &-60 29 18 & 605$\pm$3&	30 &73$\pm$2&5.9 & 5.35 & \ref{lvhis8}\\
HIPASS J0705-58&AM 0704-582      &07 05 18 &-58 31 19 & 564$\pm$2&	 &65&4.9 & 5.63 & \ref{lvhis9}\\
HIPASS J1337-28&ESO 444-G084     &13 37 18 &-28 02 17 & 587$\pm$3&	 &39$\pm$4&5.1 & 6.21 & \ref{lvhis10}\\
HIPASS J0731-68&ESO  59-G001     &07 31 20 &-68 11 19 & 530$\pm$3&	 &20$\pm$18&4.5 & 6.39 & \ref{lvhis11}\\
HIPASS J0333-50&IC 1959          &03 33 15 &-50 25 17 & 640$\pm$4&	147 &88$\pm$4&8.2 & 6.55 & \ref{lvhis12}\\
HIPASS J1403-41&NGC 5408         &14 03 21 &-41 22 26 & 506$\pm$3&	 62 &55$\pm$8&4.9 & 6.87 & \ref{lvhis13}\\
HIPASS J1337-42&NGC 5237         &13 37 47 &-42 50 51 & 361$\pm$4&	 128 &35$\pm$0&3.7 & 7.92 & \ref{lvhis14}\\
HIPASS J1348-53&ESO 174-G?001    &13 48 01 &-53 21 31 & 688$\pm$3&	 170 &76$\pm$11&  & 10.41 & \ref{lvhis15}\\
HIPASS J1057-48&ESO 215-G?009    &10 57 32 &-48 11 02 & 598$\pm$2&	 72 &64$\pm$27&5.3 & 11.86 & \ref{lvhis16}\\
HIPASS J1501-48&ESO 223-G009     &15 01 08 &-48 17 04 & 588$\pm$2&	135 &44$\pm$19&6.0 & 11.92 & \ref{lvhis17}\\
HIPASS J2202-51&IC 5152          &22 02 41 &-51 17 37 & 122$\pm$2& 	100 &51$\pm$4&1.8 & 12.78 & \ref{lvhis18}\\
HIPASS J0256-54&ESO 154-G023     &02 56 55 &-54 34 58 & 574$\pm$2&	39 & $\ge 88$&6.8 & 13.30 & \ref{lvhis19}\\
HIPASS J1305-49&NGC 4945         &13 05 24 &-49 29 35 & 563$\pm$3&	43 &85$\pm$4&4.1 & 14.32 & \ref{lvhis20}\\
HIPASS J1321-36&NGC 5102         &13 21 55 &-36 38 03 & 468$\pm$2&	48 &70$\pm$6&3.4 & 14.59 & \ref{lvhis21}\\
HIPASS J0047-25&NGC 253         &00 47 31 &-25 17 22 & 243$\pm$2	&	52 &83$\pm$0&3.1  & 21.50 & \ref{lvhis22}\\
HIPASS J1413-65&Circinus        &14 13 27 &-65 18 46 & 434$\pm$3&	40 &64$\pm$4&4.2 & 22.72 & \ref{lvhis23}\\
HIPASS J0317-66&NGC 1313         &03 17 57 &-66 33 30 & 470$\pm$2&	 &36$\pm$7&4.0 & 42.54 & \ref{lvhis24}\\
          \hline
     \end{tabular}
\begin{minipage}{145mm}
   \scriptsize{
{\bf (1)(2):} HIPASS and NASA Extra-galactic Database (NED) names;
{\bf (3)(4):} HI centroid derived using a Gaussian fit to moment 0 in \cite{2004AJ....128...16K};
{\bf (5)(6)(7):} Systemic velocities, optical position angle ($\phi$), and optical inclination ($i$) taken from \cite{2015MNRAS.452.3139K};
{\bf (8):} Distance taken from NED as given in \cite{2015MNRAS.452.3139K};
{\bf (9):} Number of beams across the morphological major axis derived from an ellipse fit to the velocity field;
{\bf (10):} Figure number in this paper.}
\end{minipage}
\end{table*}

$\bullet$ $\Delta \phi$, $\Delta v_{\rm SYS}$, $\Delta v_{\rm ROT}$ and centre
position offset: As found in the panels (b), (c), (d), and (f), {\sc 2dbat} recovers
the input $\phi$, $v_{\rm SYS}$ and centre position ($x_{\rm C}$, $y_{\rm C}$) of most of the
artificial galaxies with $10^{\circ} < i < 70^{\circ}$ with good accuracy: $\Delta \phi
< 2^{\circ}$; $\Delta v_{\rm SYS} < 1$ channel resolution; $\Delta$$x_{\rm C} < 1.0$ beam;
$\Delta$$y_{\rm C} < 1.0$ beam regardless of $\rm N^{\rm semi-mx}$. This shows
that {\sc 2dbat} is able to provide reliable estimates of these weakly
correlated parameters in the 2D tilted-ring analysis even for edge-on-like
galaxies with $i > 80^{\circ}$.

$\bullet$ $\Delta v_{\rm ROT}$: Likewise, for most galaxies, the rotation velocity offsets, $\Delta v_{\rm ROT}$ which are the weighted
means of the residuals between the input and derived ones over all the radii given their errors are in general within 10\% of the input
maximum rotation velocities. The majority positive $\Delta v_{\rm ROT}$ values indicate
that the derived rotation velocities are lower than the input ones.
This shows that the 2D analysis is affected by beam
smearing which tends to lower the intrinsic rotation velocities.
The outliers showing more than 10\% deviations correspond to face-on galaxies,
where small inclination offsets can lead to large offsets in
$v_{\rm ROT}$. Meanwhile, the edge-on-like galaxies with $i>70^{\circ}$ also show large $\Delta
v_{\rm ROT}$ mainly due to the poor sampling of the velocity field combined with
the projection effect which results in significant offsets in inclination.
This indicates that {\sc 2dbat} can be applied to galaxies with $20^{\circ} < i < 70^{\circ}$
and at least four resolution elements across the semi-major axis. 

As shown in Figs.~\ref{Fig03} to \ref{Fig05}, regarding the residual maps between
the input and model velocity fields (i.e., line-of-sight velocities)
reconstructed using the best fit parameters, they are mostly smaller than the channel resolution
($\sim$4 \kms) for all the artificial galaxies, except for some localized
regions where S/N is low. This confirms that
the 2D fits themselves were made without any failure as also indirectly supported by
the well-defined Gaussian distribution of the posteriors in the correlation plots.
Regarding the processing time of 2DBAT, it takes about 10 minutes for a laptop with 
a standard four-cores processor to fit the kinematic model with constant PA/INCL to
the artificial dwarf galaxy shown in Fig.~\ref{Fig03}. For the sampling option with a grid
spacing of a half beam size, the rotation curves are in general agreement with those
derived from FAT whose execution time on a similar specified laptop is about 30 minutes.

In summary, for the artificial galaxies with intermediate inclinations
($20\--70^{\circ}$) that are resolved by more than four beams across the
semi-major axis, the 2D tilted-ring parameters recovered by {\sc 2dbat} in
a fully automated manner show good agreement with the input ones.

\subsection{Real galaxies} \label{real_galaxies} 

We continue to test the performance of {\sc 2dbat} using sample galaxies taken
from LVHIS \citep{Koribalski_2010}. LVHIS is a large H{\sc i} survey for
a sample of 82 nearby ($<$ 10 Mpc), gas-rich galaxies undertaken with the
Australia Telescope Compact Array (ATCA) which aims to investigate fundamental
H{\sc i} properties and kinematics of the galaxies by providing
a comprehensive H{\sc i} galaxy atlas.

Of the parent LVHIS sample, we select 24 galaxies which are
resolved by more than two independent beams across their major axes and show systematic rotation
in their velocity fields. The optical inclinations of the galaxies range
from $20^{\circ}$ to $88^{\circ}$. Although some of these are falling outside the
reliable fit range of {\sc 2dbat} (i.e., $\rm N^{\rm semi-mx} > 4$ and $20^{\circ} < i < 70^{\circ}$)
as discussed in Section~\ref{model_fit_results}, we include them to test how well {\sc 2dbat} is able to
perform in marginal cases. These LVHIS galaxies were also manually fitted by {\sc rotcur} and used to
test the performance of {\sc fat} by \cite{2015MNRAS.452.3139K}. 
The velocity fields of the resolved galaxies
from WALLABY are expected to be more or less like those of the LVHIS
sample galaxies, in terms of the spatial ($20$\--$60$\arcsec) and spectral ($4$
\kms) resolution as well as the number of resolved elements across the major
axis. The basic observational properties of the galaxies, sorted by the number of
beams across the morphological major axis are listed in Table~\ref{table:3}.
We refer to Koribalski et al. (submitted) for more details of the H{\sc i} observations and data reduction. 

In exactly the same way as the artificial galaxies were analyzed in
Section~\ref{model_galaxies}, we extract the Hermite $h_3$ velocity fields
from the data cubes of the sample galaxies as well as moment maps (0th and
2nd), and perform a 2D tilted-ring analysis using {\sc 2dbat} given two
regularisation modes of (constant or high-order) $\phi$ and $i$. For the higher
level of regularisation, we manually specify the order of B-splines of $\phi$ and $i$ for
each galaxy, depending on the number of resolved elements across the major axis
and the level of radial change in the ring parameters. The number of knots and
the orders of B-spline chosen are denoted in the panel (b) of
Figs~\ref{lvhis1} to \ref{lvhis24}.

We make a direct comparison between the {\sc 2dbat} fit results and the
ones derived from a 2D tilted-ring analysis using {\sc rotcur} in
GIPSY. The manual {\sc rotcur} fits were made by regularizing $\phi$ and $i$
with polynomials depending on the degree of their radial scatters on
a galaxy-by-galaxy basis. These manual fits are also used in \cite{2015MNRAS.452.3139K}
for a detailed comparison with the 3D method.
The extracted velocity fields, moment maps and the reconstructed velocity fields
from the {\sc 2dbat}'s best fits are presented in the Appendix,
Figs.~\ref{lvhis1} to \ref{lvhis24}, in the same format as for the artificial
galaxies. The beam size of the observations is indicated
by an ellipse in the bottom-right corner of panels (a).

As with the artificial galaxies, we present the comparisons between the ring
parameters derived using both {\sc rotcur} and {\sc 2dbat} given the
regularisation of $\phi$ and $i$ in Figs.~\ref{Fig08} and \ref{Fig09}, respectively.
We also overplot the fit results made with all
ring parameters free as given by grey solid dots. These unsupervised fits allow
us to check how the radial scatter of the individual ring parameters behaves in
general. In this paper, through the comparison with {\sc rotcur}, we focus on
how the {\sc 2dbat} fit results are comparable to those manually derived using a standard
method of 2D tilted-ring analysis. For a detailed discussion of the comparison of the {\sc
rotcur} results with a 3D method ({\sc tirific}) as well as another 2D
method ({\sc diskfit}), we refer to \cite{2015MNRAS.452.3139K}.

\subsubsection{Fit results} \label{lvhis_fit_results}

\begin{figure*} \epsscale{1.0}
\includegraphics[angle=0,width=1.0\textwidth,bb=30 10 470 430,
clip=]{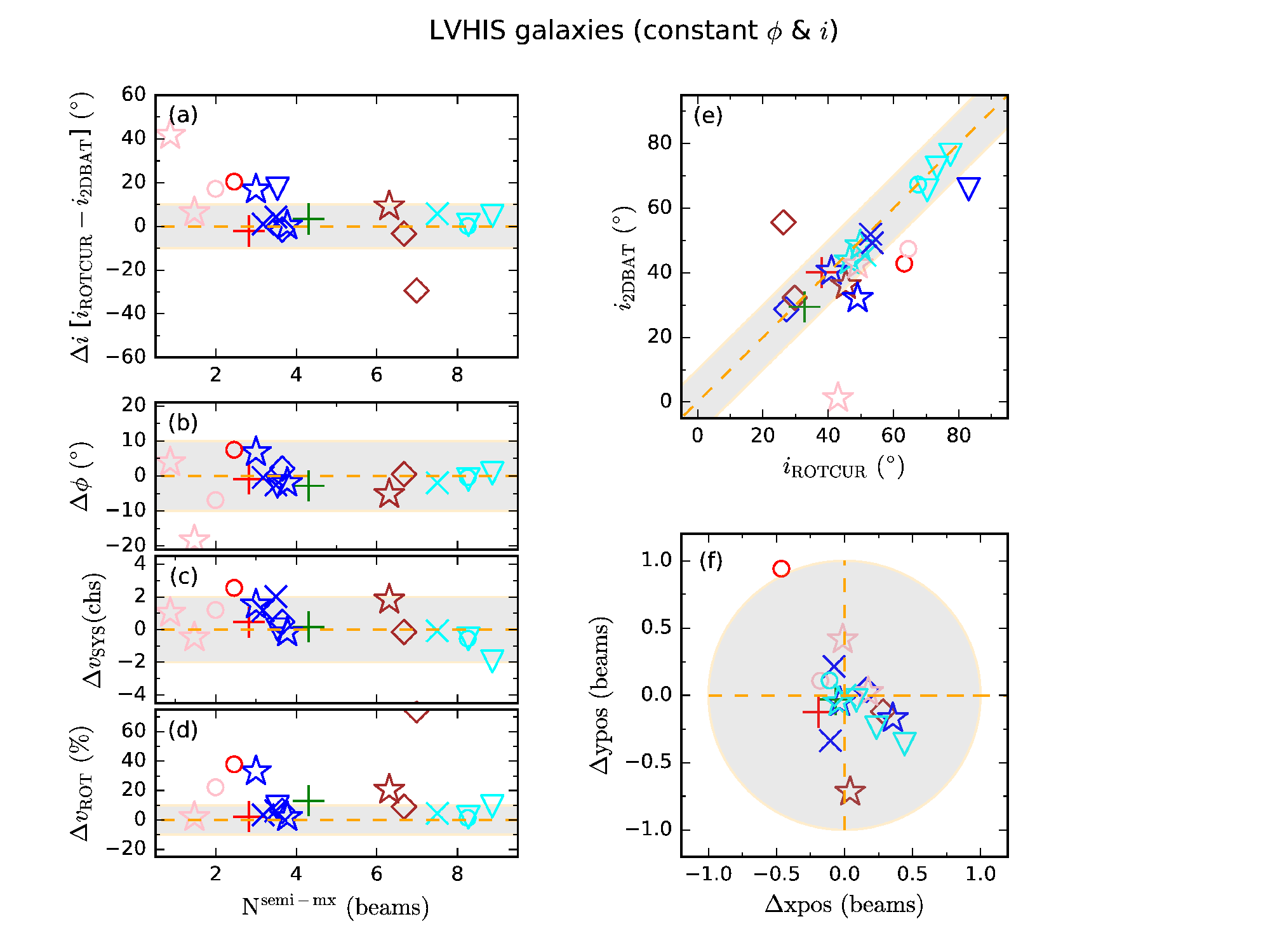} \caption{Comparison of the fit results for the LVHIS galaxies from
    {\sc 2dbat} and {\sc rotcur} (constant $\phi$ and {\it i}): (a) inclination $i$ offset against the number of
    resolved elements $\rm N^{\rm semi-mx}$ across the semi-major axis. Different symbols denote the
    range of the derived inclinations ($20^{\circ}\--70^{\circ}$) in steps of
    $10^{\circ}$. The grey shaded region indicates $\pm10^{\circ}$ offset;
    (b) position angle offset, $\Delta \phi$ ($\phi_{\rm ROTCUR}-\phi_{\rm
    2DBAT}$ in degrees against $\rm N^{\rm semi-mx}$; (c) Systemic velocity
    offset $\Delta v_{\rm SYS}$ ($v_{\rm SYS}^{\rm ROTCUR}-v_{\rm SYS}^{\rm
    2DBAT}$) in units of the channel resolution against $\rm N^{\rm semi-mx}$;
    (d) rotation velocity offset $\Delta v_{\rm ROT}$ against $\rm N^{\rm
    semi-mx}$. $\Delta v_{\rm ROT}$ is calculated as
    $(v_{\rm ROT}^{\rm ROTCUR}-v_{\rm ROT}^{\rm 2DBAT})/v_{\rm ROT}^{\rm Max}$ where $v_{\rm
    ROT}^{\rm Max}$ is the maximum rotation velocities derived from {\sc
    rotcur};
    (e) one-to-one comparison between the inclinations derived
    using {\sc 2dbat} and {\sc rotcur}. The grey shaded region indicates
    $\pm10^{\circ}$ offset with respect to the line of equality (dashed line);
    (f) centre position offset in beam size. The grey shaded region
    indicates one beam size centred on the kinematic centre positions derived
    using {\sc rotcur}.
\label{Fig08}}
\end{figure*}

\begin{figure*} \epsscale{1.0}
\includegraphics[angle=0,width=1.0\textwidth,bb=30 10 470 430,
clip=]{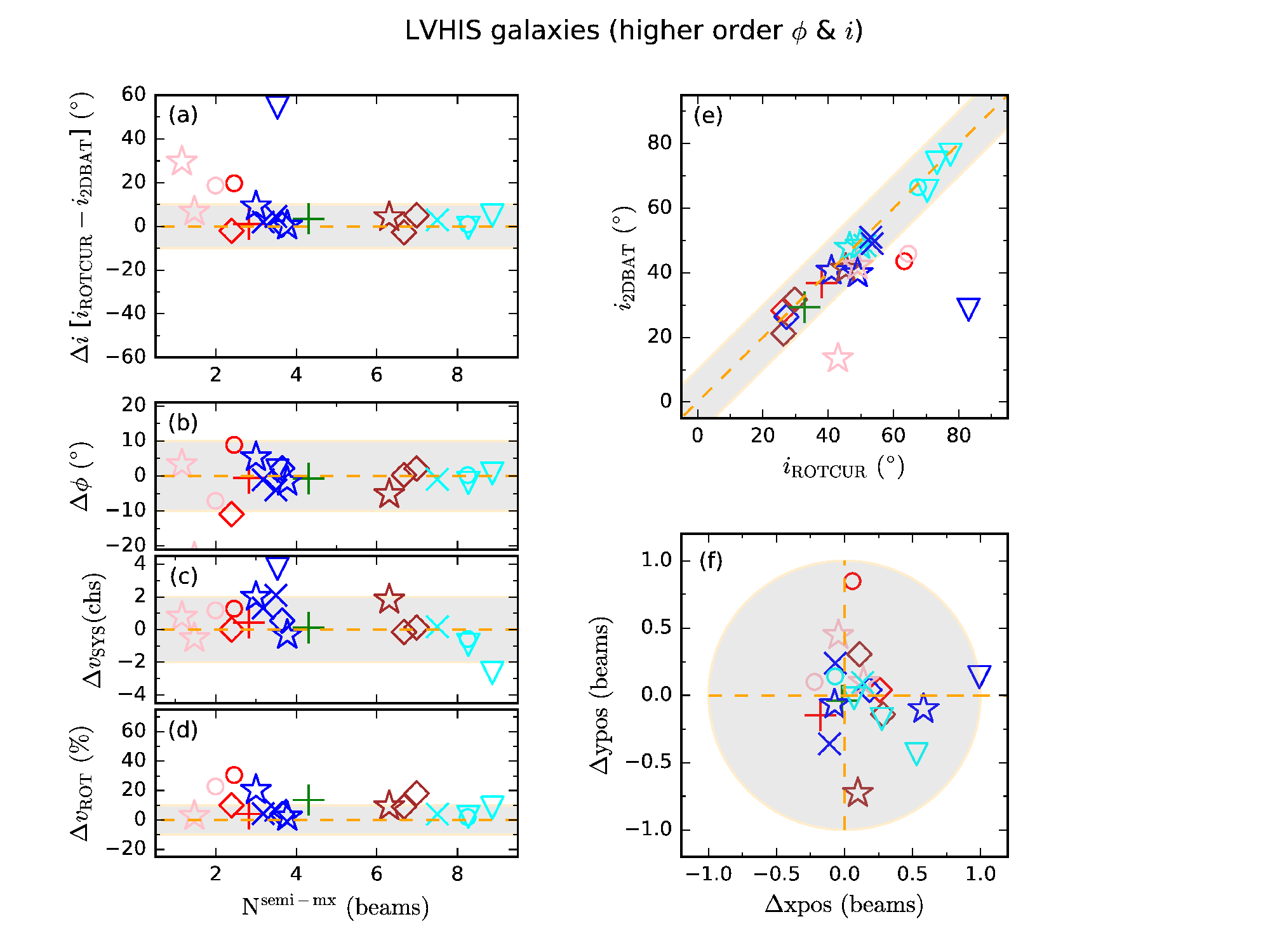} \caption{Comparison of the fit results for the LVHIS galaxies from
    {\sc 2dbat} and {\sc rotcur} (higher order $\phi$ and {\it i}): (a) inclination $i$ offset against the number of
    resolved elements $\rm N^{\rm semi-mx}$ across the semi-major axis. Different symbols denote the
    range of the derived inclinations ($20^{\circ}\--70^{\circ}$) in steps of
    $10^{\circ}$. The grey shaded region indicates $\pm10^{\circ}$ offset;
    (b) position angle offset, $\Delta \phi$ ($\phi_{\rm ROTCUR}-\phi_{\rm
    2DBAT}$ in degrees against $\rm N^{\rm semi-mx}$; (c) Systemic velocity
    offset $\Delta v_{\rm SYS}$ ($v_{\rm SYS}^{\rm ROTCUR}-v_{\rm SYS}^{\rm
    2DBAT}$) in units of the channel resolution against $\rm N^{\rm semi-mx}$;
    (d) rotation velocity offset $\Delta v_{\rm ROT}$ against $\rm N^{\rm
    semi-mx}$. $\Delta v_{\rm ROT}$ is calculated as
    $(v_{\rm ROT}^{\rm ROTCUR}-v_{\rm ROT}^{\rm 2DBAT})/v_{\rm ROT}^{\rm Max}$ where $v_{\rm
    ROT}^{\rm Max}$ is the maximum rotation velocities derived from {\sc
    rotcur};
    (e) one-to-one comparison between the inclinations derived
    using {\sc 2dbat} and {\sc rotcur}. The grey shaded region indicates
    $\pm10^{\circ}$ offset with respect to the line of equality (dashed line);
    (f) centre position offset in beam size. The grey shaded region
    indicates one beam size centred on the kinematic centre positions derived
    using {\sc rotcur}.
\label{Fig09}}
\end{figure*}

$\bullet$ $\Delta i:$ As shown in the panel (a) of Fig.~\ref{Fig08}, there may
be a trend of increasing $\Delta i$ with decreasing $\rm N^{\rm semi-mx}$
although it is less clear compared to the artificial galaxies.
As examples, HIPASS J1441-62 ($\Delta i \sim +40^{\circ}$; open star; Fig.~\ref{lvhis1}),
HIPASS J0320-52 ($\Delta i \sim +15^{\circ}$; open circle; Fig.~\ref{lvhis3}),
HIPASS J1337-39 ($\Delta i \sim +20^{\circ}$; open circle; Fig.~\ref{lvhis4}),
HIPASS J1219-79 ($\Delta i \sim +15^{\circ}$; open star; Fig.~\ref{lvhis5}), and
HIPASS J1047-38 ($\Delta i \sim +20^{\circ}$; upside-down triangle; Fig.~\ref{lvhis6}) with
$\rm N^{\rm semi-mx} < 4$ lie outside
the grey shaded region of $\pm10^{\circ}$ in the panel (a) of Figs.~\ref{Fig08} and \ref{Fig09}.
However, they have intermediate inclinations of $\sim$$40^\circ$ according to the manual fit
using {\sc rotcur} somewhat dependent on subjective model choices for the ring parameters.
As shown in Figs.~\ref{lvhis1}, \ref{lvhis3}, \ref{lvhis4}, \ref{lvhis5} and \ref{lvhis6}, their velocity fields are poorly sampled, smoothing the kinematic structure around the central region.
These galaxies are likely to be affected by beam smearing. However, according to the Gaussian-like posterior distributions of
the ring parameters as shown in the figures, and the corresponding small amplitudes ($<1$ channel resolution) of the residual 
maps between the input and model velocity fields, {\sc 2dbat} seems to provide reasonable fits.
The subjective choices of the regularisation in the course of manual tilted-ring
analysis using {\sc rotcur} can induce large values for $\Delta i$.

% Fig.8. 2DBAT LVHIS results : constant pa + incl

% Fig.9. 2DBAT LVHIS results : varying pa + incl

As more extreme examples, HIPASS J1047-38 ($\Delta i \sim -20^{\circ}$; open circle; see Fig.~\ref{lvhis6}),
HIPASS J1305-40 ($\Delta i \sim +20^{\circ}$; open star; ), and
HIPASS J1441-62 ($\Delta i \sim -30^{\circ}$; open star) with $\rm N^{\rm semi-mx} < 2$ lie outside
the grey shaded region of $\pm10^{\circ}$ in the panel (a) of Figs.~\ref{Fig08} and \ref{Fig09}.
However, they have intermediate inclinations of $\sim$$65^\circ$ according to the manual fit
using {\sc rotcur} although again somewhat dependent
on subjective model choices for the ring parameters.
As shown in Figs.~\ref{lvhis6}, \ref{lvhis16} and \ref{lvhis20}, their velocity fields are poorly sampled, smoothing the kinematic structure around the central region. These galaxies are even more likely to
be affected by beam smearing. However, according to the Gaussian-like posterior distributions of
the ring parameters as shown in Figs.~\ref{lvhis6}, \ref{lvhis16} and \ref{lvhis20},
and the corresponding small amplitudes ($<1$ channel resolution) of the residual maps between the
input and model velocity fields, {\sc 2dbat} again seems to provide reasonable fits.

On the other hand, the amplitudes of $\Delta i$ are in general reduced if the higher order
regularisation for $\phi$ and $i$ is used. This is because the high-order regularisation
mode usually provides a better kinematic description of most galaxies than the constant
regularisation mode in terms of the fit quality. As shown in Fig.~\ref{lvhis17}, HIPASS J1501-48 is one
such galaxy where significant radial variations of $\phi$ and $i$ are shown in their velocity fields.
In a more quantitative sense, this is also supported by the smaller value of BIC in the high-order
regularisation mode. We emphasize that the number
of knots and orders of B-spline for $\phi$ and $i$ were chosen to avoid any
significant overfit in the high-order regularisation mode.

The inclinations of HIPASS J1047-38 (ESO 318-G013) and HIPASS J0333-50 derived
using both {\sc rotcur} and {\sc 2dbat} are comparable with each other but show
significant difference ($>30^{\circ}$) compared to the ones calculated from
their optical axis ratios. There may be intrinsic difference between the kinematic and
photometric geometries of the galaxies, resulting in such large inclination
offsets. Or it could be due to the low spatial sampling of the velocity fields
given that the kinematic $\phi$ are close to the optical ones. As shown in the
panels (a) of Figs.~\ref{lvhis6} and \ref{lvhis12}, the iso-velocity contours seen in the velocity
fields are predominantly parallel to the minor axes. 
Such iso-velocity contours often originate from either kinematic or observational
characteristics of galaxies, such as solid body-like rotations
\citep{1974gegr.book..291W}, co-rotating disks of edge-on galaxies, significant expansion velocities
\citep{2003A&A...409..879B}, beam smearing effects
\citep{2002ASPC..275..217T}, or bar streaming motions
\citep{2004ApJ...605..183W}. For the case of HIPASS J1047-38 (ESO 318-G013)
and HIPASS J0333-50, the beam smearing effect appears to be mainly responsible
for the parallel iso-velocity contours given the small number of resolved
elements ($\rm N^{\rm semi-mx} < 3$). As a way to circumvent the beam smearing
effect and correct for rotation velocities, the fitting can be made after
fixing $\phi$ and $i$ (including the centre position if needed) with the
photometric ones derived from optical or infrared observations (e.g.,
\citealt{2003MNRAS.340...12W}). As discussed earlier, this is a fundamental
limitation of 2D tilted-ring analysis and is a situation in which the 3D
approach may be preferred. For instances, for HIPASS J1047-38 and J0333-50, the 3D fit results
from FAT give inclinations of 80 and 85 degrees, respectively, which are more
comparable to the optical ones. We refer to \cite{2015MNRAS.452.3139K} for more
discussion on the comparison of the marginally resolved LVHIS galaxies between
2D and 3D tilted-ring analyses.

$\bullet$ $\Delta \phi$, $\Delta v_{\rm SYS}$, $\Delta v_{\rm ROT}$ and centre
position: As shown in the panels (b), (c), (d) and (f) in Figs.~\ref{Fig08} and
\ref{Fig09}, {\sc 2dbat}, in general, provides
{\it comparable} fit results to the manual fits using {\sc rotcur}, giving small offsets in $\phi$
($< 2^{\circ}$), $v_{\rm SYS}$ ($< 1$ channel resolution) and centre position
(within one beam size) except for a few outliers showing $\sim$$10^{\circ}$ offsets mainly in $\phi$. 
For the outlying galaxies, the fits with higher order regularisation of $\phi$ and $i$
give smaller offsets in $\Delta\phi$. As shown in Fig.~\ref{lvhis17}, the significant radial variation in $\phi$ of HIPASS J1501-48 is better modeled by a cubic B-spline which is also comparable to that
derived using {\sc rotcur} in manual.

$\bullet$ $\Delta v_{\rm ROT}$: The corresponding $\Delta v_{\rm ROT}$,
given both the constant and higher order
regularisation for $\phi$ and $i$, are mostly within 10\% of the maximum
rotation velocities of the galaxies with $\rm N^{\rm semi-mx} > 5$. As found in the case of the artificial galaxies,
there is a similar trend, with offsets at small $\rm N^{\rm semi-mx}$,
particularly $\rm N^{\rm semi-mx} < 4$.
As discussed earlier, inclination offsets caused by the poor sampling of the
velocity fields are the major factor that induces such offsets in $v_{\rm ROT}$.
As presented in the residual maps between the input and model velocity fields of Figs.~\ref{lvhis1} to \ref{lvhis24}, the residual LOS velocities
are mostly smaller than 2 \kms\ on average which is less than a half of the channel resolution for
the LVHIS data. This indicates that the fits themselves were made without any
convergence failure, regardless of the degeneracy between $v_{\rm ROT}$ and $i$ found in
the less-resolved and less-inclined galaxies. 

In summary, as found in the performance test using the artificial galaxies, 
{\sc 2dbat} provides reliable 2D tilted-ring parameters and rotation velocities in a fully
automated manner for the LVHIS sample galaxies with intermediate inclinations ($20^{\circ}\--70^{\circ}$)
resolved by more than four beams across their semi-major axis.
The best fit values and models of centre
position, $v_{\rm SYS}$, $\phi$ and $i$ are representative enough to account for their radial
scatter seen when a fit is made with all the ring parameters free. 
Moreover, they are also largely comparable to those derived in manual
using {\sc rotcur}. 
For most galaxies, higher-order regularisation of $\phi$ and $i$ with the
respective B-splines provides a better kinematic description
while taking into account the level of radial variations in terms
of the calculated BIC statistics. 
On the other hand, for either the less-resolved ($<$ four beams across the
semi-major axis), less or highly-inclined ($<30^{\circ}$ or $>70^{\circ}$) galaxies, the $\phi$
and $i$ derived using {\sc 2dbat} tend to deviate from the ones calculated from the
optical axis ratios although some are comparable to the {\sc rotcur} results.
The enhanced degeneracy between rotation velocity and inclination in
such less-resolved, less or highly-inclined galaxies is most likely to be the major
reason for the large deviations.

\section{Conclusions} \label{concolusions}
In this paper we present a newly developed 2D tilted-ring fitting algorithm
based on a Bayesian MCMC technique which allows us to derive 2D tilted-ring
parameters and rotation curves of disk galaxies in a fully automated manner.  In
the algorithm, the ring parameters except for rotation velocity are grouped in
two sub-groups, (1) kinematic centre and systemic velocity, and (2) position
angle, inclination, and expansion velocity, which are regularised by single
values and B-spline functions, respectively. The Einasto halo model
rotation velocity comprising three free parameters is used for parameterizing 
the rotation velocity which is then used together with the other ring
parameters for building a 2D kinematic disk model. The disk model is then fitted
to the entire region of the input velocity field (without dividing them into individual tilted-rings
as in the traditional tilted-ring analysis) in a Bayesian framework at one time.
After determining the geometrical ring parameters of the disk model, such as kinematic centre,
position angle and inclination, the final rotation velocities are fitted to
the tilted-rings defined with the derived geometrical parameters.

For the 2D Bayesian model fitting, we have developed a standalone software
written in C, the so-called {\sc 2dbat} which employs {\sc multinest}
\citep{2008MNRAS.384..449F,2009MNRAS.398.1601F,2013arXiv1306.2144F}, a Bayesian
inference tool library implementing the nested sampling algorithm.  {\sc
multinest} has been found to be efficient and robust in calculating the
posterior distribution and the evidence for a given likelihood function, even in
high dimensions, and successfully used in a wide range of astrophysical
inference problems.  The most important advantage of {\sc 2dbat} based on the
Bayesian MCMC analysis is that only broadly defined ranges are required for the
prior of each ring parameter, which makes the fitting procedure
fully automated. 

To improve the fit quality and reduce the processing time, it
includes some pre-processing steps, such as (1) masking outlying pixels out in
the input velocity field and (2) providing initial priors. {\sc 2dbat} then
derives the best fits of the ring parameters by calculating the maximum
likelihood estimates of 2D kinematic models for a given 2D velocity field.
To further minimise the processing time, it is written in MPI, which
ensures the parallel implementation of the {\sc multinest} on either
a multi-core single or cluster system. 

We test {\sc 2dbat} on the Hermite
$h_3$ velocity fields of 24 LVHIS sample galaxies \citep{Koribalski_2010} as well as 52 artificial
galaxies presented in \cite{2015MNRAS.452.3139K} using two regularisation regimes
(constant or high-order $\phi$ and $i$), and derive (1) 2D tilted-ring
parameters, (2) rotation curves and (3) model velocity
fields. The fit results are then compared with those that were used to construct
the artificial galaxies and those derived using {\sc rotcur} in GIPSY by hand for the
LVHIS sample galaxies, respectively.
From this, we found that, for the galaxies with moderate inclinations
($20^{\circ}\--70^{\circ}$) resolved by more than four beams
across the semi-major axis, {\sc 2dbat} is able to provide robust and acceptable fits of 2D
kinematic models in a fully automated manner which are well consistent with
either the input models or the ones derived manually. {\sc 2dbat}
is limited in breaking the degeneracy between rotation velocity and inclination in
the 2D tilted-ring model for poorly sampled ($<4$ beams) galaxies as well as
galaxies outside the range $20^{\circ}\--70^{\circ}$. These suffer the greatest
from the beam smearing effect.
This limitation of 2D tilted-ring analysis would be improved by expanding the current 2D
parameter space of {\sc 2dbat} to the 3D one in a Bayesian framework.

Together with {\sc fat} which is based on {\sc tirific},
{\sc 2dbat} will be useful for robust kinematic analysis
of a large number of galaxies from the upcoming SKA pathfinder galaxy
surveys, such as ASKAP WALLABY and also from other spectral line observations
including optical integral field unit (IFU) spectroscopic or CO observations. 

\acknowledgements This research was conducted by the Australian Research Council Centre of Excellence for All-sky Astrophysics (CAASTRO), through project number CE110001020.

\bibliography{ms}

\clearpage

\onecolumngrid
\renewcommand{\thesection}{Appendix}
\label{Appendix}

\newcounter{appendix_section}
\setcounter{appendix_section}{1}
\setcounter{equation}{0}
\setcounter{figure}{0}
\renewcommand{\thesection}{A.\arabic{appendix_section}}
\renewcommand{\theequation}{A-1.\arabic{equation}}
\renewcommand{\thefigure}{A-2.\arabic{figure}}

\section{Error propagation of the Einasto halo rotation velocity}

In this Appendix, we provide an error propagation for the three parameters of the Einasto halo rotation velocity in Eq.~\ref{eq:14}, which is given as,

\begin{align}
	v_{\rm E}(r) &= \sqrt{\frac{GM_{\rm E}(r)}{r}} \nonumber \\
					  &= \sqrt{4\pi G n \frac{r^{3}_{-2}}{r} \rho_{-2} e^{2n}
(2n)^{-3n} \gamma (3n, \frac{r}{r_{-2}})}.
	\label{eqa:1}
\end{align}

We compute the total uncertainty in $v_{\rm E}(r)$, $\sigma_{v_{\rm E}}$ which is propagated from the 1$\sigma$ errors ($\sigma_{n}$, $r_{\rm -2}$, and $\sigma_{\rho_{\rm -2}}$) of the thee parameters derived from {\sc 2dbat} as follows:

\begin{align}\label{eqa:2}
\sigma_{v_{\rm E}} &= \sqrt{{(\sigma_{n}} \frac{\partial v_{\rm
E}}{\partial n})^{2} + (\sigma_{r_{\rm -2}} \frac{\partial v_{\rm
E}}{\partial r_{\rm -2}})^{2} + (\sigma_{\rho_{\rm -2}} \frac{\partial v_{\rm
E}}{\partial \rho_{\rm -2}})^{2}
+ 2 \frac{\partial v_{\rm E}}{\partial n} \frac{\partial v_{\rm E}}{\partial r_{-2}}\sigma_{n r_{-2}}
+ 2 \frac{\partial v_{\rm E}}{\partial r_{-2}} \frac{\partial v_{\rm E}}{\partial \rho_{-2}}\sigma_{r_{-2}\rho_{-2}}
+ 2 \frac{\partial v_{\rm E}}{\partial n} \frac{\partial v_{\rm E}}{\partial \rho_{-2}}\sigma_{n \rho_{-2}}}
\end{align}

\noindent where $\sigma_{n}$, $\sigma_{r_{-2}}$, and $\sigma_{\rho_{-2}}$ are the standard deviations of
the fitted $n$, $r_{-2}$ and $\rho_{-2}$ of the Einasto halo rotation velocity from the Bayesian analysis.
$\sigma_{n r_{-2}}$, $\sigma_{r_{-2}\rho_{-2}}$, and $\sigma_{n \rho_{-2}}$ are the covariances between the
parameters which are given by
\begin{equation}\label{eqa:3}
	\sigma_{n r_{-2}} = \frac{1}{N}\sum_{i=1}^{N}(n^{i}-n)(r_{-2}^{i}-r_{-2}),
\end{equation}

\begin{equation}\label{eqa:4}
	\sigma_{r \rho_{-2}} = \frac{1}{N}\sum_{i=1}^{N}(r_{-2}^{i}-r_{-2})(\rho_{-2}^{i}-\rho_{-2}),
\end{equation}

\begin{equation}\label{eqa:5}
	\sigma_{n \rho_{-2}} = \frac{1}{N}\sum_{i=1}^{N}(n^{i}-n)(\rho_{-2}^{i}-\rho_{-2}).
\end{equation}

In addition, the partial derivatives of $v_{\rm E}$ with respect to $n$, $r_{-2}$, and $\rho_{-2}$ are given by
\begin{equation}\label{eqa:6}
\frac{\partial v_{\rm E}}{\partial n} = \frac{1}{2}v_{\rm E}^{-\frac{1}{2}} 4\pi G \frac{r_{-2}^{3}}{r} \rho_{-2} e^{2n} (2n)^{-3n} \times
\biggl[\gamma(3n, \frac{r}{r_{-2}}) + 2n\gamma(3n, \frac{r}{r_{-2}}) - 3n(1+{\rm log}\,2+{\rm log}\,n)\gamma(3n, \frac{r}{r_{-2}})
+ n\frac{\partial \gamma(3n, \frac{r}{r_{-2}})}{\partial n}\biggr]
\end{equation}

\begin{equation}\label{eqa:7}
\frac{\partial v_{\rm E}}{\partial r_{-2}} = \frac{1}{2}v_{\rm E}^{-\frac{1}{2}} 4\pi G \frac{r_{-2}^{2}}{r} \rho_{-2} e^{2n} (2n)^{-3n} \times
\biggl[3\gamma(3n, \frac{r}{r_{-2}}) + r_{-2}\frac{\partial \gamma(3n, \frac{r}{r_{-2}})}{\partial r_{-2}}\biggr]
\end{equation}

\begin{equation}\label{eqa:8}
\frac{\partial v_{\rm E}}{\partial \rho_{-2}} = \frac{1}{2}v_{\rm E}^{-\frac{1}{2}} 4\pi G \frac{r_{-3}^{2}}{r} e^{2n} (2n)^{-3n}\gamma(3n, \frac{r}{r_{-2}})
\end{equation}

\noindent where $\gamma$ is the lower incomplete gamma function given by,
\begin{equation}\label{eqa:9} 
	\gamma (3n, x) = \int_{0}^{x} dt\,e^{-t} t^{3n-1},
\end{equation}

\noindent and the partial derivatives of $\gamma(3n, \frac{r}{r_{-2}})$ with respect to $r_{-2}$ and $n$ are: 

\begin{eqnarray}\label{eqa:10}
\frac{\partial \gamma(3n, \frac{r}{r_{-2}})}{\partial r_{-2}} &=& \frac{\partial}{\partial r_{-2}} \int_{0}^{\frac{r}{r_{-2}}} dt\,e^{-t}\,t^{3n-1} \\ \nonumber
&=& -\frac{r}{r_{-2}^{2}}\biggl(\frac{r}{r_{-2}}\biggr)^{3n-1}\,e^{-\frac{r}{r_{-2}}}
\end{eqnarray}

\begin{eqnarray}\label{eqa:11}
\frac{\partial \gamma(3n, \frac{r}{r_{-2}})}{\partial n} &=& \frac{\partial}{\partial n} \int_{0}^{\frac{r}{r_{-2}}} dt\,e^{-t} t^{3n-1} \\ \nonumber
&=& \int_{0}^{\frac{r}{r_{-2}}} dt\,3\,{\rm log}\,t\,e^{-t}\,t^{3n-1}.
\end{eqnarray}

The value of partial derivative, $\frac{\partial \gamma(3n, \frac{r}{r_{-2}})}{\partial n}$ for the derived $n$ and $r_{-2}$ at a galaxy radius $r$ can be
then computed numerically.

%%%%%%%%%%%%%%%%%%%%%%%%%%%%%%%%%%%%%%%%%%%%%%%%%%%%%%%%%%%%%%%%%%%%%%%%%%%%%%
%%%%%%%%%%%%%%%%%%%%%%%%%%%%%%%%%%%%%%%%%%%%%%%%%%%%%%%%%%%%%%%%%%%%%%%%%%%%%%
%%%%%%%%%%%%%%%%%%%%%%%%%%%%%%%%%%%%%%%%%%%%%%%%%%%%%%%%%%%%%%%%%%%%%%%%%%%%%%

\setcounter{appendix_section}{2}
\section{{\sc 2dbat} analysis for LVHIS sample galaxies}\label{A-2}

In this appendix, we present the 2D tilted-ring analysis using {\sc 2dbat} for the 24 LVHIS
sample galaxies. For each galaxy, we show: {\bf (a)} ATCA H{\sc i} Hermite $h_3$ velocity field (VF),
error (VF\--error), moment maps (MOM0 \& MOM2), model (MODEL) and residual (RES) velocity fields.
The beam size is indicated by the ellipse in the bottom-right corner of each panel. The BIC values
derived from the Bayesian fits for the model velocity fields are denoted in the panels of model velocity
fields, respectively; {\bf (b)} 2D tilted-ring analysis\-- the rotation curves derived using the Hermite
$h_3$ velocity field in the two regularisation modes (green
open squares: constant, brown open circles: higher-order B-splines,
grey filled circles: fit results with all the ring parameters free,
orange cross mark: manual fit results derived using {\sc rotcur} which were also used for
testing the performance of {\sc fat} in \cite{2015MNRAS.452.3139K}).
See Section~\ref{model_fit_results} for more details; {\bf (c)} Correlations of the ring parameters\-- the black
contours and histograms show the posterior constraints from the Bayesian
analysis. The best fits are indicated by red lines.

%\section{{\sc 2dbat} fit results of LVHIS sample galaxies}
%: 10. HIPASS J1441-62.atlas.ps
\begin{figure*} \epsscale{1.0}
\includegraphics[angle=0,width=1.0\textwidth,bb=100 160 520 610,clip=]
{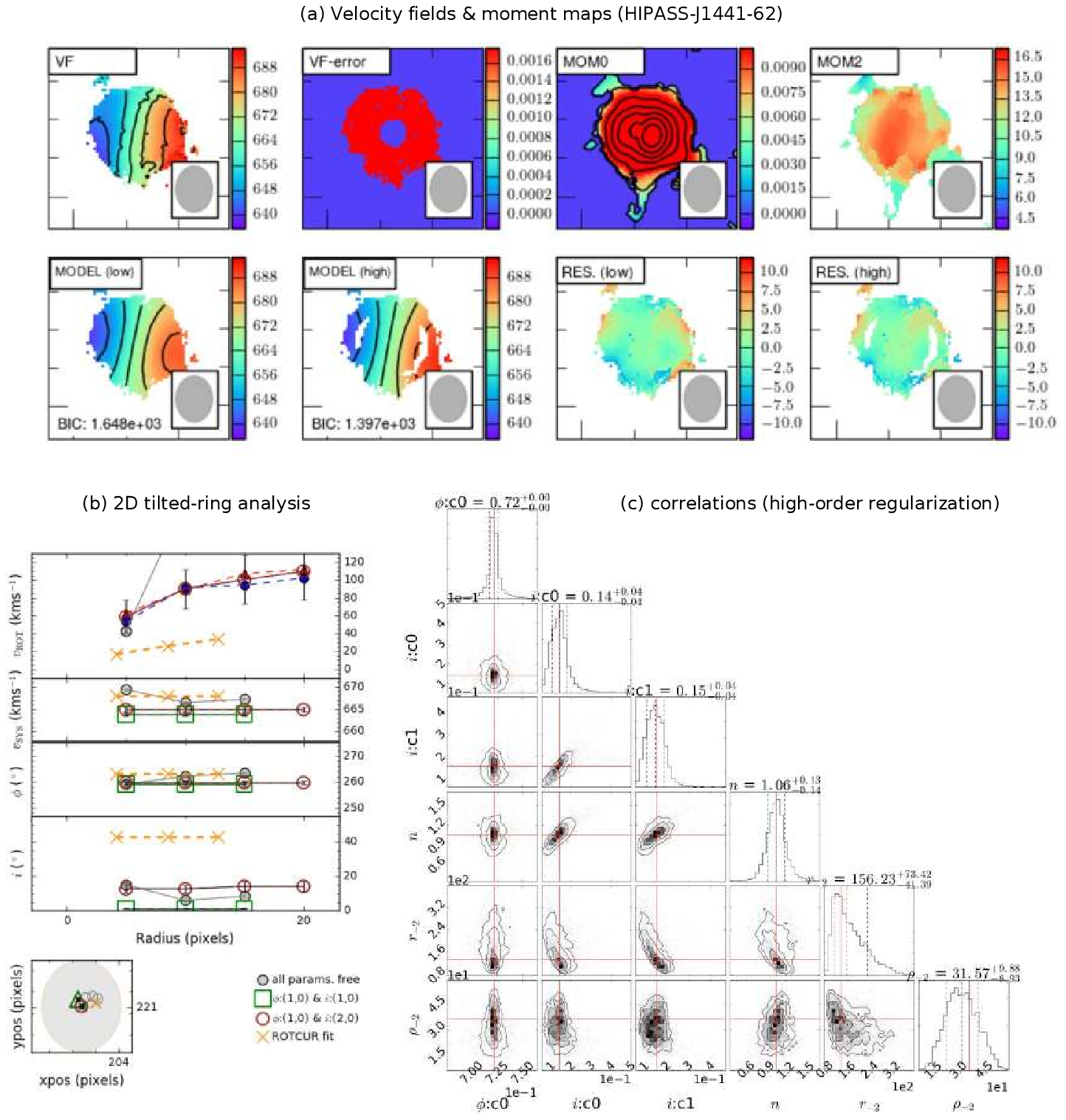}
\caption{{\sc 2dbat} analysis for HIPASS J1441-62: Contours in {\bf (a)} are spaced by 10 \kms\ on the velocity fields,
and 0.1 $\rm mJy\,beam^{-1}$ on the moment 0. The pixel scale in {\bf (b)} is 5 \arcsec. See Appendix section~\ref{A-2} for details.
\label{lvhis1}}
\end{figure*}

% 11. HIPASS J1305-40
\begin{figure*} \epsscale{1.0}
\includegraphics[angle=0,width=1.0\textwidth,bb=100 160 520 610,clip=]
{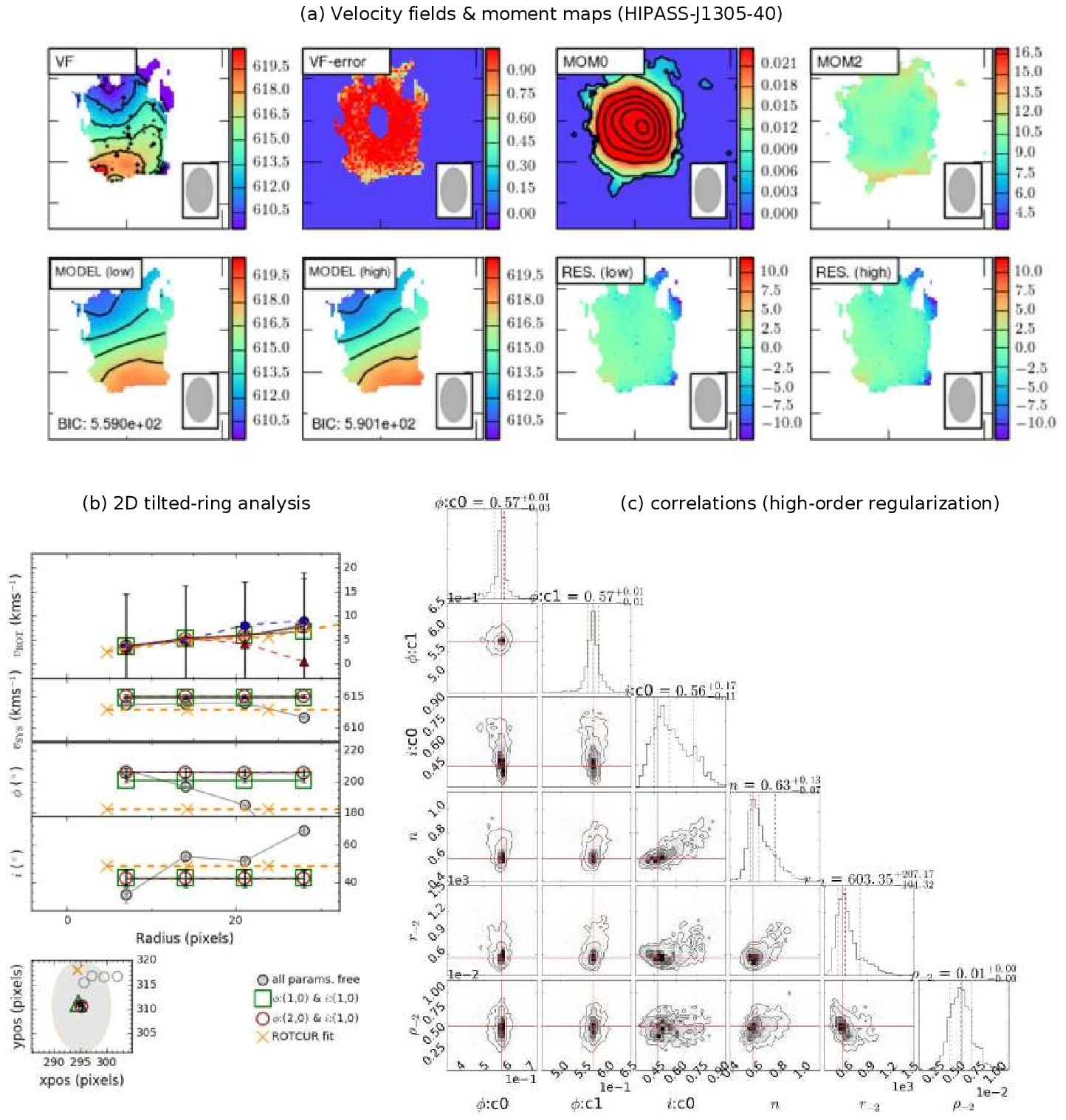}
\caption{{\sc 2dbat} analysis for HIPASS J1305-40:
Contours in {\bf (a)} are spaced by 2 \kms\ on the velocity fields,
and 0.1 $\rm mJy\,beam^{-1}$ on the moment 0. The pixel scale in {\bf (b)} is 4 \arcsec. See Appendix section~\ref{A-2} for details.
\label{lvhis2}}
\end{figure*}

%: 12. HIPASS J0320-52.atlas.ps
\begin{figure*} \epsscale{1.0}
\includegraphics[angle=0,width=1.0\textwidth,bb=100 160 520 610,clip=]
{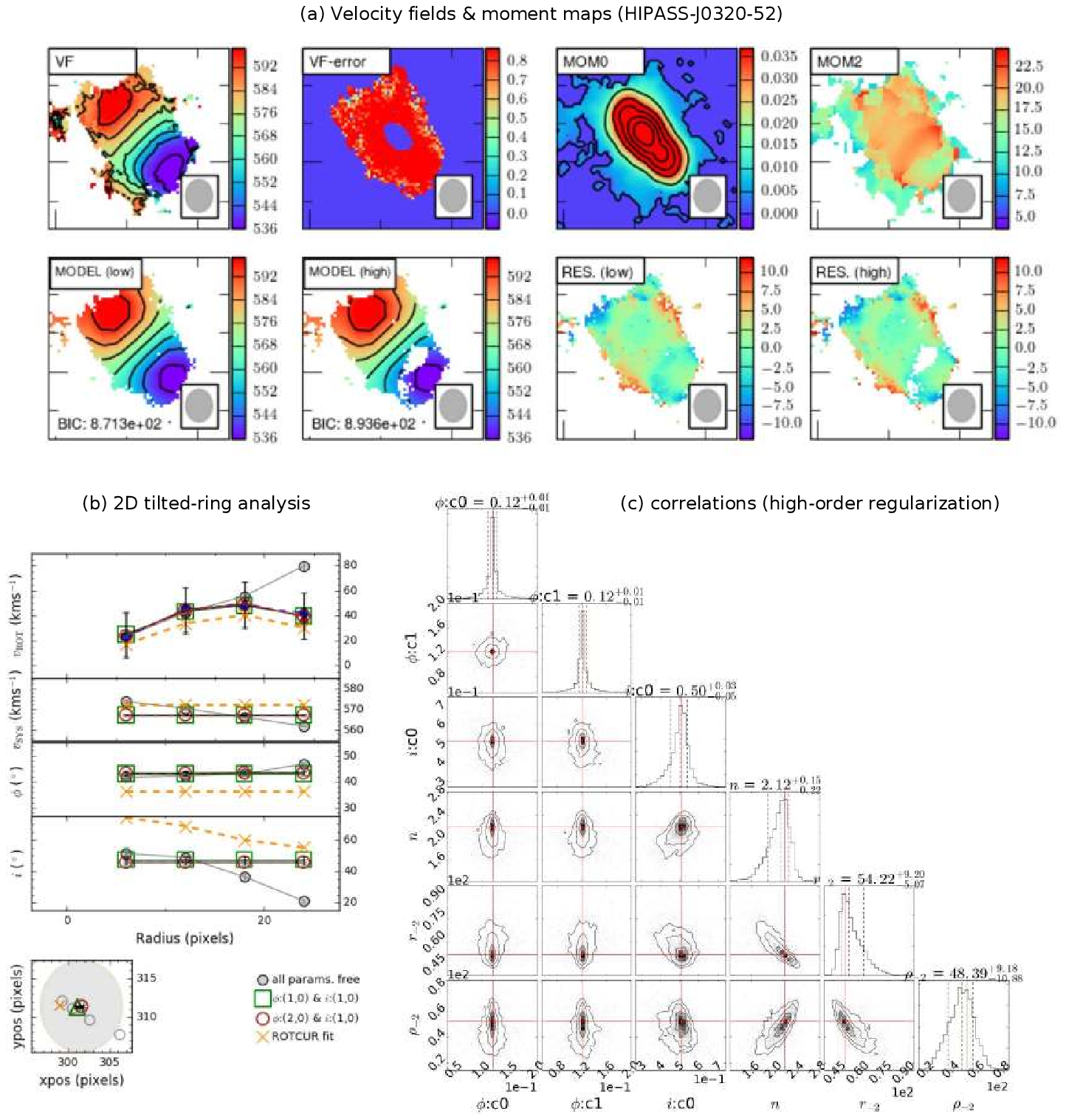}
\caption{{\sc 2dbat} analysis for HIPASS J0320-52:
Contours in {\bf (a)} are spaced by 10 \kms\ on the velocity fields,
and 0.1 $\rm mJy\,beam^{-1}$ on the moment 0. The pixel scale in {\bf (b)} is 5 \arcsec. See Appendix section~\ref{A-2} for details.
\label{lvhis3}}
\end{figure*}

%: 13. HIPASS J1337-39.atlas.ps
\begin{figure*} \epsscale{1.0}
\includegraphics[angle=0,width=1.0\textwidth,bb=100 160 520 610, clip=]
{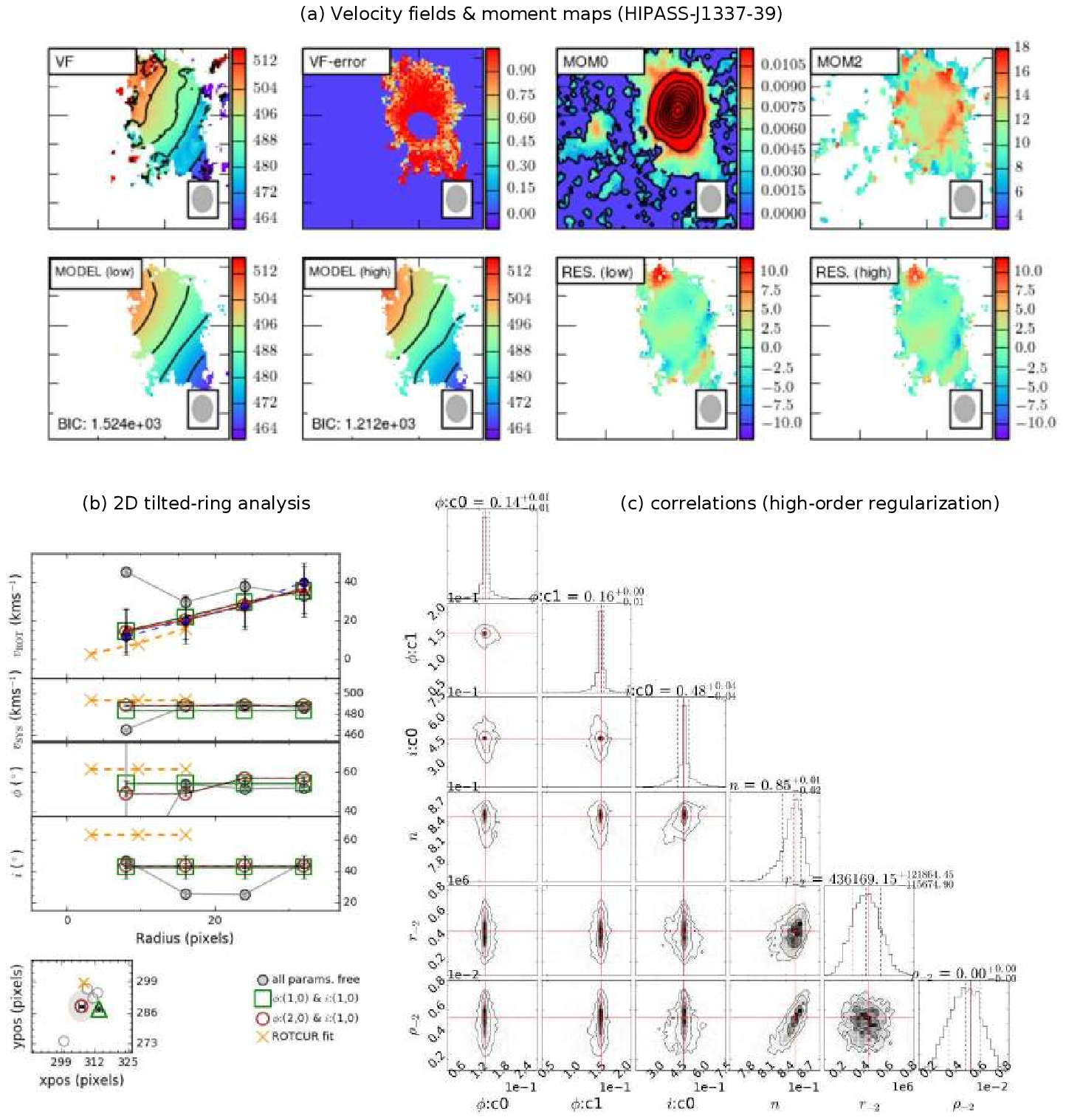}
\caption{{\sc 2dbat} analysis for HIPASS J1337-39:
Contours in {\bf (a)} are spaced by 10 \kms\ on the velocity fields,
and 0.1 $\rm mJy\,beam^{-1}$ on the moment 0. The pixel scale in {\bf (b)} is 5 \arcsec. See Appendix section~\ref{A-2} for details.
\label{lvhis4}}
\end{figure*}

%: 14. HIPASS J1219-79.atlas.ps
\begin{figure*} \epsscale{1.0}
\includegraphics[angle=0,width=1.0\textwidth,bb=100 160 520 610,clip=]
{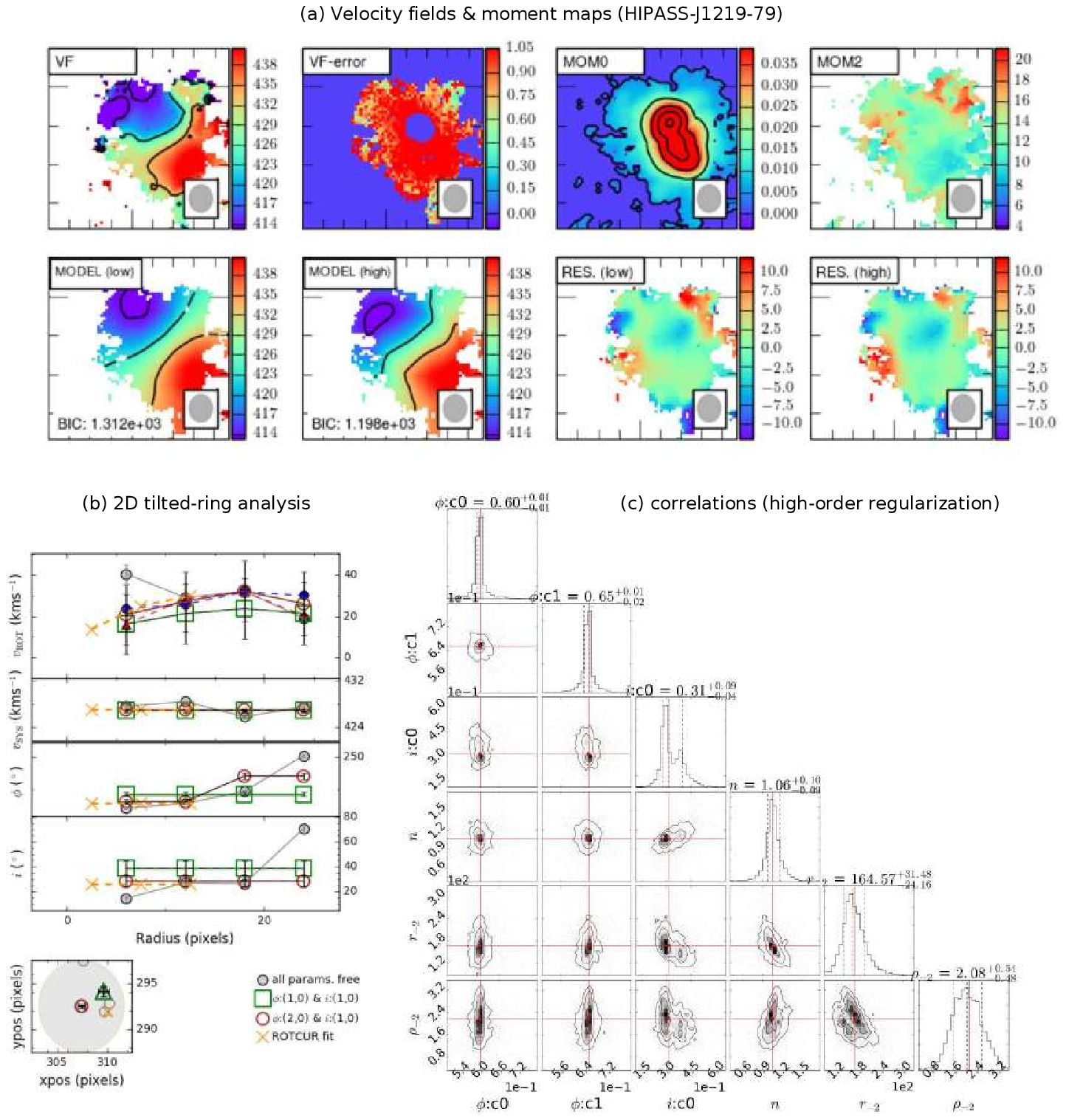}
\caption{{\sc 2dbat} analysis for HIPASS J1219-79:
Contours in {\bf (a)} are spaced by 10 \kms\ on the velocity fields,
and 0.1 $\rm mJy\,beam^{-1}$ on the moment 0. The pixel scale in {\bf (b)} is 5 \arcsec. See Appendix section~\ref{A-2} for details.
\label{lvhis5}}
\end{figure*}

%: 15. HIPASS J1047-38.atlas.ps
\begin{figure*} \epsscale{1.0}
\includegraphics[angle=0,width=1.0\textwidth,bb=100 160 520 610,clip=]
{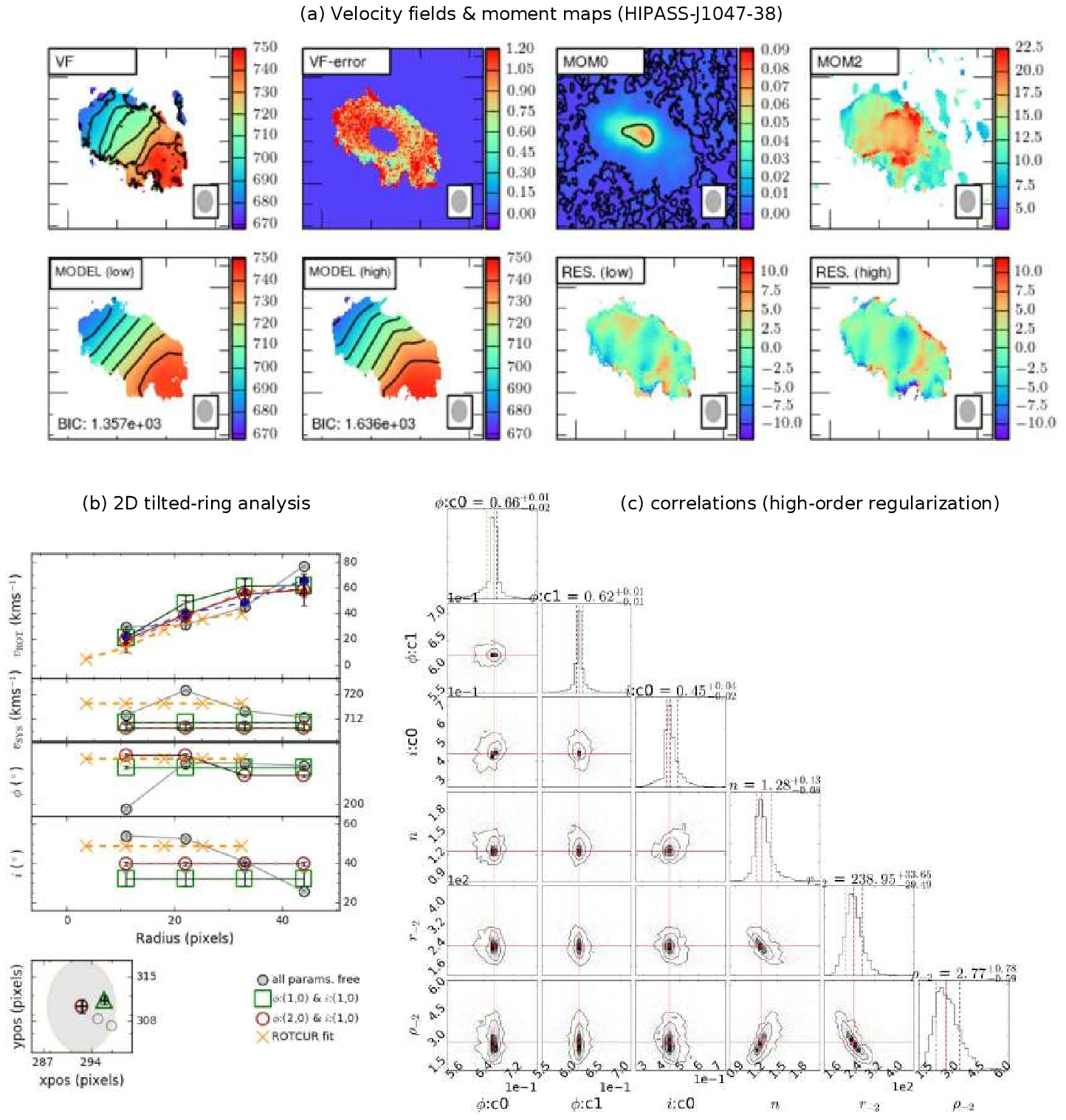}
\caption{{\sc 2dbat} analysis for HIPASS J1047-38:
Contours in {\bf (a)} are spaced by 10 \kms\ on the velocity fields,
and 0.1 $\rm mJy\,beam^{-1}$ on the moment 0. The pixel scale in {\bf (b)} is 5 \arcsec. See Appendix section~\ref{A-2} for details.
\label{lvhis6}}
\end{figure*}

%: 16. HIPASS J1428-46.atlas.ps
\begin{figure*} \epsscale{1.0}
\includegraphics[angle=0,width=1.0\textwidth,bb=100 160 520 610,clip=]
{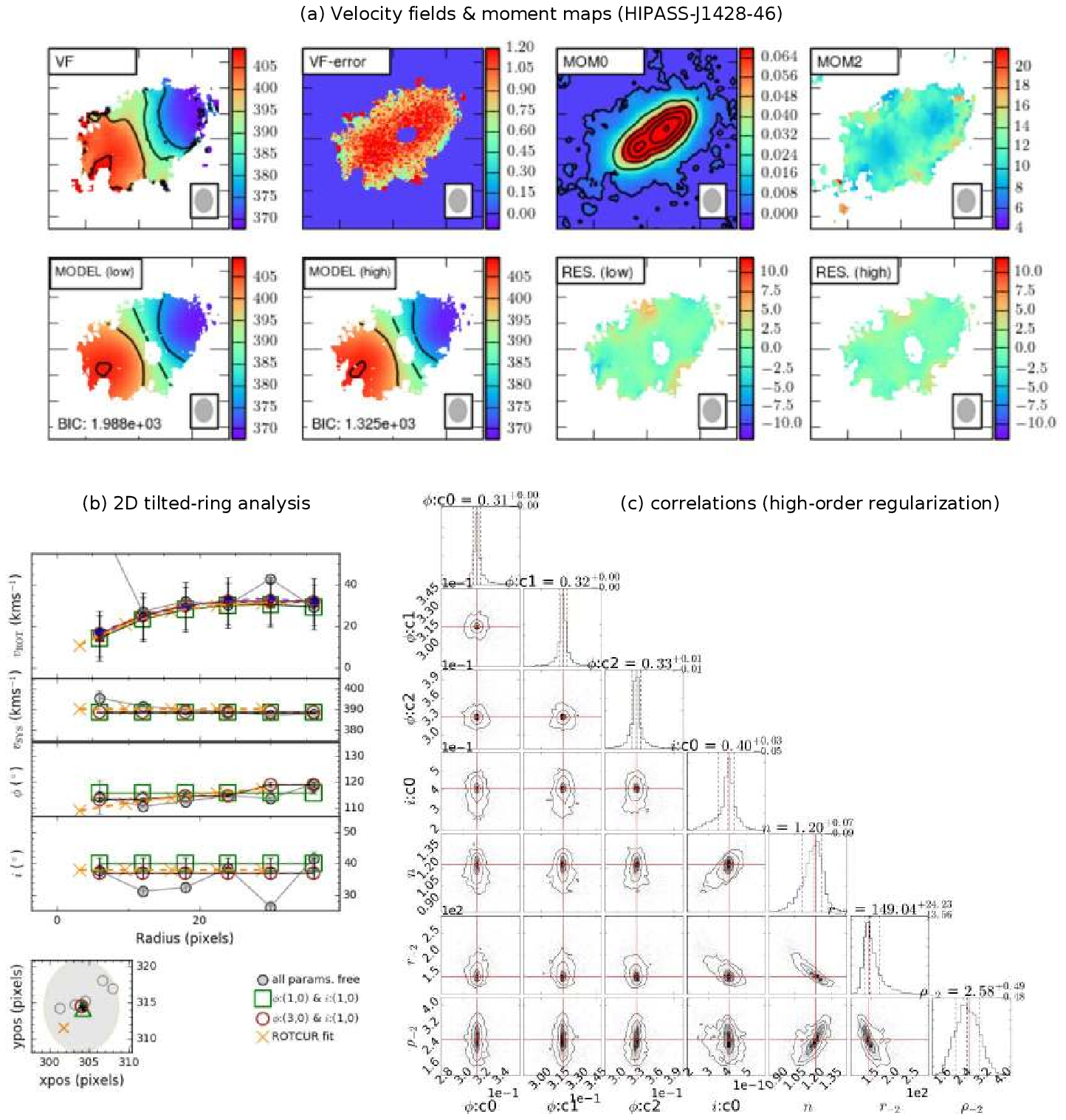}
\caption{{\sc 2dbat} analysis for HIPASS J1428-46:
Contours in {\bf (a)} are spaced by 10 \kms\ on the velocity fields,
and 0.1 $\rm mJy\,beam^{-1}$ on the moment 0. The pixel scale in {\bf (b)} is 5 \arcsec. See Appendix section~\ref{A-2} for details.
\label{lvhis7}}
\end{figure*}

%: 17. HIPASS J1620-60.atlas.ps
\begin{figure*} \epsscale{1.0}
\includegraphics[angle=0,width=1.0\textwidth,bb=100 160 520 610,clip=]
{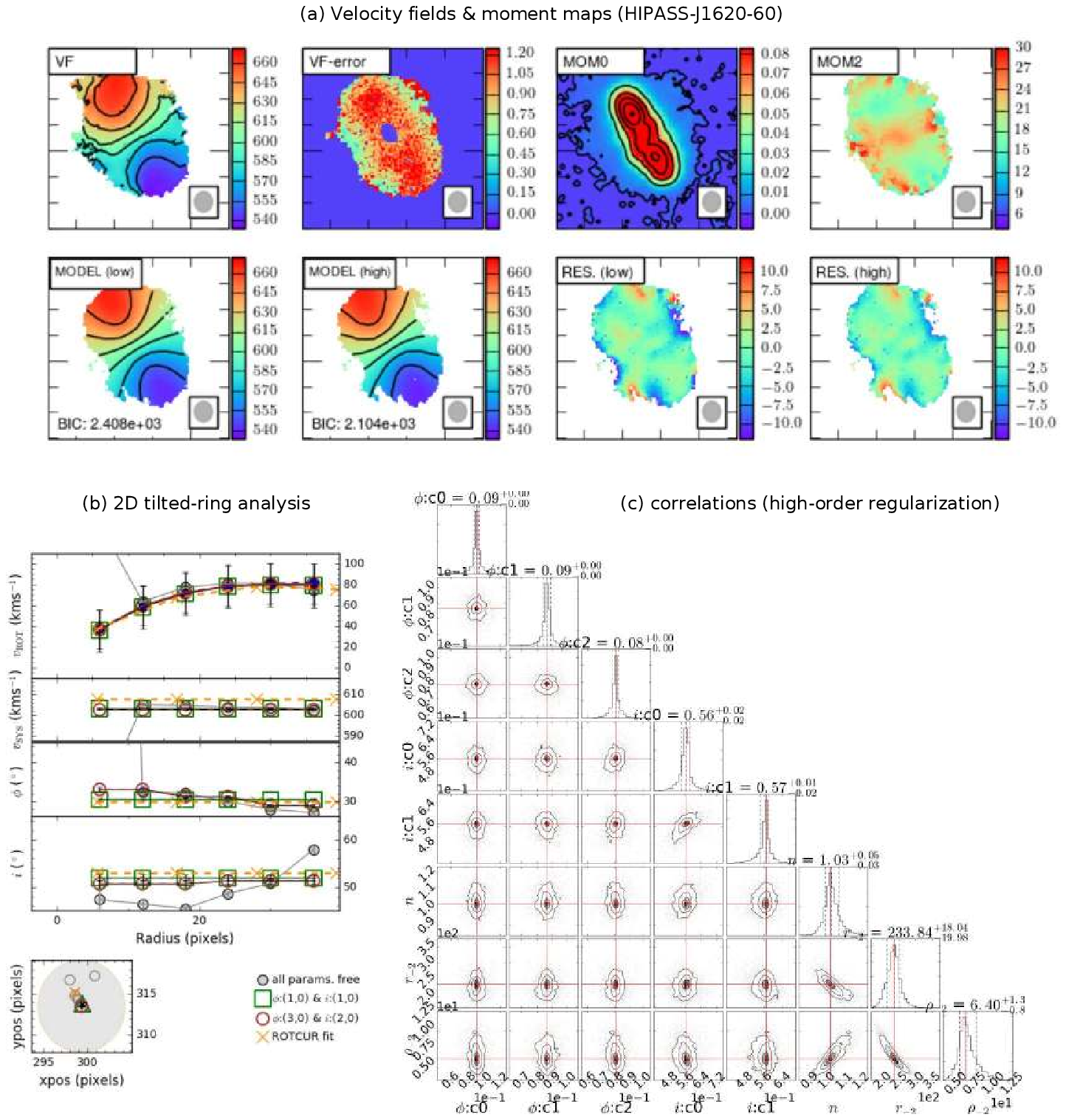}
\caption{{\sc 2dbat} analysis for HIPASS J1620-60:
Contours in {\bf (a)} are spaced by 20 \kms\ on the velocity fields,
and 0.1 $\rm mJy\,beam^{-1}$ on the moment 0. The pixel scale in {\bf (b)} is 5 \arcsec. See Appendix section~\ref{A-2} for details.
\label{lvhis8}}
\end{figure*}

%: 18. HIPASS J0705-58.atlas.ps
\begin{figure*} \epsscale{1.0}
\includegraphics[angle=0,width=1.0\textwidth,bb=100 160 520 610 ,clip=]
{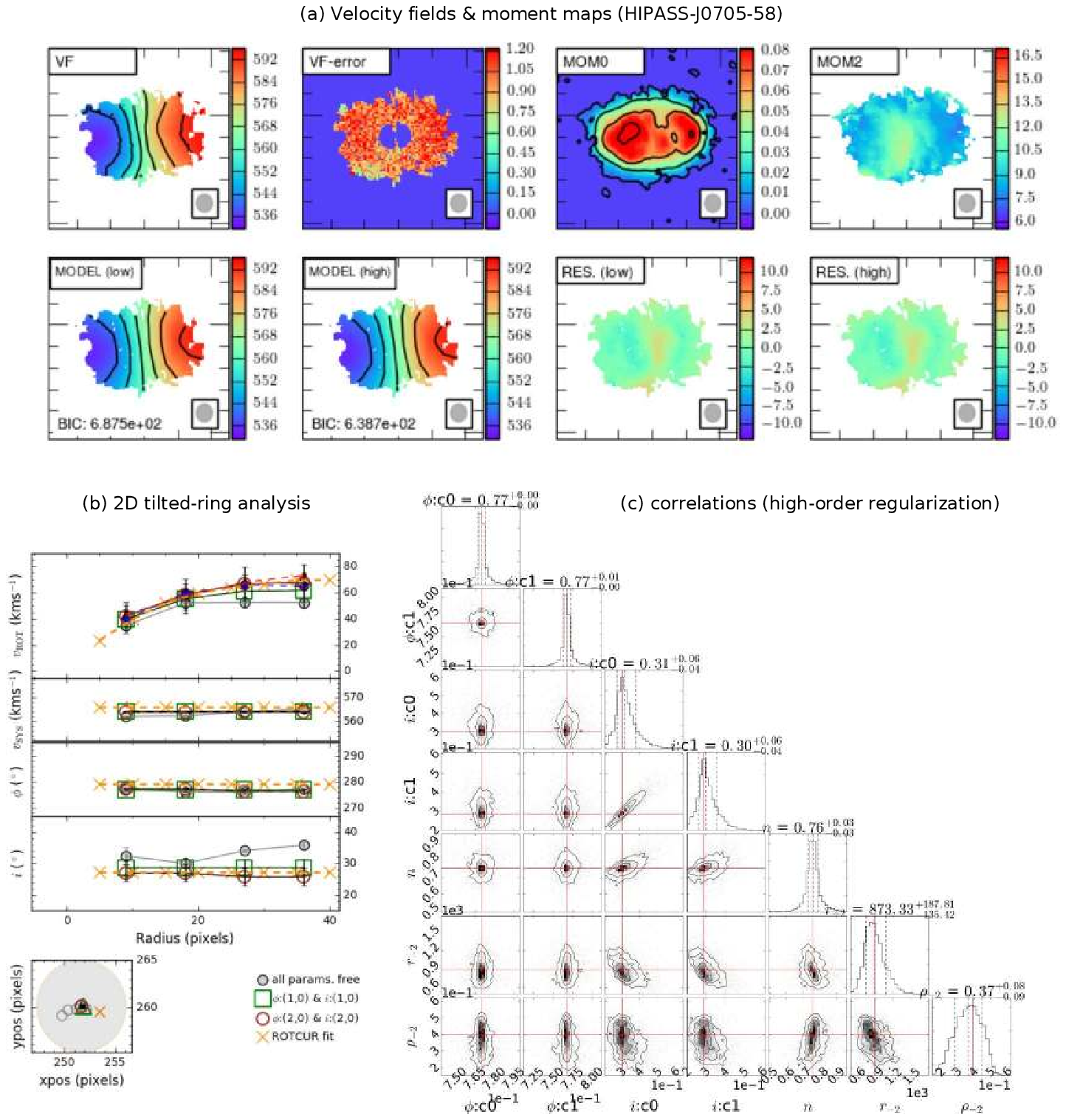}
\caption{{\sc 2dbat} analysis for HIPASS J0705-58:
Contours in {\bf (a)} are spaced by 10 \kms\ on the velocity fields,
and 0.1 $\rm mJy\,beam^{-1}$ on the moment 0. The pixel scale in {\bf (b)} is 6 \arcsec. See Appendix section~\ref{A-2} for details.
\label{lvhis9}}
\end{figure*}

%: 19. HIPASS J1337-28.atlas.ps
\begin{figure*} \epsscale{1.0}
\includegraphics[angle=0,width=1.0\textwidth,bb=100 160 520 610,clip=]
{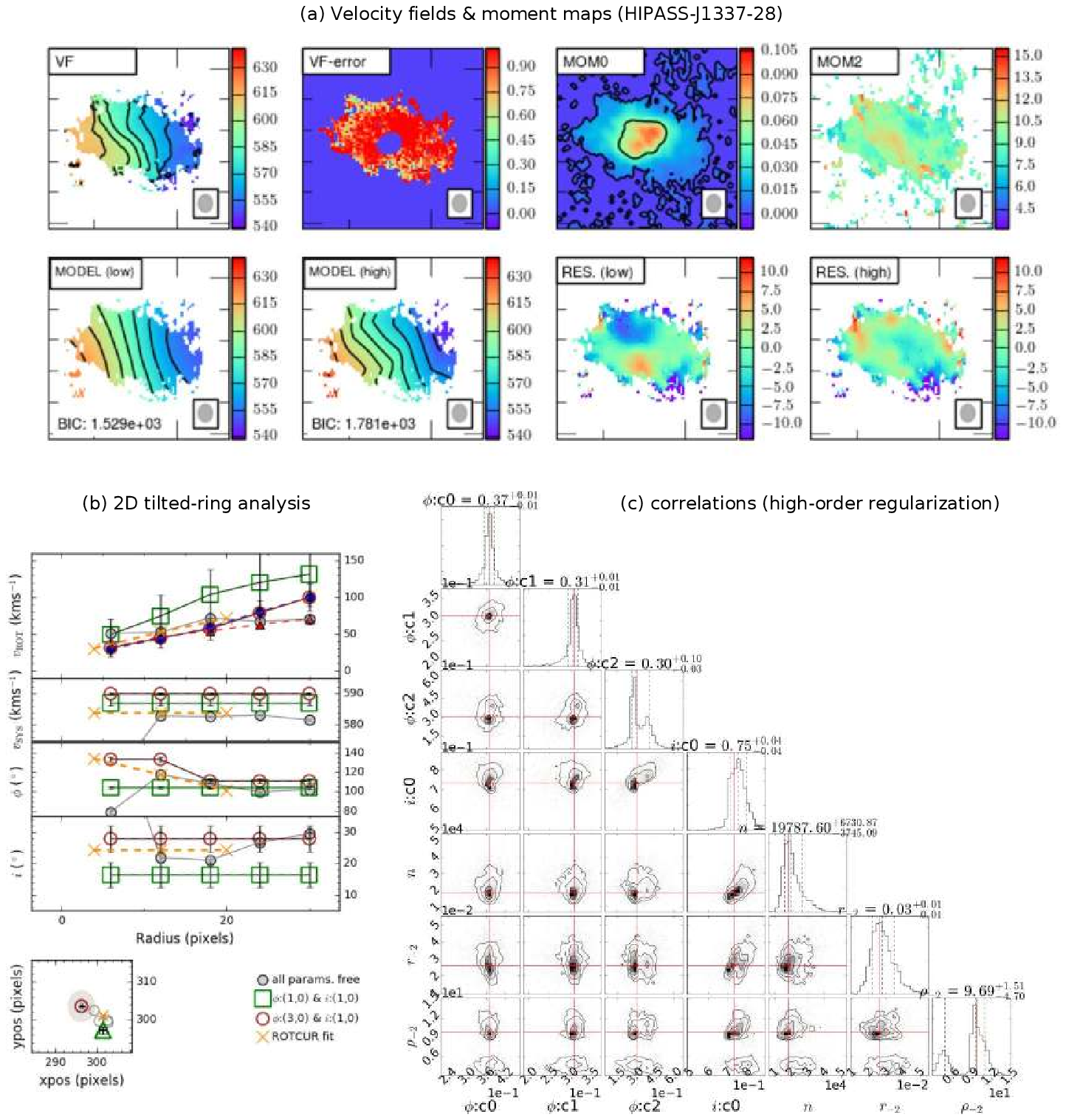}
\caption{{\sc 2dbat} analysis for HIPASS J1337-28:
Contours in {\bf (a)} are spaced by 10 \kms\ on the velocity fields,
and 0.1 $\rm mJy\,beam^{-1}$ on the moment 0. The pixel scale in {\bf (b)} is 5 \arcsec. See Appendix section~\ref{A-2} for details.
\label{lvhis10}}
\end{figure*}

%: 20. HIPASS J0731-68.atlas.ps
\begin{figure*} \epsscale{1.0}
\includegraphics[angle=0,width=1.0\textwidth,bb=100 160 520 610,clip=]
{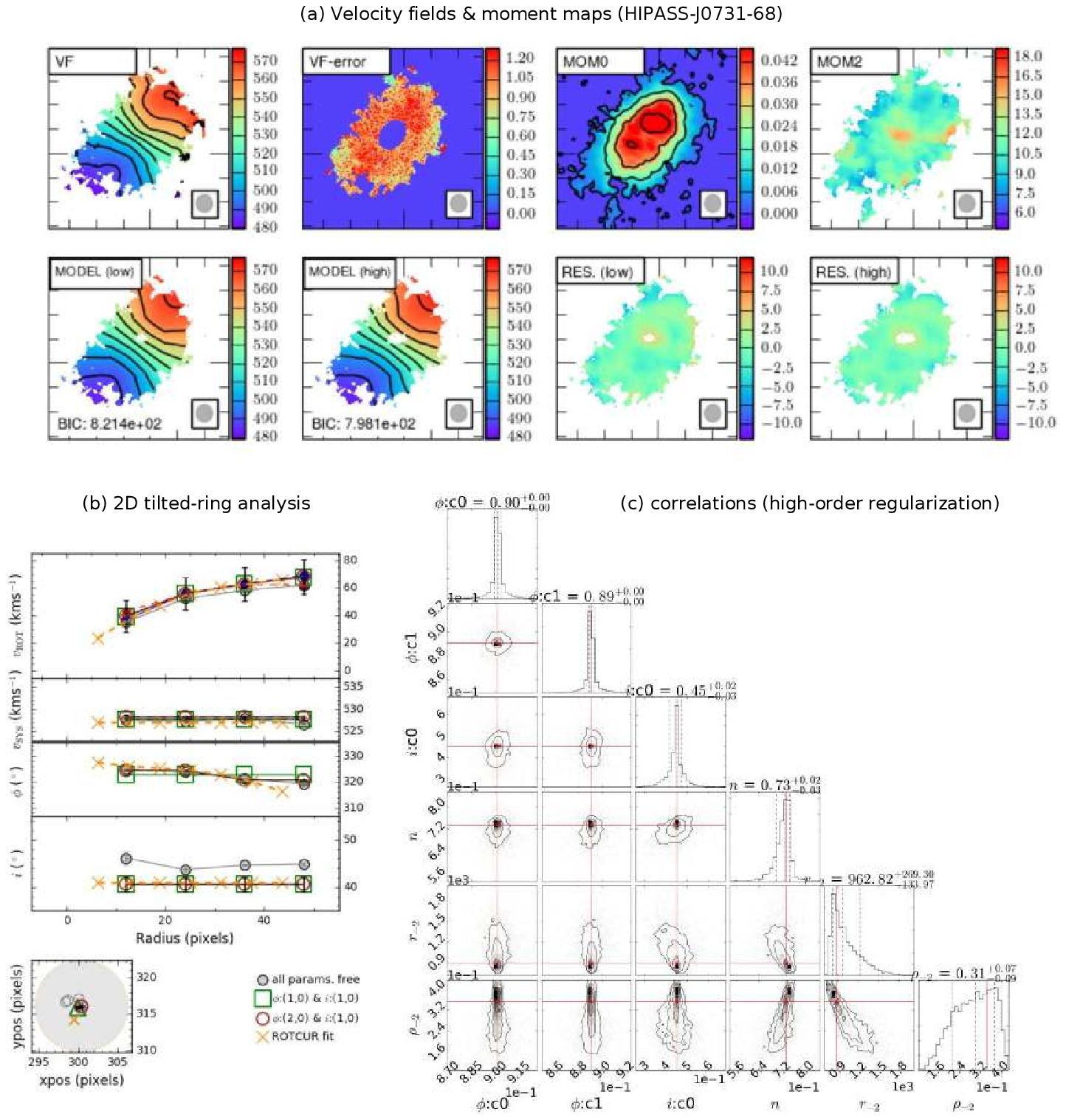}
\caption{{\sc 2dbat} analysis for HIPASS J0731-68:
Contours in {\bf (a)} are spaced by 10 \kms\ on the velocity fields,
and 0.1 $\rm mJy\,beam^{-1}$ on the moment 0. The pixel scale in {\bf (b)} is 4 \arcsec. See Appendix section~\ref{A-2} for details.
\label{lvhis11}}
\end{figure*}

%: 21. HIPASS J0333-50.atlas.ps
\begin{figure*} \epsscale{1.0}
\includegraphics[angle=0,width=1.0\textwidth,bb=100 160 520 610,clip=]
{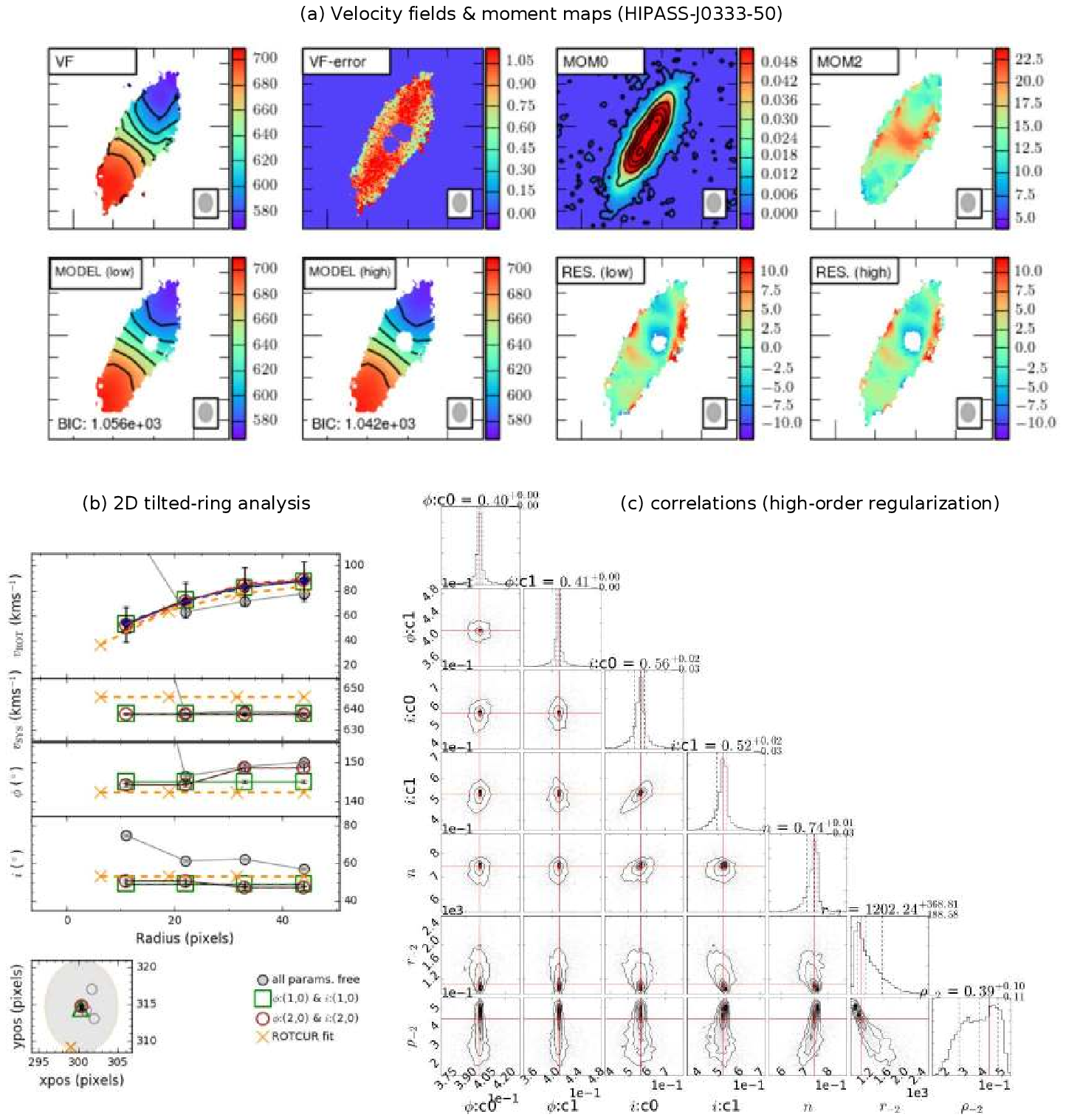}
\caption{{\sc 2dbat} analysis for HIPASS J0333-50:
Contours in {\bf (a)} are spaced by 20 \kms\ on the velocity fields,
and 0.1 $\rm mJy\,beam^{-1}$ on the moment 0. The pixel scale in {\bf (b)} is 5 \arcsec. See Appendix section~\ref{A-2} for details.
\label{lvhis12}}
\end{figure*}

%: 22. HIPASS J1403-41.atlas.ps
\begin{figure*} \epsscale{1.0}
\includegraphics[angle=0,width=1.0\textwidth,bb=100 160 520 610,clip=]
{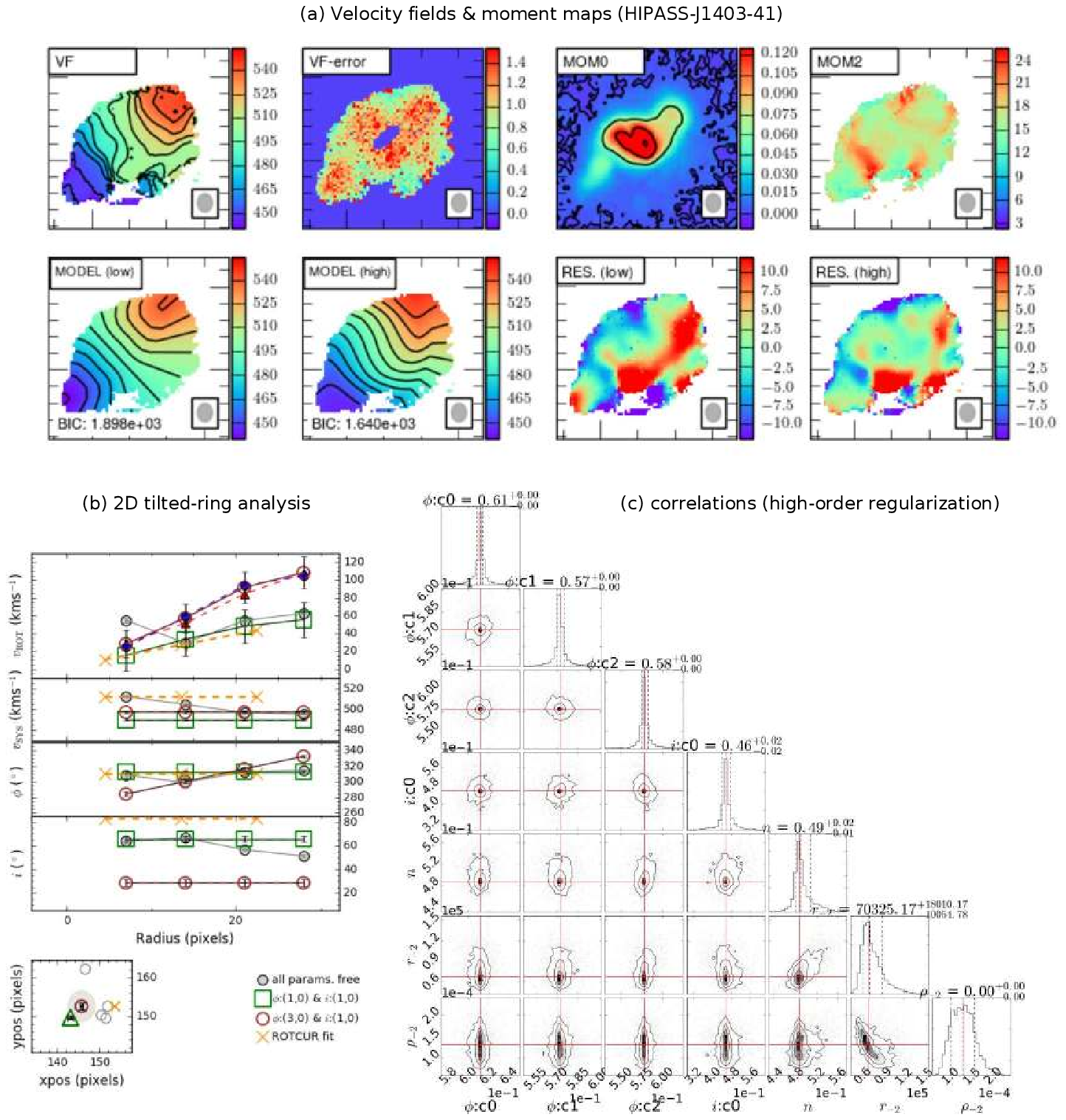}
\caption{{\sc 2dbat} analysis for HIPASS J1403-41:
Contours in {\bf (a)} are spaced by 10 \kms\ on the velocity fields,
and 0.1 $\rm mJy\,beam^{-1}$ on the moment 0. The pixel scale in {\bf (b)} is 10 \arcsec. See Appendix section~\ref{A-2} for details.
\label{lvhis13}}
\end{figure*}

%: 23. HIPASS J1337-42.atlas.ps
\begin{figure*} \epsscale{1.0}
\includegraphics[angle=0,width=1.0\textwidth,bb=100 160 520 610,clip=]
{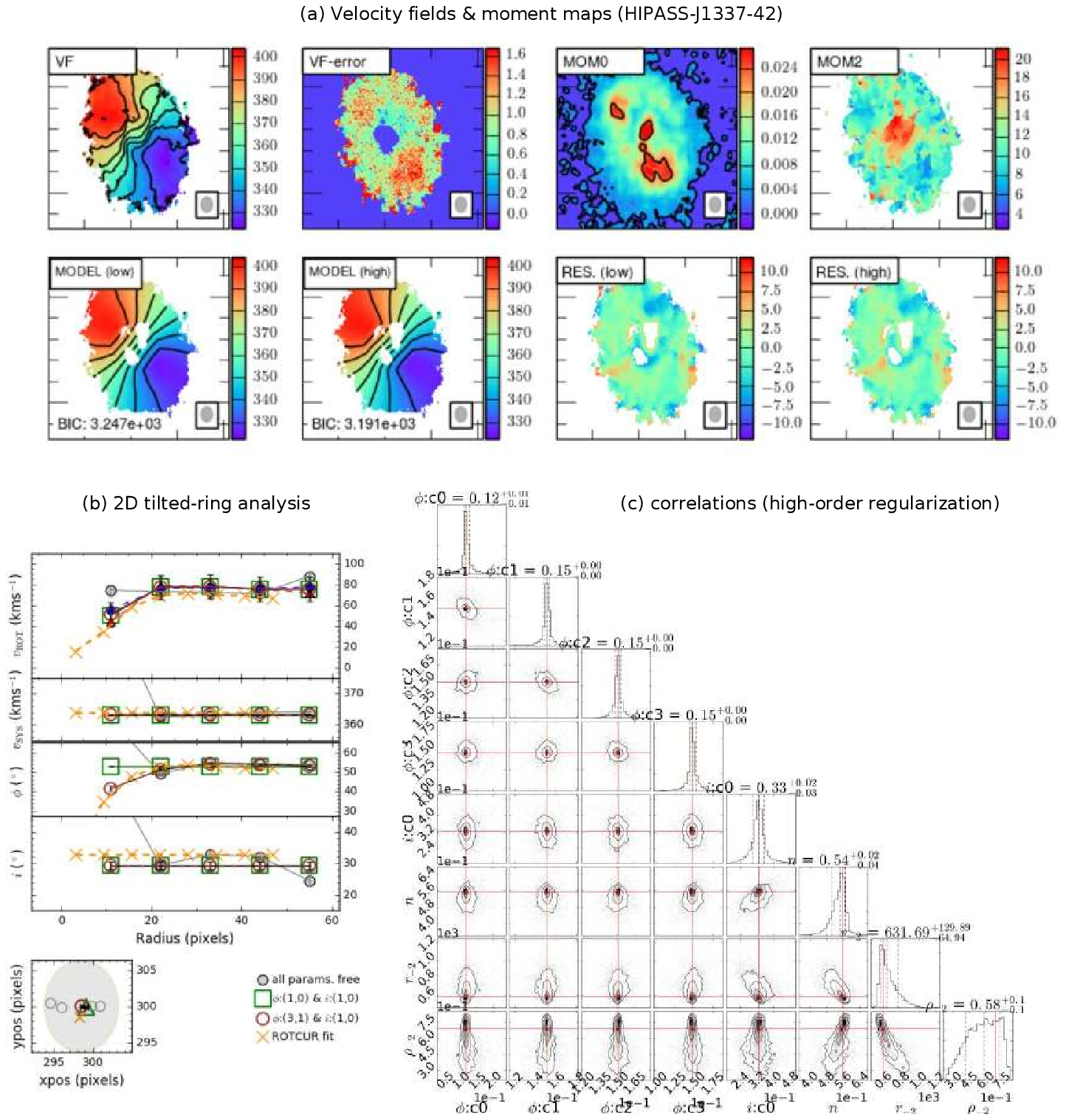}
\caption{{\sc 2dbat} analysis for HIPASS J1337-42:
Contours in {\bf (a)} are spaced by 10 \kms\ on the velocity fields,
and 0.1 $\rm mJy\,beam^{-1}$ on the moment 0. The pixel scale in {\bf (b)} is 4 \arcsec. See Appendix section~\ref{A-2} for details.
\label{lvhis14}}
\end{figure*}

%: 24. HIPASS J1348-53.atlas.ps
\begin{figure*} \epsscale{1.0}
\includegraphics[angle=0,width=1.0\textwidth,bb=100 160 520 610,clip=]
{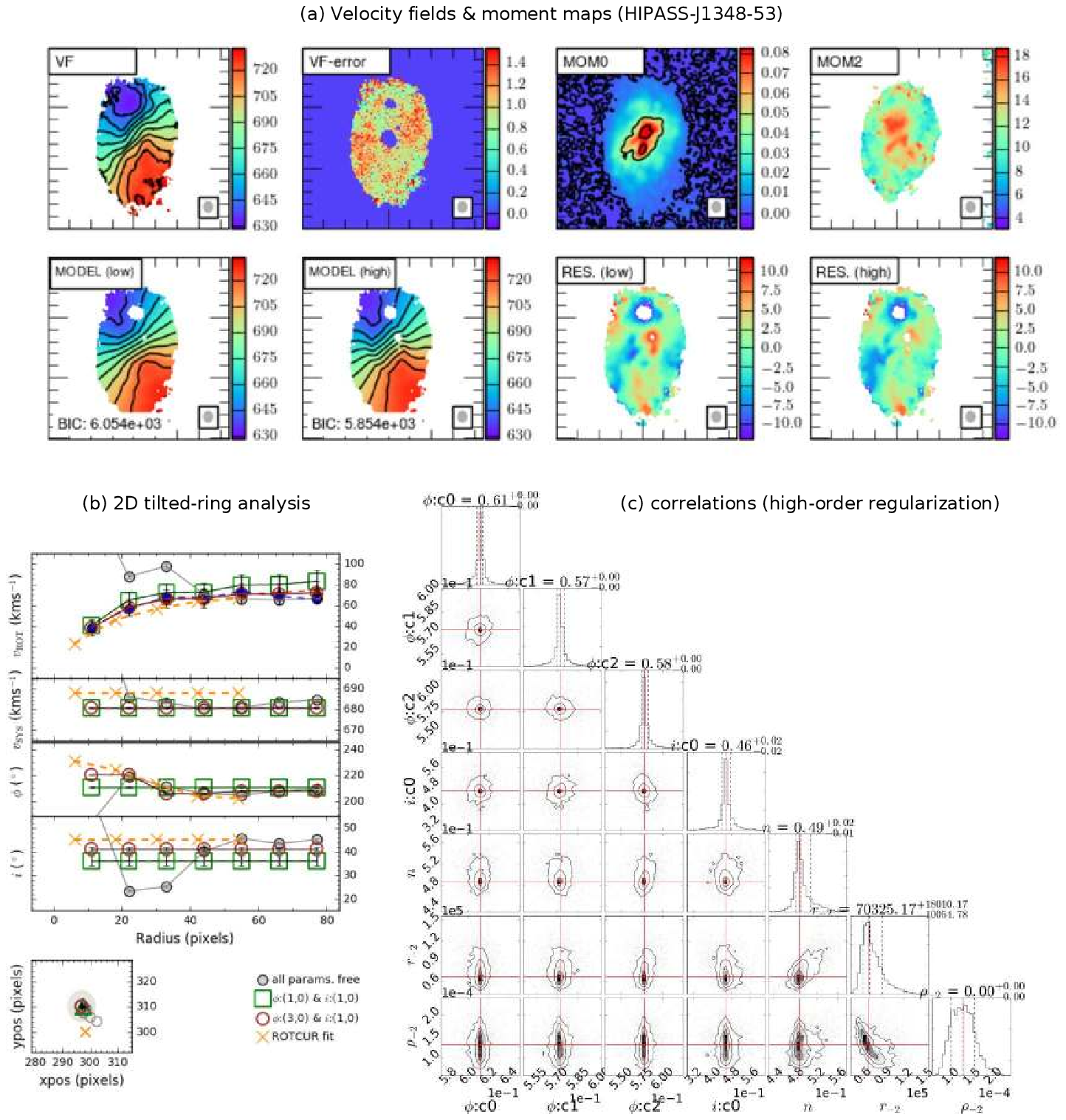}
\caption{{\sc 2dbat} analysis for HIPASS J1348-53:
Contours in {\bf (a)} are spaced by 10 \kms\ on the velocity fields,
and 0.1 $\rm mJy\,beam^{-1}$ on the moment 0. The pixel scale in {\bf (b)} is 5 \arcsec. See Appendix section~\ref{A-2} for details.
\label{lvhis15}}
\end{figure*}

%: 25. HIPASS J1057-48.atlas.ps
\begin{figure*} \epsscale{1.0}
\includegraphics[angle=0,width=1.0\textwidth,bb=100 160 520 610,clip=]
{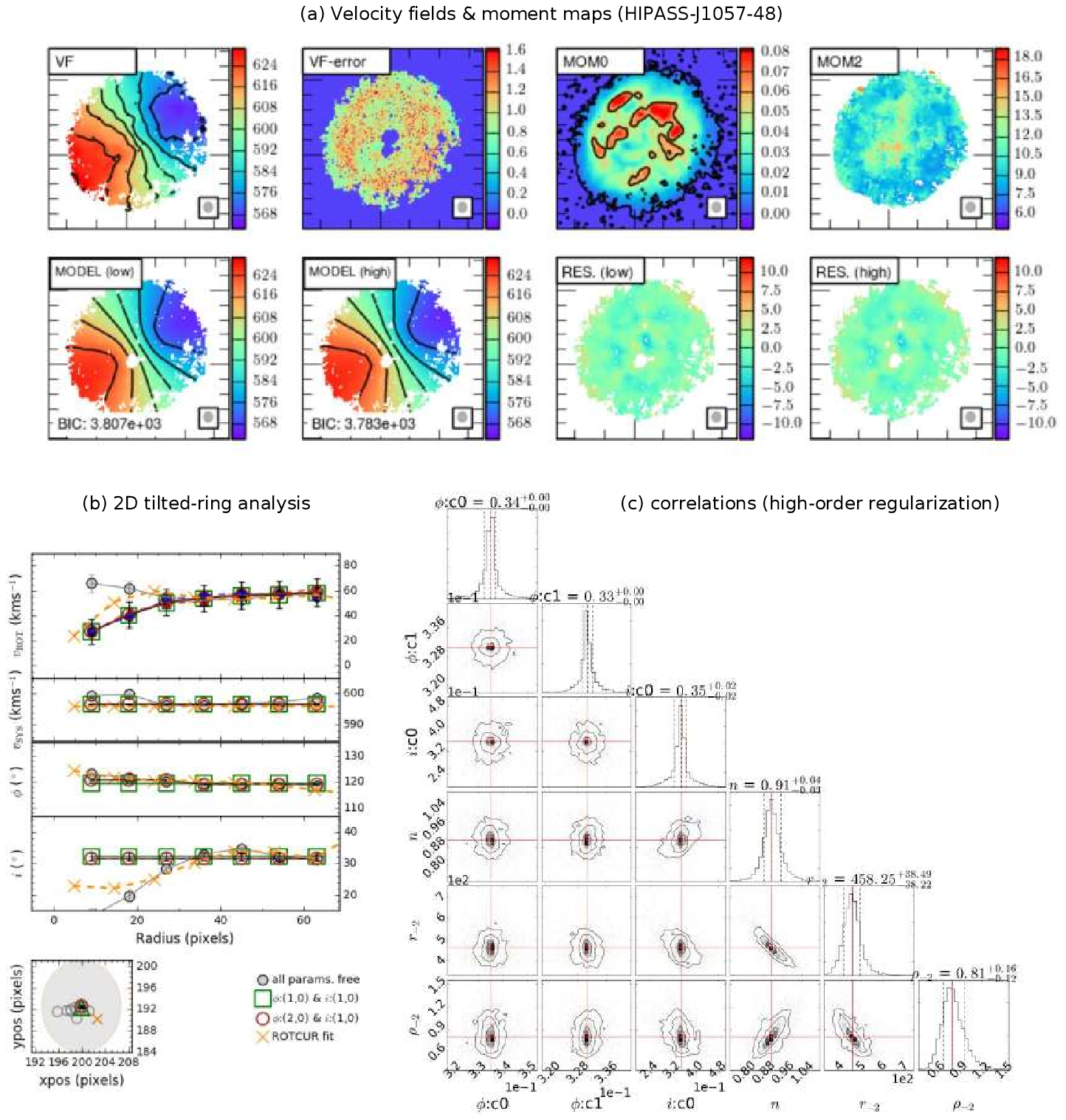}
\caption{{\sc 2dbat} analysis for HIPASS J1057-48:
Contours in {\bf (a)} are spaced by 10 \kms\ on the velocity fields,
and 0.1 $\rm mJy\,beam^{-1}$ on the moment 0. The pixel scale in {\bf (b)} is 5 \arcsec. See Appendix section~\ref{A-2} for details.
\label{lvhis16}}
\end{figure*}

%: 26. HIPASS J1501-48.atlas.ps
\begin{figure*} \epsscale{1.0}
\includegraphics[angle=0,width=1.0\textwidth,bb=100 160 520 610,clip=]
{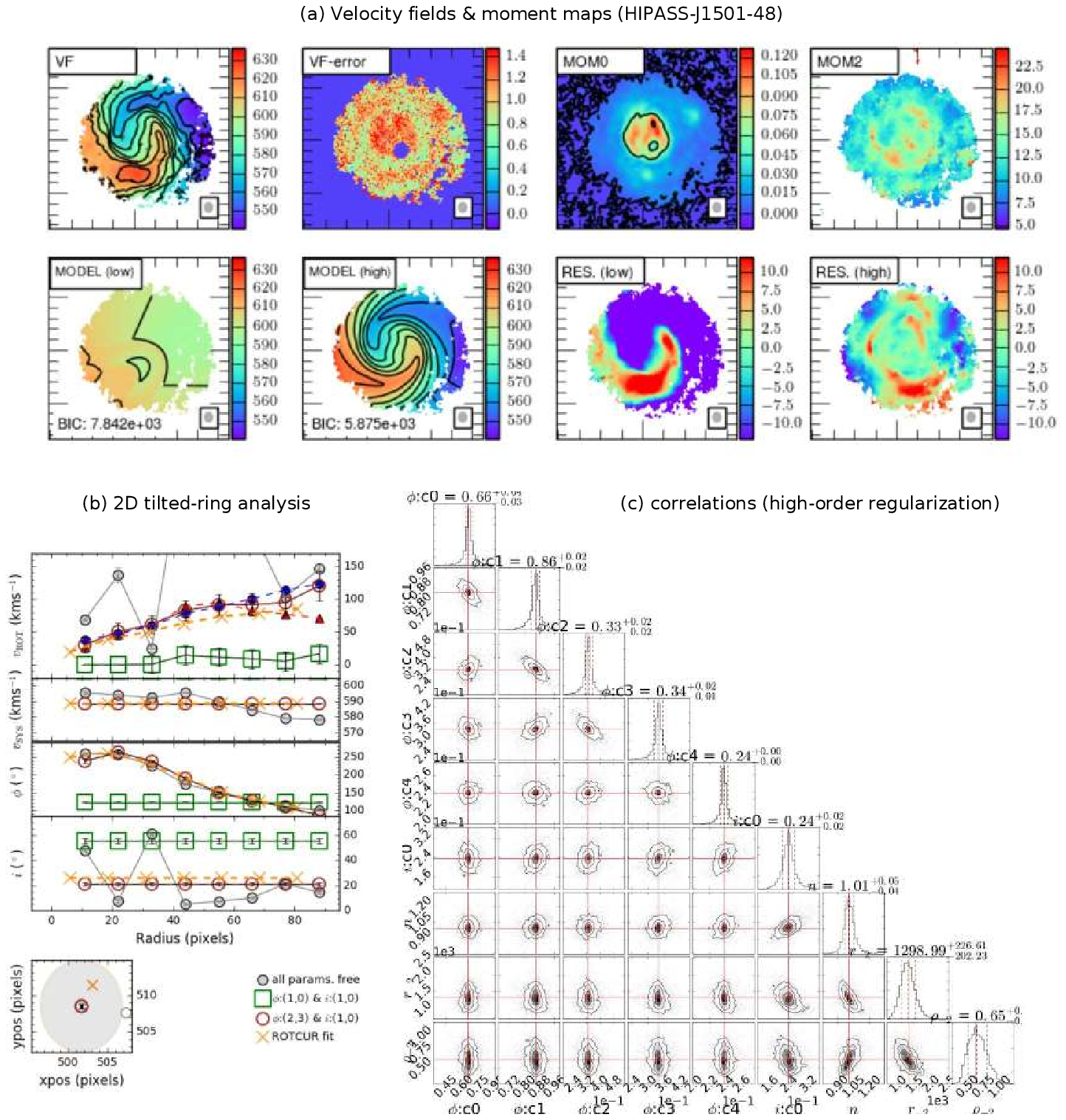}
\caption{{\sc 2dbat} analysis for HIPASS J1501-48:
Contours in {\bf (a)} are spaced by 10 \kms\ on the velocity fields,
and 0.1 $\rm mJy\,beam^{-1}$ on the moment 0. The pixel scale in {\bf (b)} is 5 \arcsec. See Appendix section~\ref{A-2} for details.
\label{lvhis17}}
\end{figure*}

%: 27. HIPASS J2202-51.atlas.ps
\begin{figure*} \epsscale{1.0}
\includegraphics[angle=0,width=1.0\textwidth,bb=100 160 520 610,clip=]
{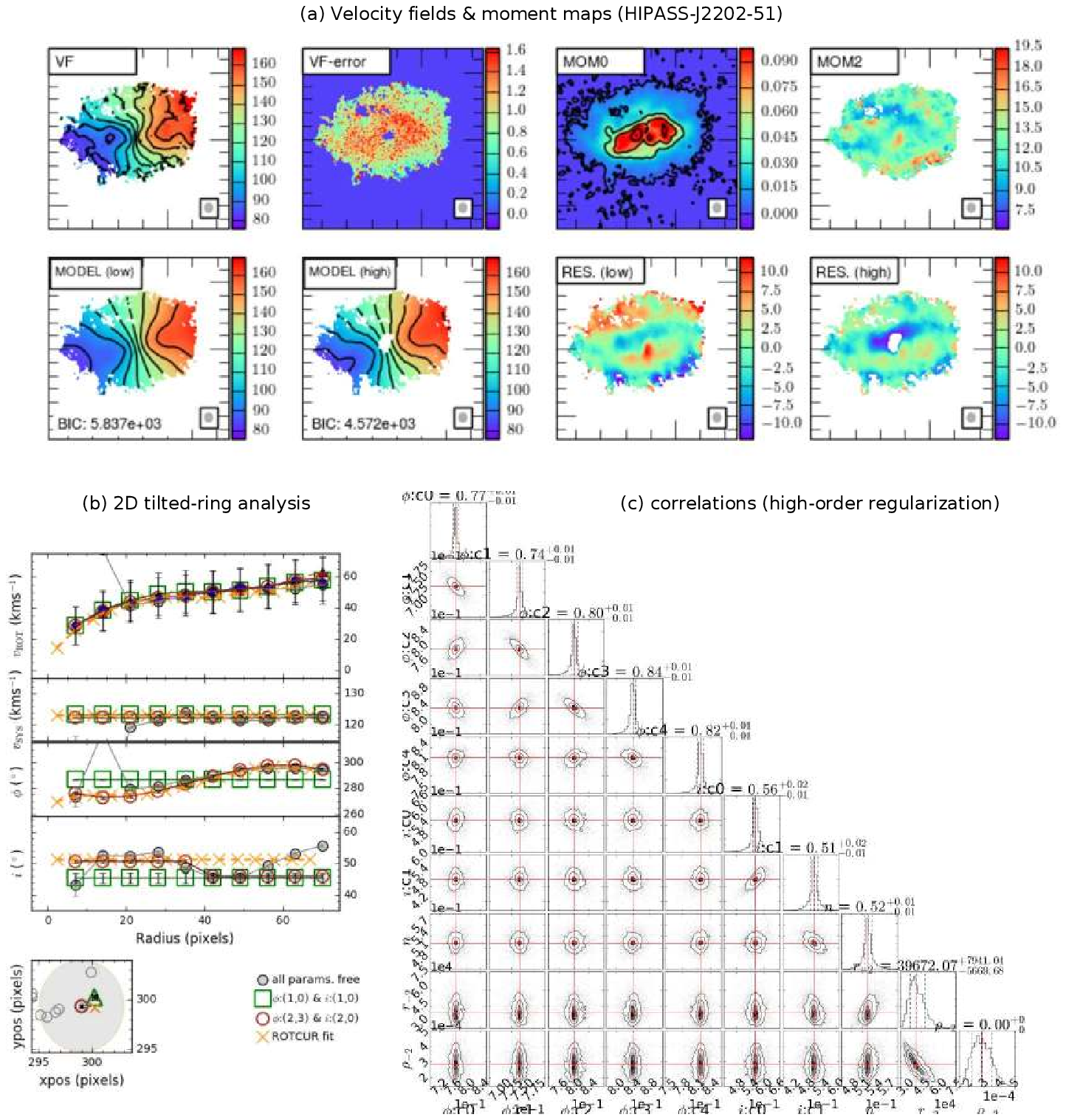}
\caption{{\sc 2dbat} analysis for HIPASS J2202-51:
Contours in {\bf (a)} are spaced by 10 \kms\ on the velocity fields,
and 0.1 $\rm mJy\,beam^{-1}$ on the moment 0. The pixel scale in {\bf (b)} is 5 \arcsec. See Appendix section~\ref{A-2} for details.
\label{lvhis18}}
\end{figure*}

%: 28. HIPASS J0256-54.atlas.ps
\begin{figure*} \epsscale{1.0}
\includegraphics[angle=0,width=1.0\textwidth,bb=100 160 520 610,clip=]
{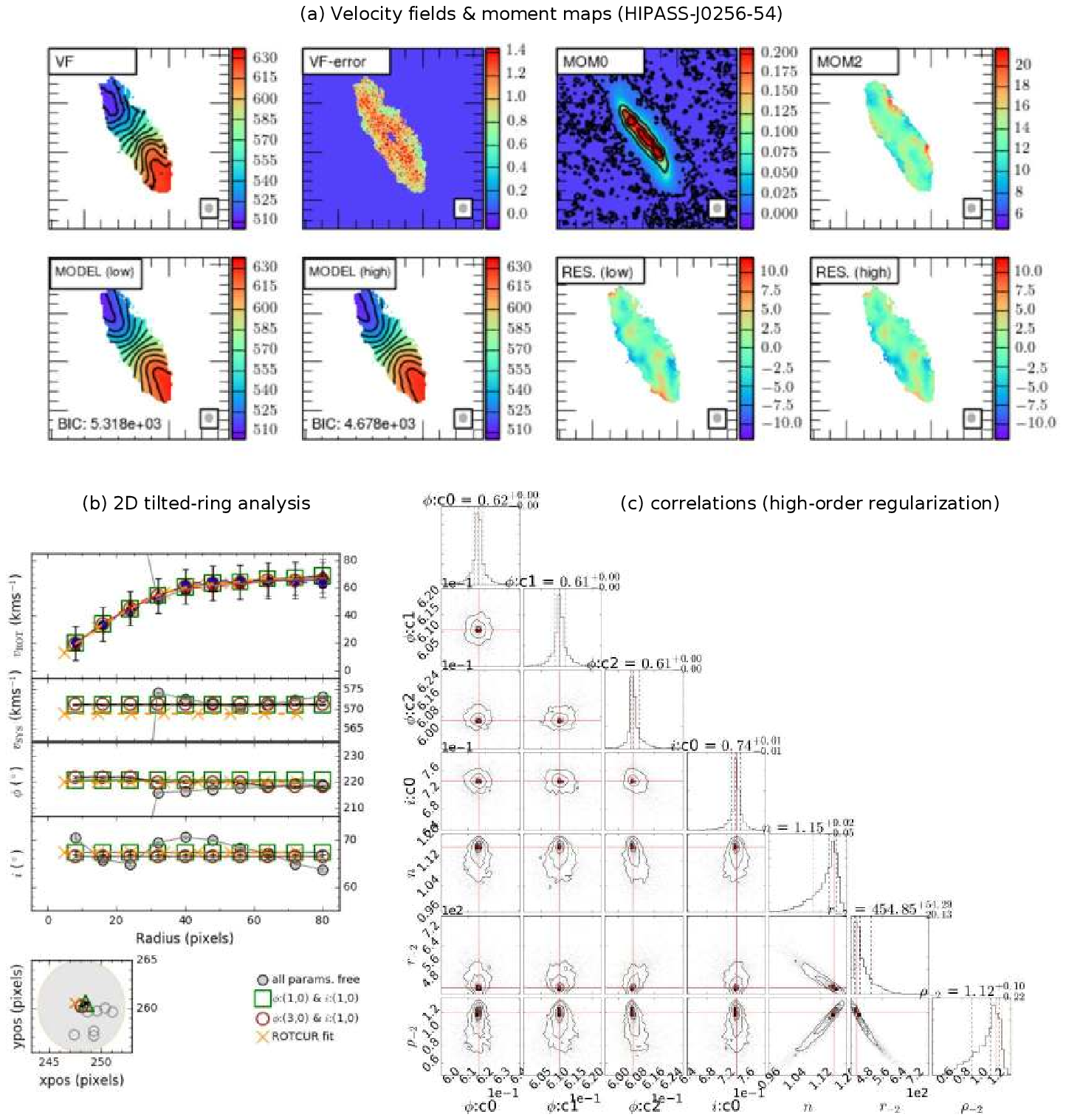}
\caption{{\sc 2dbat} analysis for HIPASS J0256-54:
Contours in {\bf (a)} are spaced by 10 \kms\ on the velocity fields,
and 0.1 $\rm mJy\,beam^{-1}$ on the moment 0. The pixel scale in {\bf (b)} is 6 \arcsec. See Appendix section~\ref{A-2} for details.
\label{lvhis19}}
\end{figure*}

%: 29. HIPASS J1305-49.atlas.ps
\begin{figure*} \epsscale{1.0}
\includegraphics[angle=0,width=1.0\textwidth,bb=100 160 520 610,clip=]
{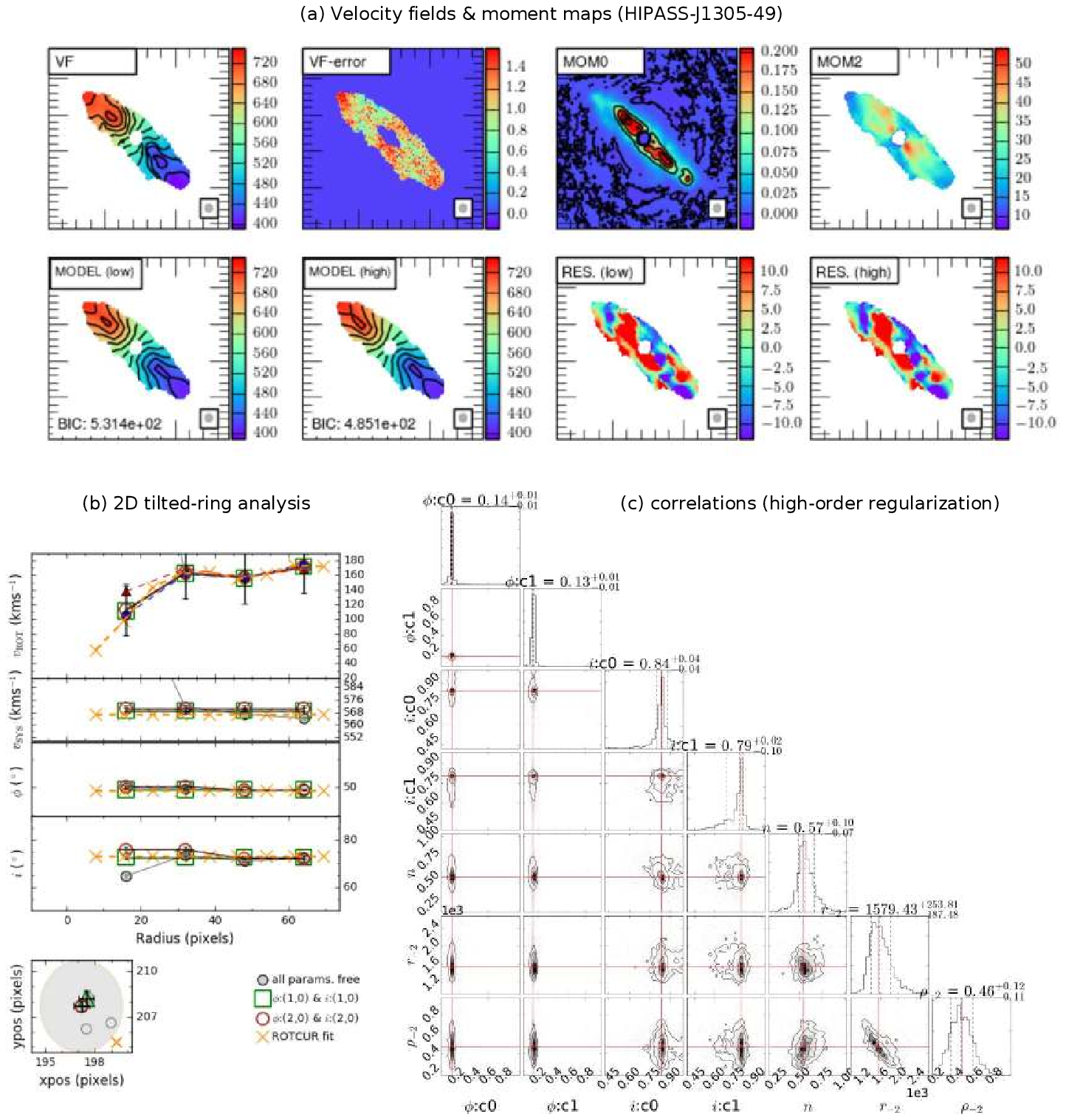}
\caption{{\sc 2dbat} analysis for HIPASS J1305-49:
Contours in {\bf (a)} are spaced by 30 \kms\ on the velocity fields,
and 0.1 $\rm mJy\,beam^{-1}$ on the moment 0. The pixel scale in {\bf (b)} is 10 \arcsec. See Appendix section~\ref{A-2} for details.
\label{lvhis20}}
\end{figure*}

%: 30. HIPASS J1321-36.atlas.ps
\begin{figure*} \epsscale{1.0}
\includegraphics[angle=0,width=1.0\textwidth,bb=100 160 520 610,clip=]
{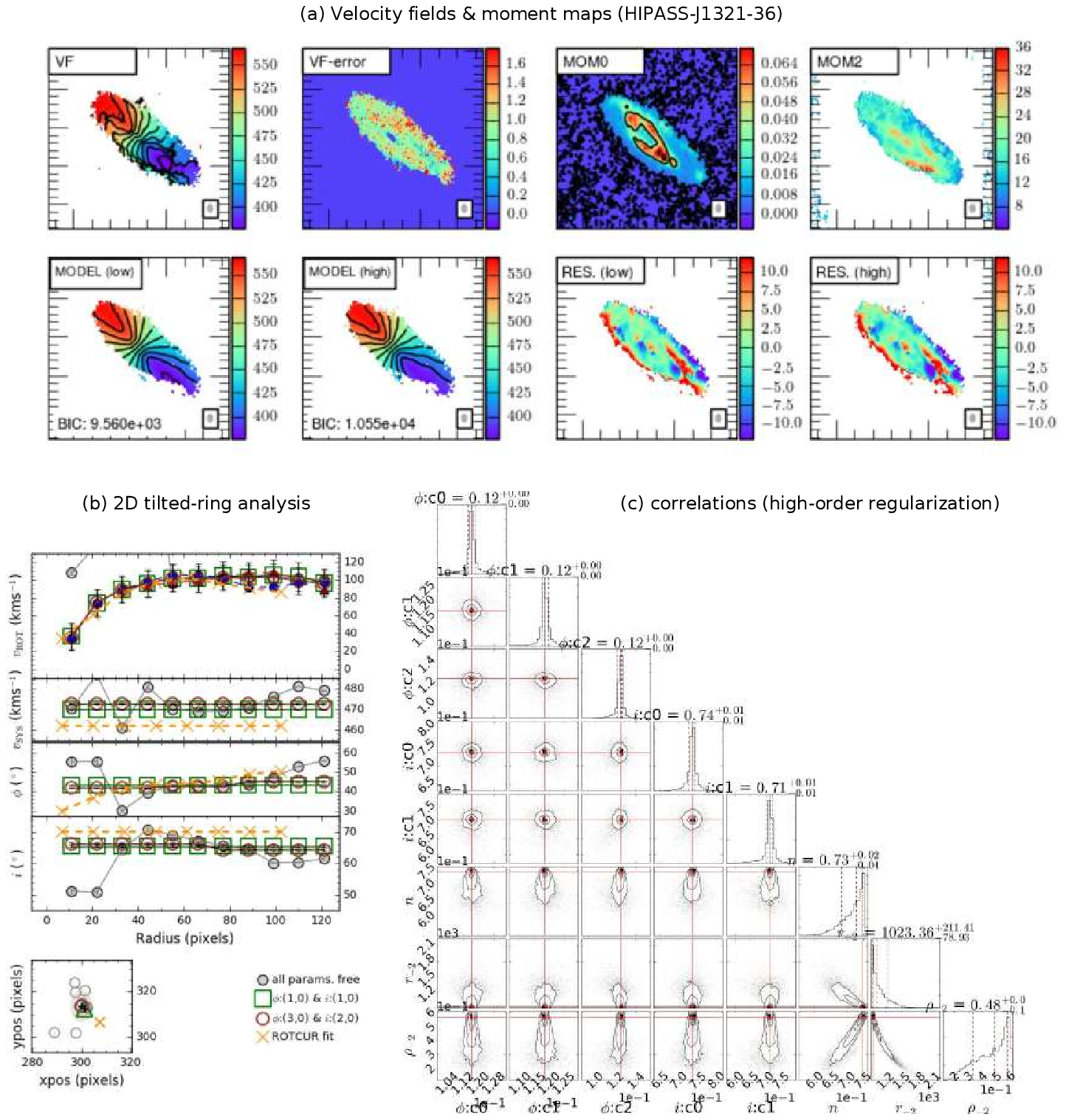}
\caption{{\sc 2dbat} analysis for HIPASS J1321-36:
Contours in {\bf (a)} are spaced by 20 \kms\ on the velocity fields,
and 0.1 $\rm mJy\,beam^{-1}$ on the moment 0. The pixel scale in {\bf (b)} is 5 \arcsec. See Appendix section~\ref{A-2} for details.
\label{lvhis21}}
\end{figure*}

%: 31. HIPASS J0047-25.atlas.ps
\begin{figure*} \epsscale{1.0}
\includegraphics[angle=0,width=1.0\textwidth,bb=100 160 520 610,clip=]
{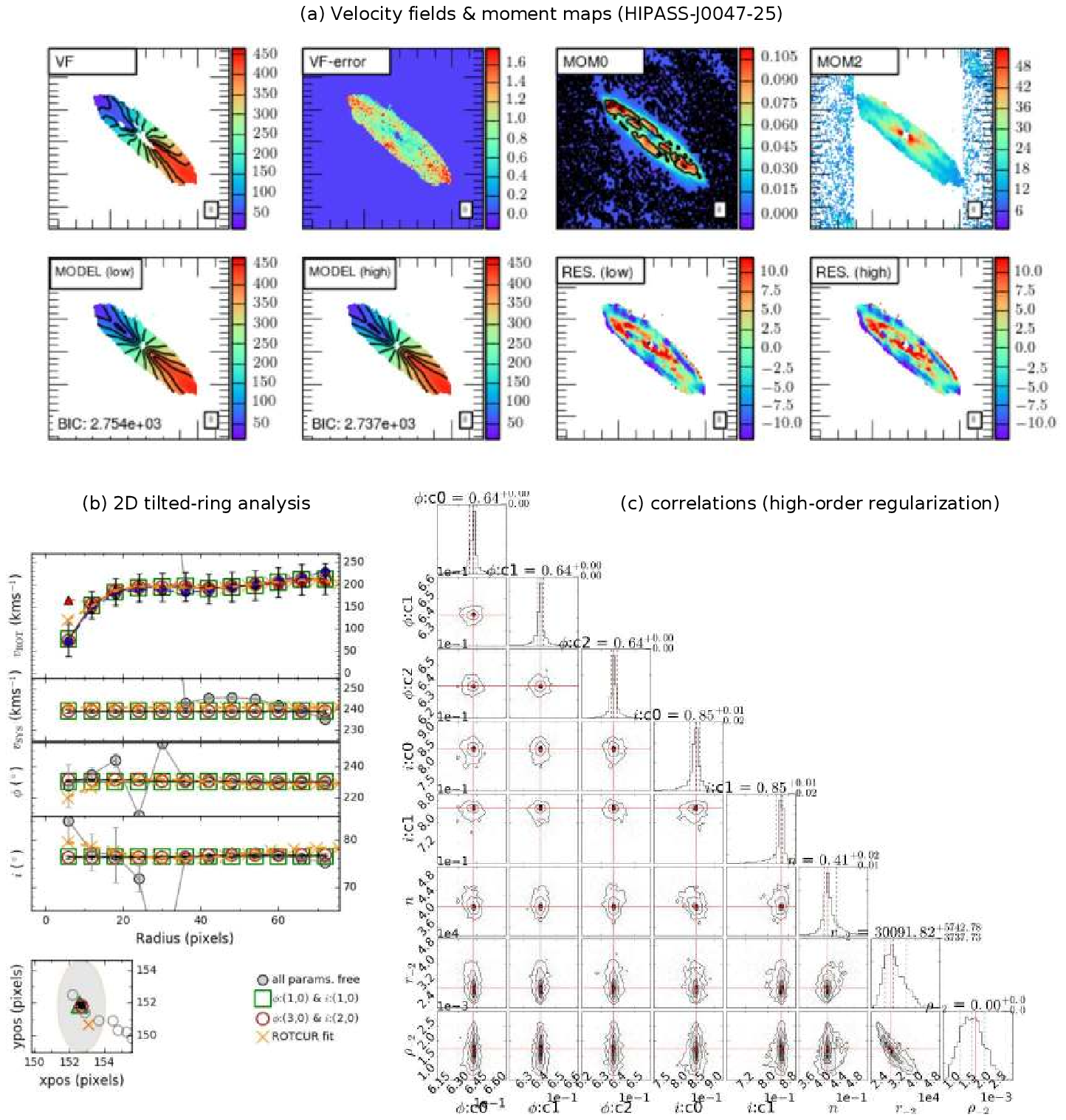}
\caption{{\sc 2dbat} analysis for HIPASS J0047-25:
Contours in {\bf (a)} are spaced by 40 \kms\ on the velocity fields,
and 0.1 $\rm mJy\,beam^{-1}$ on the moment 0. The pixel scale in {\bf (b)} is 10 \arcsec. See Appendix section~\ref{A-2} for details.
\label{lvhis22}}
\end{figure*}

%: 32. HIPASS J1413-65.atlas.ps
\begin{figure*} \epsscale{1.0}
\includegraphics[angle=0,width=1.0\textwidth,bb=100 160 520 610,clip=]
{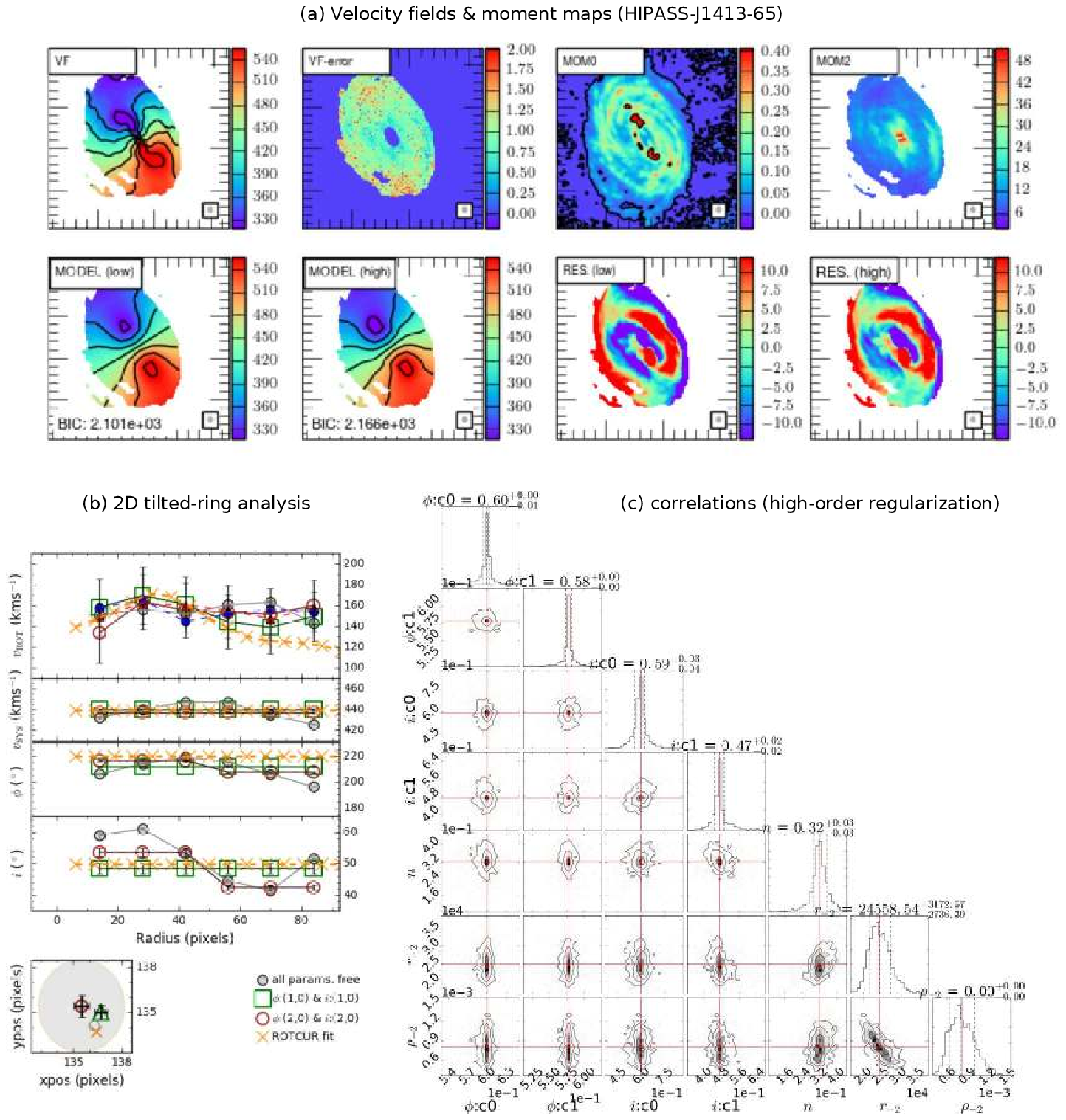}
\caption{{\sc 2dbat} analysis for HIPASS J1413-65:
Contours in {\bf (a)} are spaced by 40 \kms\ on the velocity fields,
and 0.1 $\rm mJy\,beam^{-1}$ on the moment 0. The pixel scale in {\bf (b)} is 20 \arcsec. See Appendix section~\ref{A-2} for details.
\label{lvhis23}}
\end{figure*}

%: 33. HIPASS J0317-66.atlas.ps
\begin{figure*} \epsscale{1.0}
\includegraphics[angle=0,width=1.0\textwidth,bb=100 160 520 610,clip=]
{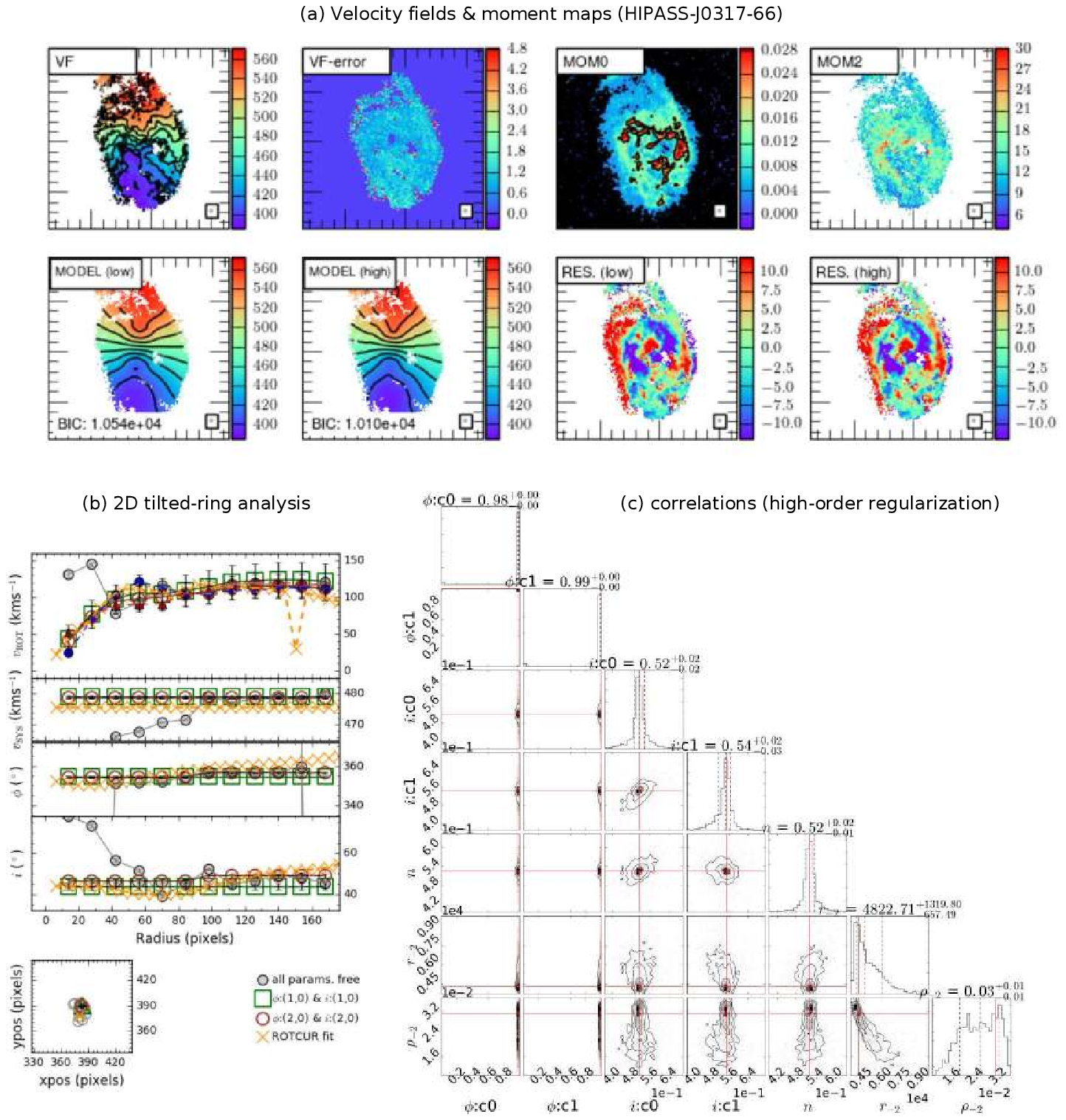}
\caption{{\sc 2dbat} analysis for HIPASS J0317-66:
Contours in {\bf (a)} are spaced by 20 \kms\ on the velocity fields,
and 0.1 $\rm mJy\,beam^{-1}$ on the moment 0. The pixel scale in {\bf (b)} is 2.75 \arcsec. See Appendix section~\ref{A-2} for details.
\label{lvhis24}}
\end{figure*}

\end{document}